\documentclass[12pt]{article}
\pdfoutput=1
\usepackage[nosort]{cite}
\usepackage[normalem]{ulem}

\usepackage[dvipsnames]{xcolor}
\usepackage{epsfig}
\usepackage{amsfonts}
\usepackage{amscd}
\usepackage{latexsym}
\usepackage{amsmath,amssymb}
\usepackage{verbatim}
\usepackage{setspace}
\usepackage{color}
\usepackage{fancyhdr}
\usepackage{cite}
\usepackage{hyperref}
\usepackage{tikz}
\usepackage{slashed}
\usepackage{multirow}
\usetikzlibrary{calc}
\usetikzlibrary{topaths}
\usetikzlibrary{decorations}
\usetikzlibrary{decorations.pathmorphing}
\usetikzlibrary{arrows,decorations.markings,cd}
\usetikzlibrary{calc,arrows,cd,decorations.markings,snakes}
\tikzset{
->-/.style args={#1rotate#2}{decoration={markings, mark=at position #1 with {\arrow[scale=1.5,rotate = #2 ]{stealth}}}, postaction={decorate}}
}
\usetikzlibrary{shapes.geometric}
\usetikzlibrary{knots}
\usepackage{tikz-3dplot}

\usepackage{float}
\usepackage{draft}
\usepackage{hyperref}
\usepackage{graphicx,color,subfig}
\usepackage{cite}
\usepackage{mciteplus}
\usepackage{skak}
\usepackage{bbm}

\usepackage{epsfig}
\usepackage{amsfonts}
\usepackage{amscd}
\usepackage{latexsym}
\usepackage{amsmath,amssymb}
\usepackage{verbatim}
\usepackage{setspace}
\usepackage[dvipsnames]{xcolor}
\usepackage{fancyhdr}
\usepackage{cite}
\usepackage{hyperref}
\usepackage{slashed}
\usepackage{multirow}
   \usepackage{draft}
\usepackage{hyperref}
\usepackage{graphicx,color,subcaption}
\usepackage{cite}
\usepackage{skak}
\usepackage{bbm}
\usepackage[english]{babel}
\usepackage{amsthm}
\usepackage{soul}
\usepackage{upgreek}
\usepackage[scr]{rsfso}

\usepackage{makecell}

\usepackage{tikz}

\numberwithin{equation}{section}

\def\TRR{T}

\def\Todd{T}

\def\D{\mathcal{D}}

\def\R{\mathcal{R}}

\def\T{\mathcal{T}}
\def\Q{\mathcal{Q}}
\def\X{\mathcal{X}}
\def\U{\mathcal{U}}
\def\V{\mathcal{V}}

\def\sC{\mathsf{C}}
\def\sP{\mathsf{P}}
\def\sT{\mathsf{T}}
\def\sR{\mathsf{R}}

\def\sQ{\mathsf{Q}}
\def\sU{\mathsf{U}}
\def\sV{\mathsf{V}}

\def\PP{{\tilde\Pi}}

\def\scP{\mathscr{P}}
\def\scR{\mathscr{R}}
\def\scT{\mathscr{T}}

\colorlet{mylinkcolor}{NavyBlue}
\colorlet{mycitecolor}{Aquamarine}
\colorlet{myurlcolor}{Aquamarine}
\newcommand\myshade{90}

\hypersetup{
  linkcolor  = mylinkcolor!\myshade!black,
  citecolor  = mycitecolor!\myshade!black,
  urlcolor   = myurlcolor!\myshade!black,
  colorlinks = true,
}
 
\begin{document}
\begin{titlepage}

\hfill  MIT-CTP/5902\\

\title{LSM and CPT}

\author{Nathan Seiberg$^\sC$, Shu-Heng Shao$^\sP$, and Wucheng Zhang$^\sT$}

 \address{${}^\sC$ School of Natural Sciences, Institute for Advanced Study, Princeton, NJ}
 \address{${}^\sP$ Center for Theoretical Physics - a Leinweber Institute, Massachusetts Institute of Technology, Cambridge, MA}
 \address{${}^\sT$ Department of Physics, Princeton University, Princeton, NJ}

\abstract{
\noindent We study a number of 1+1d lattice models with anti-unitary symmetries that simultaneously reflect space and reverse time. Some of these symmetries are anomalous, leading to Lieb-Schultz-Mattis-type constraints, thus excluding a trivially gapped phase. Examples include a mod 8 anomaly in the Majorana chain and various mod 2 anomalies in the spin chain.  In some cases, there is an exact, non-anomalous lattice symmetry that flows in the continuum to CPT. In some other cases, the CPT symmetry of the continuum theory is emergent or absent.  Depending on the model, the anomaly of the lattice model is matched in the continuum in different ways.  In particular, it can be mapped to an emergent anomaly of an emanant symmetry.}

 \end{titlepage}
\tableofcontents

\section{Introduction}

This note studies various aspects of CPT-like symmetries on the lattice and their anomalies.  These anomalies provide powerful constraints on the low-energy behavior of the system.

Both the CPT symmetry and anomalies are well-understood.  However, in order to set the stage for our later discussion, we will start the introduction with some general comments about these topics.

\subsection{Introductory comments about 't Hooft anomalies}\label{sec:introanomaly}

Different people follow different definitions of 't Hooft anomalies \cite{tHooft:1979rat}. 
Below is an incomplete list of different notions of anomalies that have appeared in the literature.
\begin{itemize}
    \item The standard definition in the study of continuum quantum field theory involves placing the system on a general Euclidean spacetime with background gauge fields for the internal symmetries and finding that the partition function is not well-defined. Comments:
    \begin{itemize}
    \item Anomalies involving time-reversal or spatial-reflection are probed with a nonorientable spacetime \cite{Witten:2016cio}.
    \item The anomaly is characterized by an ``anomaly theory.'' This is a classical field theory of the background fields in one higher dimension, and the anomalous system resides on its boundary.\footnote{The way the anomaly is characterized by a higher-dimensional field theory goes back to the '80s.  See e.g., \cite{Callan:1984sa} and references therein.}  This classical theory is also referred to as an invertible theory \cite{Freed:2004yc,Freed:2019jzd}.
    \item A special case of it is when the Euclidean spacetime is $\Sigma\times S^1$.  Then, we can view  $\Sigma$ as space and $S^1$ as Euclidean time.  In this case, the partition function is a trace over the Hilbert space of the system on $\Sigma$.  Then, some of the anomalies are manifest as the fact that the operator algebra is realized projectively.
    \item 't Hooft anomaly is often phrased as ``obstruction to gauging.''  Gauging is the result of summing over the background gauge fields, and if the partition function as a function of them is ill-defined, it is clear that the symmetry cannot be gauged.  Such a description in terms of obstruction to gauging is quite confusing for spacetime symmetries like translation, rotation, space-reflection, or time-reversal symmetries.\footnote{In this discussion, we ignore gravitational 't Hooft anomalies.}
    \end{itemize}
    \item Another definition of anomaly is as an obstruction to a symmetric, trivially-gapped phase.  
    While this obstruction is conventionally viewed as a consequence of anomalies, it has been advocated, e.g., in \cite{Thorngren:2019iar,Choi:2023xjw,Chatterjee:2024gje,Shirley:2025yji} that it should be viewed as its defining property.
    A characteristic example of the use of such an obstruction in condensed matter physics is Lieb–Schultz–Mattis (LSM)-like constraints \cite{Lieb:1961fr,Affleck:1986pq,Oshikawa:2000lrt,Hastings:2003zx}.  (See \cite{Affleck:1988nt,Tasaki:2022gka} for reviews.)  
    \item In condensed matter physics, it is common to consider a nontrivial, bulk Symmetry Protected Topological (SPT) phase with an anomalous theory on its boundary.  In this perspective, the focus is on the bulk theory rather than on its anomalous boundary.  This perspective is clearly related to the anomaly theory mentioned above.  One puzzle with this view is that it is not clear why the lattice SPT phase should be characterized by a Lorentz-invariant continuum field theory.
\end{itemize}  
These definitions are related, but they are not obviously the same. 
They are particularly subtle when we discuss lattice models, and the symmetries involved are crystalline symmetries.  This topic has been discussed by many authors, and in particular by \cite{ThorngrenPRX2018}, who put forward an interesting conjecture.  See \cite{Seifnashri:2025vhf,Kapustin:2025nju,Kawagoe:2025ldx,Tu:2025bqf,Shirley:2025yji}, for very recent discussions on the relations between different notions of anomalies on the lattice and in the continuum.  These papers also refer to many earlier papers.

One of the characteristics of the anomaly is its order.  We take $N_f$ copies of the system, labeled by $A=1,\cdots, N_f$.  If the system has an internal symmetry generated by $\sU$, we take $\sU^A$ to act on species $A$ and we impose invariance only under the diagonal symmetry $\sU=\sU^1 \sU^2 \cdots \sU^{N_f}$. Then, the order of the anomaly is the lowest $N_f$ such that the system is anomaly-free. 

In this note, we focus on one-dimensional lattice Hamiltonian systems. The lattice has $L$ sites, and we take time to be continuous. 
In this context, we will use a combination of the perspectives above to constrain the order of the anomaly.
\begin{itemize}
\item Lower-bound on the order.  We simply look for a projective representation of the symmetry algebra. In this discussion, we view the 1+1d system as an effective 0+1d quantum mechanical system where locality does not play any role. 
This can be repeated for various-sized systems, i.e., different values of $L$.  It is often the case that there is an anomaly for some values of $L$, but not for others.  
\item Stronger lower-bound on the order.  We introduce twists by global symmetries, equivalently, background gauge fields for these symmetries, and look for a projective representation in the twisted Hilbert space.  Unlike the previous technique, here, we rely on the locality of the system, and therefore, we can find a stronger bound. In this context, one can view the change in $L$ mentioned above as a ``twist in the translation symmetry.''  However, as reviewed in \cite{Cheng:2015kce,Cho:2017fgz,Metlitski:2017fmd,ThorngrenPRX2018,Manjunath2020, SongPRR2021, Manjunath2021,Ye:2021nop,Cheng:2022sgb,Seiberg:2023cdc,Seifnashri:2023dpa,Seiberg:2024gek}, this interpretation is quite subtle.
\item Upper-bound on the order.  We take several copies of the system and find an interaction that preserves the symmetries and leads to a trivially gapped state.  
\end{itemize}

Below, we will use these techniques and refer to them as the lower-bound and the upper-bound.

\subsection{Introductory comments about CPT of the continuum theory}\label{sec:introCPT}

We will focus on anomalies involving space-reflection and time-reversal.  In the continuum, we will denote these operations as $\sR$ and $\sT$ respectively.  Combining them with charge conjugation $\sC$, the famous CPT theorem states that every Lorentz invariant field theory is invariant under
\ie\label{Thetadefi}
\Theta=\sC\sR\sT\,.
\fe
In Appendix \ref{CPTreview}, we will review the CPT theorem.\footnote{Following \cite{Witten:2016cio}, we refer to the theorem as the CPT theorem, but denote the transformation as $\Theta=\sC\sR\sT$ and refer to it as the CRT transformation, i.e., we use $\sR$ for reflection in one spatial coordinate.  This point is not essential in our 1+1d examples, where parity and spatial reflection are the same.  See Appendix \ref{CPTreview} for a more detailed discussion.}  

Importantly,  $\sC$, $\sR$, and $\sT$ might not be symmetries of the problem.  And even if they are, there is freedom in how they are defined.  But in a relativistic theory, the combination $\Theta$ in \eqref{Thetadefi} is always present, and it is canonical.  
For both bosonic and fermionic theories, it satisfies
\ie
\Theta^2=1\,,
\fe
and it is anomaly-free.
As we will review in Appendix \ref{CPTreview}, in Euclidean signature, the canonical $\Theta$ corresponds to $\pi$-rotation in spacetime.  Consequently, unlike $\sR$ and $\sT$, the symmetry $\Theta$ is orientation preserving.  

Below, we will use two important facts about the canonical symmetry operator $\Theta$.  First, it commutes with all the internal symmetry operators $\sU$ of the problem
\ie\label{ThetaUc}
\sU\Theta=\Theta \sU\,.
\fe
Note that this relation prevents us from redefining $\sU$ by a phase. 
Second, while the symmetry generated by $\Theta$ is anomaly-free, in some cases, it can exhibit interesting anomalies involving other symmetries.  

Consider, for example, a $\bZ_2^\sU$ global symmetry in a bosonic 1+1d system.  The pure $\bZ_2^\sU$ anomaly is classified by $H^3(\mathbb{Z}^\sU_2,U(1))=\bZ_2$ \cite{Freed:1987qk}.  One way to detect this anomaly is to consider a $\sU$ defect.  We denote the global symmetry transformation $\sU$ in the presence of this defect by $\sU_\sU$.  Then, the pure $\bZ_2^\sU$ anomaly is characterized by the anomalous phase in \cite{Chang:2018iay,Lin:2019kpn}
\ie\label{Uanomal}
\sU_\sU^2=\epsilon\qquad, \qquad \epsilon=\pm 1\,.
\fe
The value of $\epsilon=\pm 1$ is determined by $H^3(\mathbb{Z}^\sU_2,U(1))=\bZ_2$.

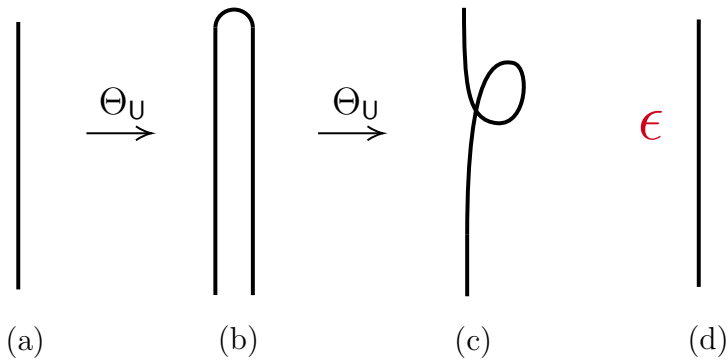
\begin{figure}[t]
    \centering
    \tikzset{every picture/.style={line width=0.75pt}} 

\begin{tikzpicture}[x=0.75pt,y=0.75pt,yscale=-0.8,xscale=0.8]

\draw [line width=1.5]    (16,16.78) -- (16,185.51) ;
\draw [line width=1.5]    (140.26,20.32) -- (140.26,189.05) ;
\draw [line width=1.5]    (163.89,20.38) -- (163.89,189.11) ;
\draw  [draw opacity=0][line width=1.5]  (140.26,20.32) .. controls (140.31,14.02) and (145.62,8.93) .. (152.12,8.95) .. controls (158.63,8.97) and (163.89,14.08) .. (163.89,20.38) -- (152.08,20.4) -- cycle ; \draw  [line width=1.5]  (140.26,20.32) .. controls (140.31,14.02) and (145.62,8.93) .. (152.12,8.95) .. controls (158.63,8.97) and (163.89,14.08) .. (163.89,20.38) ;  
\draw [line width=1.5]    (299,151.07) -- (299,189.84) ;
\draw [line width=1.5]    (299,151.07) .. controls (299,45.04) and (317,40.29) .. (328,42.66) .. controls (339,45.04) and (337,82.23) .. (318,80.64) .. controls (299,79.06) and (297,56.9) .. (297,7.84) ;
\draw [line width=1.5]    (445,15.2) -- (445,183.93) ;
\draw    (59,87) -- (98,87) ;
\draw [shift={(100,87)}, rotate = 180] [color={rgb, 255:red, 0; green, 0; blue, 0 }  ][line width=0.75]    (9.84,-4.41) .. controls (6.25,-2.07) and (2.97,-0.6) .. (0,0) .. controls (2.97,0.6) and (6.25,2.07) .. (9.84,4.41)   ;
\draw    (205,87) -- (244,87) ;
\draw [shift={(246,87)}, rotate = 180] [color={rgb, 255:red, 0; green, 0; blue, 0 }  ][line width=0.75]    (9.84,-4.41) .. controls (6.25,-2.07) and (2.97,-0.6) .. (0,0) .. controls (2.97,0.6) and (6.25,2.07) .. (9.84,4.41)   ;
\draw [draw opacity=0]   (8,245) -- (468,245) ;

\draw (404.67,71.02) node [anchor=north west][inner sep=0.75pt]  [font=\Huge,color={rgb, 255:red, 208; green, 2; blue, 27 }  ,opacity=1 ]  {$\epsilon $};
\draw (6,207) node [anchor=north west][inner sep=0.75pt]   [align=left] {(a)};
\draw (140,206) node [anchor=north west][inner sep=0.75pt]   [align=left] {(b)};
\draw (289,207) node [anchor=north west][inner sep=0.75pt]   [align=left] {(c)};
\draw (436,206) node [anchor=north west][inner sep=0.75pt]   [align=left] {(d)};
\draw (65,56.4) node [anchor=north west][inner sep=0.75pt]  [font=\large]  {$\Theta _{\mathsf{U}}$};
\draw (212,56.4) node [anchor=north west][inner sep=0.75pt]  [font=\large]  {$\Theta _{\mathsf{U}}$};

\end{tikzpicture}
    \caption{A pictorial way to compute  $\Theta_\sU^2$.  In panel (a), we show a $\bZ_2^\sU$ defect.   Applying $\Theta_\sU$, which reverses space and time, leads to the defect in panel (b). Applying $\Theta_\sU$ again leads to the defect in panel (c). Alternatively, $\Theta_\sU^2$, which is a $2\pi$-rotation in Euclidean spacetime, maps the configuration in panel (a) directly to the configuration in panel (c). Certain topological manipulations of the defect, such as those described in \cite{Bhardwaj:2017xup,Chang:2018iay,Lin:2019kpn}, bring the defect back to a straight line, with a factor of $\epsilon = \pm1$, reflecting the anomaly.  A precise discussion of this manipulation can depend on a possible counterterm at the intersection point. }
\label{Thetasquare}
\end{figure}

Since $\sU$ is of order 2, the system with the defect still has a $\Theta$ symmetry.  
However, the operator $\Theta$ might be modified due to the presence of a $\sU$ defect.  We denote the modified operator as  $\Theta_\sU$. 
The discussion in \cite{Hason:2020yqf,Cordova:2019wpi} essentially means that even though in the absence of defects, $\Theta^2=1$, in the presence of a $\sU$ defect, $\Theta_\sU$ satisfies
\ie\label{ThetaUsq}
\Theta_\sU^2=\epsilon\,.
\fe
In particular, a nontrivial anomaly of $\sU$, which is characterized by $\epsilon=-1$ in \eqref{Uanomal}, leads to an anomalous phase  in the algebra of $\Theta_\sU$ \eqref{ThetaUsq}. 
This algebra implies that states in the $\mathbb{Z}_2^\sU$ twisted Hilbert space are in Kramers' doublets.  See Figure \ref{Thetasquare} for a presentation of the computation of $\Theta_\sU^2$.\footnote{It is instructive to compare this discussion with \cite{Hason:2020yqf,Cordova:2019wpi}.  The authors of \cite{Cordova:2019wpi} study a conformal field theory (CFT) with a topological defect, which is constrained by conformal symmetry and modular covariance.  On the other hand, the authors of \cite{Hason:2020yqf} study a broken gapped phase with a domain wall in non-compact space.  Since the system is gapped, they can discuss the effective theory on the co-dimension one wall.  Also, unlike \cite{Hason:2020yqf} and our discussion, the symmetry operators are not modified from their form without the wall. To relate the setup with the wall to ours, one can compactify the infinite space and add a defect.  Then, the existence of the added defect forces a modification of the symmetry operators.  Now, the domain wall is the lowest energy state in the twisted Hilbert space of the system in the broken phase.}

One might be nervous that this discussion depends on a possible counterterm at the crossing point in Figure \ref{Thetasquare}.  In the lattice analysis below, this counterterm is fixed, and we will determine the counterpart of the projective algebra \eqref{ThetaUsq}.  In particular, the projective algebra \eqref{ThetaUsq} is unambiguous.
We will also see how similar expressions can be used to probe lattice anomalies and the continuum limit of different lattice systems.

\subsection{Going to the lattice}

Our goal is to start with a lattice model and relate it to a continuum description.  To do that, we should first identify the lattice counterparts of the continuum transformations discussed above.

Even though they might not be symmetries of our system, we will consider a unitary operator $\R$ that acts as a site-centered reflection and an anti-unitary operator $\T$ that acts as time-reversal.  Therefore, in the continuum limit,
\ie\label{RTflow}
\R\to \sR\qquad, \qquad \T\to \sT\,.
\fe

Unlike a continuum relativistic system, which must have a symmetry operator $\Theta$, the lattice model does not have to have such a symmetry.  Yet, in some of our examples, we will present a lattice symmetry transformation $\Pi$ that reverses the spatial and the time direction\footnote {The reason for using the notation $\Pi$ is that in the continuum, this symmetry is a $\pi$-rotation in Euclidean spacetime.} and flows in the continuum limit to $\Theta$ 
\ie\label{Piflow}
\Pi \to \Theta\,.
\fe
In some of the examples, $\Pi$ reflects space as a site-centered reflection and includes a factor of $\R\T$.  In other examples, it acts as a link-centered reflection $T\R$ with $T$ the lattice translation and includes a factor of $T\R\T$. 

The 1+1d lattice examples we study below are:
\begin{itemize}
    \item The Majorana chain, which flows to the free Majorana CFT  in Section \ref{sec:maj}.
   \item The Heisenberg spin chain in its anti-ferromagnetic and its ferromagnetic phases in Section \ref{Spinchain}.
    \item The Levin-Gu edge model \cite{Levin:2012yb} in Section \ref{Levin-Gu}. 
   \item The Ising model in Section \ref{sec:noninvertible}.
 
\end{itemize}

The presence of a symmetry like $\Pi$ and its role in various anomalies are different in the various models.

As we will see, some of these models have a lattice symmetry $\Pi$ that flows to $\Theta$.  Other models have such a $\Pi$ that flows to $\Theta$ in some continuum limit, but not in others.

Other models do not have such a $\Pi$ symmetry, but do have another anti-unitary symmetry $\PP$ that includes $\R\T$.  For example, in the Majorana chain, we can take $\PP=T^{-1}\Pi$ with $T$ the translation operator.  Then, we can break many of the other symmetries and preserve only $\PP$.

In all of these cases, we can look for lattice anomalies involving $\Pi$ or $\PP$, perhaps with other symmetries, e.g., lattice translations generated by $T$.  These anomalies lead to LSM-type constraints on the low-energy theory.  In particular, a nonzero lattice anomaly means that the long-distance physics cannot be trivially gapped.

\subsection{Back to the continuum: emanant symmetries and emergent anomalies}\label{sec:emanaandem}

The lattice anomalies are more powerful than merely stating that the IR theory cannot be trivially gapped.  They also constrain the IR theory to have the same anomalies as the lattice model.  This matching is particularly interesting when the lattice crystalline symmetries, e.g., lattice translations, are realized differently in the continuum.

In some cases, the discrete lattice translation symmetry, generated by $T$, leads to an emanant discrete internal symmetry \cite{Cheng:2022sgb}.  Normally, the $\bZ$ translation symmetry of the infinite chain (or the $\bZ_L$ symmetry of the finite chain with $L$ sites) leads to a continuous $\mathbb R$ (or $U(1)$) translation symmetry.  However, in some cases, in the continuum limit, the eigenvalues of the lattice operator $T$ do not approach one.  Instead, they approach an integer powers of $e^{2\pi i\over n}$ with some integer $n$.  In these cases, the IR theory has a $\bZ_n$ internal symmetry.  Unlike ordinary emergent internal symmetries, this internal symmetry is not violated by any irrelevant operator.  It is an emanant symmetry \cite{Cheng:2022sgb}.

This emanant $\bZ_n$ symmetry is not a subgroup of the exact $\bZ$ (or $\bZ_L$) lattice translation symmetry.  Instead, it is a quotient of it.  Therefore, the IR theory can be placed in a background $\bZ_n$ gauge field that does not correspond to any background of the underlying UV lattice theory.  

Thus, the  emanant symmetry in the IR  may have a nontrivial ’t Hooft anomaly even when the corresponding UV symmetry is anomaly-free. 
In such cases, although the symmetry itself is emanant, its anomaly emerges out of nowhere. 
Such anomalies are therefore referred to as emergent anomalies \cite{Metlitski:2017fmd}.  See also \cite{Wang:2017loc,Thorngren:2020wet,Seiberg:2025bqy}.

In some of the examples below, lattice translation leads in the continuum to an emanant internal $\bZ_2$ symmetry (corresponding to $n=2$ above).  This $\bZ_2$ symmetry has an emergent mod 2 or mod 8 anomaly.  
Interestingly, this emergent anomaly can be identified on the lattice as an exact anomaly involving $\Pi$.

\subsection{Outline}

In Section \ref{sec:maj}, we study the Majorana chain with a single Majorana fermion $\chi_\ell$ on every site $\ell$.  
We discuss two anti-unitary symmetries. The first one, $\Pi$, acts as $\chi_\ell \to \chi_{-\ell+1}$.  It is anomaly-free and flows in the continuum to $\Theta$. 
The second one, $\PP=T^{-1}\Pi$, acts as $\chi_\ell \to \chi_{-\ell}$.  It has a mod 8 anomaly even if all the other symmetries except fermion parity $(-1)^F$ are broken. In the continuum limit, the model has an emanant $\bZ_2$ symmetry, which we will denote as $\sC$.  (This symmetry is often denoted as  $(-1)^{F_\text{L}}$.)  The lattice mod 8 anomaly leads to an emergent anomaly for this internal symmetry.

In Section \ref{Spinchain}, we examine a bosonic system, the Heisenberg chain.  For anti-ferromagnetic couplings, we identify a lattice symmetry $\Pi$ that flows to $\Theta$.  As in the Majorana chain, we also discuss another anti-unitary symmetry $\PP$ with interesting anomalies.  Again, in the continuum of the anti-ferromagnetic system, we find an emanant $\bZ_2$ internal symmetry with an emergent anomaly, reflecting the anomalies of the lattice model. The lattice anomalies are independent of the details of the Hamiltonian.  In particular, they are the same in the ferromagnetic phase of the model.  Here, there is no emanant $\bZ_2$ symmetry and the same lattice anomalies are realized differently in the continuum theory.

In Section \ref{Levin-Gu}, we discuss the 1+1d Levin-Gu edge model on a spin chain.  Again, we have a lattice symmetry $\Pi$ that flows to $\Theta$ and another interesting lattice symmetry $\PP$ with various anomalies.  Unlike the Heisenberg chain, here the unitary transformation $\Pi\PP^{-1}$ is an anomalous internal $\bZ_2 $ symmetry on the lattice. 

Section \ref{sec:noninvertible} is devoted to the Ising chain.  Here we show that the non-invertible Kramers-Wannier duality symmetry \cite{Grimm:1992ni,Oshikawa:1996dj,Petkova:2000ip,Frohlich:2004ef,Aasen:2016dop,Chang:2018iay,Tantivasadakarn:2021vel,Lootens:2021tet,Seiberg:2023cdc,Seiberg:2024gek,10.1063/pt.fspd.veje} leads to a non-invertible  RT symmetry.

In a number of appendices, we review some known facts and provide additional technical details.  In Appendix \ref{CPTreview}, we review the famous CPT theorem, and in Appendix \ref{QMA}, we review the time reversal anomaly in fermionic quantum mechanics, which is important in Section \ref{sec:maj}.  Appendices \ref{continuumMajorana} and \ref{appsyman} summarize the results of the Majorana system in the continuum and the lattice, respectively.  These results are used in  Section \ref{sec:maj}.

In Appendix \ref{MZM}, we present an analysis of Majorana zero modes in the continuum, which is related to the discussion of the Smith isomorphism in Section \ref{SmithFermions}.  And in Appendix \ref{app:boson}, we relate the action of the lattice $\Pi$ and the continuum $\Theta$ in the Heisenberg chain of Section \ref{Spinchain}.

\section{Majorana chain}\label{sec:maj}

In this section, we discuss an anomaly involving lattice translation and a lattice CRT symmetry on the 1+1d Majorana chain.

\subsection{Lattice translation and CRT}\label{sec:majsym}

In this subsection, we consider Majorana fermions and some transformations acting on them.  We do not pick a particular Hamiltonian, and we do not refer to these transformations as symmetries.

We consider a periodic 1d lattice with $L$ sites labeled by $\ell\sim \ell+L$. 
On each site, there are $N_f$  Majorana fermion $\chi_\ell^A$ with $A=1,2,\cdots,N_f$.  
They obey the Clifford algebra 
\ie
\{ \chi_\ell^A ,\chi_{\ell'}^B\}  = 2\delta_{\ell\ell'}\delta_{AB}\,.
\fe
For most of the discussion, we will assume periodic boundary conditions, i.e., $\chi_{\ell+L}^A = \chi_\ell^A$. 

In this section, we assume $N_fL$ to be even, so that the total number of fermions is even.\footnote{There are various subtleties associated with having an odd number of fermions. See e.g., \cite{Stanford:2019vob,Delmastro:2021xox,Witten:2023snr,Seiberg:2023cdc,Freed:2024apc,Harlow:2025cqc} for recent discussions. In particular, for odd $N_fL$, the fermion parity is an outer automorphism, and there is no unitary operator on the Hilbert space implementing this symmetry transformation. As a result, the Hilbert space is not graded.  Also, the standard relation between this Hilbert space and the path integral is not maintained.  In our case, when $N_f=3,7$ mod 8 and $L$ is odd, the anti-unitary symmetry $\Pi$ defined below is also an outer automorphism. See Appendices \ref{QMA} and \ref{app:oddL}.\label{ft:oddfermion}} 
In this case, we can define the fermion parity operator:
\ie\label{FP}
 (-1)^F = 
i^{N_fL \over2} (\chi^1_1 \chi^1_2\cdots \chi^1_L)\cdots(\chi^{N_f}_1 \chi^{N_f}_2\cdots \chi^{N_f}_L)
\fe
which flips the sign of all the fermions: 
\ie\label{minusFo}
(-1)^F \chi^A_\ell (-1)^F = - \chi^A_\ell \,.
\fe
The overall phase ensures that $[(-1)^F]^2=1$. 

We also define three crystalline operators: lattice translation $T$, 
\ie
T \chi_\ell^A T^{-1} = \chi_{\ell+1}^A\, \label{eq:tranLAT}
\fe
site-centered reflection $\R$, 
\ie
\R \chi_\ell^A \R^{-1} = (-1)^\ell \chi_{-\ell}^A\,.  \label{eq:Paction}
\fe
and anti-unitary time-reversal $\T$,
\ie
\T \chi_\ell^A \T^{-1} = (-1)^\ell \chi_\ell^A\,, \quad \T i \T^{-1} = -i\,. \label{eq:Taction}
\fe 
As a Heisenberg operator, the time dependence of $\chi_\ell^A$ changes sign. 
In Appendix \ref{appsyman}, we present the explicit expression for these operators and their algebra.\footnote{When $L$ is odd, the actions of $\R$ and $\T$ are not compatible with the boundary conditions, but the combination $\R\T$ respects them.}

We will be particularly interested in the anti-unitary operator 
\ie
\Pi = T\R\T\,,
\fe
which reflects the spatial coordinate around a link and reflects  the time coordinate as well
\ie
\Pi \, \chi_\ell^A \,\Pi^{-1} = \chi_{-\ell+1}^A\,,~~~
\Pi \, i\, \Pi^{-1} =- i\,.
\fe
We will argue in Section \ref{sec:majcont} that for particular Hamiltonians, this $\Pi$ is the lattice version of the celebrated CRT symmetry of unitary, Lorentz-invariant theories, which is reviewed in Appendix \ref{CPTreview}. 
For brevity, we will refer to $\Pi$ as the lattice CRT operator.

We will focus on the algebra generated by the lattice translation $T$, the lattice CRT operator $\Pi$, and the fermion parity operator $(-1)^F$. 
When $L$ is even, they obey
\ie\label{eq:evenLTPi}
&[(-1)^F]^2=1\,,~~~~&&T^L=1\,,~~~~&&\Pi^2=1\,,\\
&(-1)^F \Pi  = \Pi (-1)^F\,,
~~~~&&T(-1)^F=(-1)^{N_f}(-1)^FT\,,~~~~
&&\Pi \, T = (-1)^{\frac{N_f(N_f-1)}{2}}T^{-1} \,\Pi\,.
\fe
When $L$ is odd (and $N_f$ is even), they obey
\ie\label{odd L algebra} 
&[(-1)^F]^2=1\,,~~~~&&T^L=1\,,~~~~
&&\Pi^2 = (-1)^{\frac{N_f(N_f-2)}{8}},\\
&(-1)^F\Pi  = (-1)^{\frac{N_f}{2}}  \Pi (-1)^F,
&&T(-1)^F=(-1)^FT\,,\quad &&\Pi T = T^{-1}\Pi\,.
\fe
The important phases above do not follow merely from the action of these operators on other operators \eqref{minusFo} - \eqref{eq:Taction}.  Instead, they are determined by the projective action of these operators on the Hilbert space.
See Appendix \ref{appsyman} for more details.

Finally, we will also be interested in the operator
\ie \label{Thetadef}
\PP=T^{-1}\Pi=\R\T\,,
\fe 
which reflects the spatial coordinate around a site and reflects  the time coordinate as well
\ie 
\PP \chi_\ell^A \PP^{-1} = \chi_{-\ell}^A\,,\qquad \PP i \PP^{-1}=-i\,. 
\fe 
For even $L$, it obeys the algebra
\ie \label{even L algebraT} 
\PP^2=(-1)^{\frac{N_f(N_f-1)}{2}}\,, \quad\quad(-1)^F \PP=(-1)^{N_f} \PP(-1)^F\,, \quad\quad \PP T=(-1)^{\frac{N_f(N_f-1)}{2}} T^{-1} \PP\,.
\fe 
For odd $L$ (and even $N_f$),  $\PP$ and $\Pi$ are unitarily equivalent, $\PP=T^{L-1\over2} \Pi T^{- {L-1\over2}}$ and therefore, $\PP$ satisfies the same algebra as $\Pi$ in \eqref{odd L algebra}
\ie\label{odd L algebraT} 
&\PP^2 = (-1)^{\frac{N_f(N_f-2)}{8}}\,,
&\quad(-1)^F\PP  = (-1)^{\frac{N_f}{2}}  \PP (-1)^F,
\quad &&\PP T = T^{-1}\PP\,.
\fe

\subsection{Lattice Hamiltonians and the continuum limit}\label{sec:majcont}
As we said above, so far, we have treated $(-1)^F$, $T$, $\R$, $\T$, and their combinations $\Pi=T\R\T$ and $\PP=\R\T$ as operators, and we discussed their algebra.  Now, we consider specific Hamiltonians, and then only a subset of these operators will be symmetries of the system.  Also, only for certain Hamiltonians does this $\Pi$ flow to the CRT symmetry of the continuum theory.

For concreteness, we start with the gapless Majorana chain with the Hamiltonian
\ie\label{HNf}
H = i  \sum_{A=1}^{N_f} \sum_{\ell=0}^{L-1} \chi_\ell^A \chi_{\ell+1}^A\,.
\fe
It is well known that in the continuum limit, this system flows to $N_f$ copies of the 1+1d Majorana conformal field theory, reviewed in Appendix \ref{continuumMajorana}, 
\ie\label{LNf}
{\cal L} = \frac{i}{2}\sum_{A=1}^{N_f}\chi_\text{L}^A (\partial_t -\partial _x )\chi_\text{L}^A
+\frac{i}{2}\sum_{A=1}^{N_f}\chi_\text{R}^A (\partial_t +\partial _x )\chi_\text{R}^A
\,.
\fe
This will allow us to compare the lattice operators discussed above with the corresponding operators in this continuum limit.

For even $L$, both the left- and right-moving continuum fermions obey periodic boundary conditions. For odd $L$, the left- and right-moving continuum fermions obey opposite boundary conditions.  
See \cite{Seiberg:2023cdc} for recent discussions.

We start with the even $L$ lattice model and its continuum limit. 
It is known that the lattice translation $T$ leads to two symmetries: continuum translation and chiral fermion parity $\sC$ (sometimes denoted as $(-1)^{F_\text{L}}$). 
The latter generates a unitary, internal $\mathbb{Z}_2$ symmetry, and is referred to as an emanant symmetry \cite{Cheng:2022sgb,Seiberg:2023cdc,Chatterjee:2024gje}. It is not a subgroup of the lattice translation symmetry, but a quotient of it.  We denote the continuum counterparts of the lattice reflection $\R$ and time-reversal $\T$ by $\sR$ and $\sT$, respectively.  
These three continuum symmetries  act on the continuum fermion fields as
\ie\label{eq:contCPT}
&\sC \,\chi_\text{L}^A(t,x) \,\sC^{-1} = - \chi_\text{L}^A(t,x)\,,&&
\sC\,\chi_\text{R}^A(t,x) \,\sC^{-1} = \chi_\text{R}^A(t,x)\,, \\
&\sR \, \chi_\text{L}^A \,(t,x)\, \sR^{-1} = \chi_\text{R}^A(t,-x)\,,&&
\sR \,\chi_\text{R}^A (t,x) \, \sR^{-1} = \chi_\text{L}^A(t,-x)\,,\\
&\sT\, \chi_\text{L}^A (t,x)\, \sT^{-1} = \chi_\text{R}^A(-t,x) \,,&&
\sT\, \chi_\text{R}^A (t,x)\, \sT^{-1} = \chi_\text{L}^A(-t,x)\,. 
\fe
Therefore, the lattice  symmetry $\Pi = T\R\T$ becomes the canonical $\Theta=\sC\sR\sT$ in the continuum limit, which acts as: 
\ie\label{Thetachi}
\Theta \,\chi_\text{L}^A (t,x) \,\Theta^{-1} = -\chi_\text{L}^A (-t,-x) \,,~~~
\Theta \,\chi_\text{R}^A (t,x) \, \Theta^{-1} = \chi_\text{R}^A (-t,-x) \,.
\fe
Note that $\Theta^2=1$ and see Appendix \ref{CPTreview} for more discussions. 
These identifications are summarized in Table \ref{tab:majshort}.

\begin{table}[t]
    \centering
     \begin{tabular}{|c|c||c|c|}
     \hline
        \text{Lattice} &\text{Order of}  &\text{Continuum} & \text{Order of} \\
          \text{symmetries} &\text{anomalies}  &\text{symmetries} & \text{anomalies}\\
        \hline
       $T$  & 2 & $\sC$ & 8 \\
        $\R$ & 2  & $\sR$&2 \\
        $\T$ & 1  & $\sT$&1 \\
        $\Pi =T\R\T$ & 1  & $\Theta=\sC\sR\sT$&1 \\
        $T,\Pi$&8 & $\sC,\sC\sR\sT$ &8\\
        $\PP=\R\T$ & 8 & $\sR\sT$ & 8\\
        $T\R$ & 1  & $\sC\sR$&1 \\
        $T\T$ & 2  & $\sC\sT$&2 \\
        $T,\R$&2 & $\sC,\sR$ &8\\
        $T,\T$ & 2 & $\sC,\sT$ &8\\
        \hline
    \end{tabular}
    \caption{Dictionary between the lattice translation $T$, (site-centered) reflection $\R$, time-reversal $\T$ of the Majorana chain \eqref{HNf}, and the $\sC, \sR,\sT$ symmetries of the continuum non-chiral, free, massless Majorana fermion \eqref{LNf}. The second and fourth columns show the order of the lattice and continuum anomalies, where we also assume the fermion parity $(-1)^F$. Here, the order of the anomaly is the minimal number of copies of the system $N_f$, such that it is anomaly-free.  (Order 1 means that there is no anomaly.) In this table, we assume that $L$ is even. See Appendix \ref{app:evenL} for more details.}
    \label{tab:majshort}
\end{table}

Let us return to the lattice Hamiltonian \eqref{HNf}.  It is invariant under all the symmetries discussed in the previous subsection.  We would like to explore a smaller set of symmetries.  In particular, we would like to study the symmetry generated by $T$, $\Pi$, $(-1)^F$, or an even smaller symmetry, generated by $\PP=T^{-1}\Pi$, $(-1)^F$.  To achieve that, we deform the original Hamiltonian \eqref{HNf} with appropriate symmetry-breaking terms.

We deform by the next-to-nearest-neighbor coupling \cite{PhysRevB.73.214407,2009arXiv0905.1849S,OBrien:2017wmx}\footnote{The bosonized version of this deformation
is called the Dzyaloshinskii–Moriya interaction  \cite{RevModPhys.25.166, PhysRev.120.91,DZYALOSHINSKY1958241}. 
See Section \ref{sec:noninvertible}.}
\ie\label{eq:HNNN}
H_\text{NNN}=i\sum_{A=1}^{N_f}\sum_{\ell=0}^{L-1}\chi_\ell^A\chi_{\ell+2}^A\,,
\fe 
which is odd under $\T$ and $\R$, while still preserving $T$, $\Pi$ and $(-1)^F$.  
It flows to $i\sum_{A=1}^{N_f}(\chi^A_{\text L}\partial_x\chi^A_{\text L}+\chi^A_{\text R}\partial_x\chi^A_{\text R})$, which inherits the same transformation properties under $\sR$, $\sT$ and $\sC$. Up to normalization, this corresponds to a deformation by the energy-momentum tensor $T_{++}-T_{--}$  \cite{OBrien:2017wmx,Zou:2019dnc}. Note that this continuum theory is not Lorentz invariant, but it is $\Theta$-invariant.

In Appendix \ref{app:deformation}, we will discuss a staggered version of next-to-nearest-neighbor coupling, $H_\text{NNN}'$, which preserves $\PP$ and $T^2$, but breaks $T$ and $\Pi$.  
 We will show that this interaction is redundant and modifies the spectrum in a trivial way.

\subsection{A mod 8 anomaly involving the lattice  CRT} \label{sec:majCRT}

In this section, we discuss anomalies in the discrete group generated by  $\Pi=T\R\T$ and $T$. It is important that this discussion depends only on the structure of the Hilbert space and the action of these operators on the degrees of freedom.  The Hamiltonian $H$ is important only insofar as it determines the actual symmetry of the model; i.e., the subgroup of the group generated by $\Pi=T\R\T$ and $T$ that commutes with $H$.

\subsubsection{Anomaly for $\PP= \R\T$}\label{sec:majRT}

In this subsection, we focus on a subset of the symmetry generated by $(-1)^F$ and $\PP$. 
 
It is straightforward to see that for $N_f=0\mod 8$, the system is anomaly-free.  To do that, we deform the  $N_f=0\bmod 8$ Hamiltonian, while preserving its symmetries to
\ie
 \sum_{ABCD} \mu_{ABCD} \sum_\ell\chi_\ell^A \chi_\ell^B \chi_\ell^C \chi_\ell^D\,. \label{eq:FKchain}
\fe
This Hamiltonian acts independently on each site, such that we have $L$ decoupled systems, each with $N_f=0\bmod 8$ fermions.  This is exactly the system discussed in \cite{Fidkowski:2009dba} and reviewed in Appendix 
\ref{QMA}.  For generic $\mu_{ABCD}$, the Hilbert space in each site has a unique ground state.  As a result, the total system is trivially gapped, and therefore it is anomaly-free. After establishing this fact, the same applies to any deformation of the Hamiltonian, including \eqref{HNf}. This argument thus provides an upper bound of 8 on the order of the anomaly.

To find a lower bound on the order of the anomaly, we first view our system as a 0+1d system, ignoring its locality properties.  In this case, $\PP$ is a time-reversal symmetry in quantum mechanics whose anomaly comes with a modulo 8 classification \cite{Fidkowski:2009dba}, which we review in Appendix \ref{QMA}.  In order to find the actual anomaly, we should examine how the fermions transform under $\PP$.

Let us start with odd $L$.  Since $\PP$ maps a fermion at $\ell$ to a fermion at $-\ell$, the eigenvectors of $\PP$ are
\ie\label{eq:pairing}
\PP \left( \chi^A_\ell + \chi^A_{-\ell}\right)\PP^{-1} =  \left( \chi^A_\ell + \chi^A_{-\ell}\right)\,,~~~~
\PP \left( \chi^A_\ell - \chi^A_{-\ell}\right)\PP^{-1} = - \left( \chi^A_\ell -\chi^A_{-\ell}\right) \,,
\fe
and 
\ie\label{PPchi0}
\PP \chi_0^A \PP^{-1} = \chi_{0}^A\,.
\fe
We see that under the action of $\PP$, there are $\nu_+ = \left({L-1\over2} +1\right)N_f$ fermions that transform by $+1$, and $\nu_- = {L-1\over2}N_f$ fermions that transform by $-1$. 
Viewing this 1+1d system as a 0+1d quantum mechanical system, these fermions contribute to the time-reversal anomaly  $\nu = \nu_+-\nu_-=N_f$ mod 8. This analysis agrees with the phases in \eqref{odd L algebraT}: $\PP^2=(-1)^{\frac{N_f(N_f-2)}{8}}$ and $(-1)^F \PP=(-1)^{\frac{N_f}{2}} \PP(-1)^F$. 

We see that for odd $L$, the upper and lower bounds on the order of the anomaly coincide, and we conclude that $\PP$ has a modulo 8 anomaly.

For even $L$, we have to add to \eqref{eq:pairing} and \eqref{PPchi0}
\ie\label{PPchiL}
\PP \chi_{L\over2}^A \PP^{-1} = \chi_{L\over 2}^A\,.
\fe
In other words, there are two sets of unpaired fermions at the opposite sides in space, $\chi_0^A$ and $\chi_{\frac L2}^A$. 
Each of them contributes equally to the anomaly. 
Thus, for even $L$, the anomaly is labeled by $\nu=\nu_+-\nu_- =2N_f$ mod 8 when viewed as a 0+1d quantum mechanical system. 
This analysis agrees with the phases in \eqref{even L algebraT}: $\PP^2=(-1)^{\frac{N_f(N_f-1)}{2}}$ and $(-1)^F \PP=(-1)^{N_f} \PP(-1)^F$.  This argument provides a lower bound of 4 on the order of the anomaly of the even $L$ system.

So far, we have viewed this system as a 0+1d system without paying attention to its locality as a 1+1d system.  Intuitively, it is clear that taking this locality into account shows that the anomaly for even $L$ is also of order 8, rather than of order 4.

To see that, we note that even though, as 0+1d systems, the odd $L$ and the even $L$ systems are different, when viewed as 1+1d systems, they are related to each other. Specifically, in the infinite $L$ limit, both chains are essentially the same.  This is particularly clear when the system is gapped.  Therefore, the contribution to the anomaly from $\chi^A_0$ cannot be combined with the contribution to the anomaly from $\chi^A_{L\over 2}$.\footnote{We thank Nikita Sopenko for a helpful discussion about this point.}

\begin{figure}
    \centering
    \input{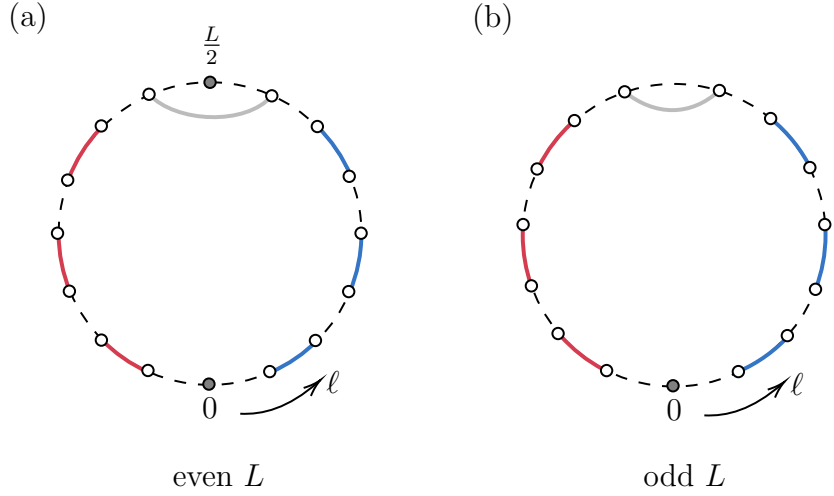}
    \caption{The lattice Hamiltonians in \eqref{HDW} preserving the anti-unitary symmetry $\PP=\R \T$. Each blue link represents a pairing of two Majorana fermions of the form $i \chi_\text{odd}^A\chi_\text{even}^A$, while each  red link represents a pairing term $i \chi_\text{even}^A\chi_\text{odd}^A$. The gray link exists when $L=4n$ or $L=4n-1$.}
    \label{fig:hdw}
\end{figure}

This discussion can be made more physical by studying the Hamiltonian
\ie\label{HDW}
&H_\text{Interface}\\
&=\begin{cases}
    i\sum_{A=1}^{N_f}  \sum_{m=1}^n \left(\chi_{2m-1}^A \chi_{2m}^A + \chi_{-2m }^A\chi_{-2m+1}^A \right) & L=4n+2\text{ or }4n+1\\
    i\sum_{A=1}^{N_f} \sum_{m=1}^{n-1} \left(\chi_{2m-1}^A \chi_{2m}^A + \chi_{-2m }^A\chi_{-2m+1}^A \right) + i\sum_{A=1}^{N_f}\chi^A_{2n-1}\chi^A_{-2n+1}& L=4n\text{ or }4n-1\,.
\end{cases}
\fe
(See Figure \ref{fig:hdw}).  It is manifestly local and $\PP$-symmetric.\footnote{The local terms in this Hamiltonian can be written in terms of the non-local pairs in \eqref{eq:pairing}:
\begin{equation}
\begin{aligned}
i\chi_{2m-1}^A \chi_{2m}^A &+ i\chi_{-2m }^A\chi_{-2m+1}^A
=\\
&\frac{i}{2}(\chi_{2m-1}^A+\chi_{-2m+1}^A)(\chi_{2m}^A-\chi_{-2m}^A)+\frac{i}{2}(\chi_{2m-1}^A-\chi_{-2m+1}^A)(\chi_{2m}^A+\chi_{-2m}^A)\,.
\end{aligned}
\end{equation}
For $L=4n$ or $4n-1$, the pairing of $\chi_{2n-1}^A\pm \chi_{-2n+1}^A$ is local and their coupling gives $i\chi^A_{2n-1}\chi^A_{-2n+1}$.}  Clearly, it gaps the system and the only light modes are the fermions $\chi_0^A$ for all $L$ and $\chi^A_{L\over 2}$ for even $L$.  For odd $L$, it is clear that the order of the anomaly is 8.  And for even $L$, the fact that these two modes are separated in space makes it impossible to lift them, and again, the anomaly is of order 8.

The anomaly between $\PP$ and $(-1)^F$ implies an LSM-type constraint: \textit{any local Hamiltonian invariant under $\PP$ and $(-1)^F$ cannot be trivially gapped.}
For instance, the Hamiltonians $H+\lambda H_\text{NNN}$ in \eqref{HNf} and \eqref{eq:HNNN} are an one-parameter family of models invariant under $\PP$ and $(-1)^F$ (as well as $T$ and $\Pi$). 
In Appendix \ref{app:deformation}, we showed that this family of models is always gapless.

What if we break $T$ and $\Pi$, while preserving $\PP$ and $(-1)^F$ (and perhaps also $T^2$)?  
The simplest nontrivial such deformation of $H+\lambda H_\text{NNN}$ is 
\ie \label{eq:majint}
\sum_{\ell=0}^{L-1} \left( \alpha_+ + \alpha_-(-1)^\ell \right)
\left(
\chi_\ell \chi_{\ell+1}\chi_{\ell+2} \chi_{\ell+4}
+\chi_\ell \chi_{\ell+2}\chi_{\ell+3} \chi_{\ell+4}
\right).
\fe
It preserves  $\PP$ and $(-1)^F$ (as well as $T^2$), but breaks $\Pi$ and $T$.
The anomaly of $\PP$ predicts that this interacting Hamiltonian cannot be trivially gapped.

In the continuum field theory of \eqref{LNf} (with $N_f=1$), one can similarly ask what the simplest deformation that preserves $\sR\sT$ but breaks   $\sC$ and $\Theta=\sC\sR\sT$ is. 
Such an operator should have an odd number of $\chi_\text{L}$ and an odd number of $\chi_\text{R}$ to break $\sC$.  Also, in order to violate $\Theta$ it should have odd Lorentz spin.  (See Appendix \ref{CPTreview}.) 
Up to a total derivative, and up to the equations of motion 
$\partial_-\chi_\text{L}= \partial_+\chi_\text{R}=0$,
the lowest dimension operators are
\ie\label{continfour}
\chi_\text{R}\partial_+^{3}\chi_\text{L}\partial_+^{2}\chi_\text{L}\partial_+\chi_\text{L}\qquad ,\qquad \chi_\text{L}\partial_-^{3}\chi_\text{R}\partial_-^{2}\chi_\text{R}\partial_-\chi_\text{R}\,.
\fe
These operators have dimension 8 and are highly irrelevant.  Therefore, they do not affect the IR theory.

\subsubsection{Anomaly between translation and $\Pi$ (the lattice CRT)}

In the previous subsection, we analyzed the symmetry generated by $\PP=T^{-1}\Pi$ and $(-1)^F$ and its anomaly.  Now, we examine a larger symmetry generated by the lattice translation $T$, the lattice CRT operator $\Pi$, and fermion parity $(-1)^F$.  The anomaly in this larger symmetry includes the anomaly in the smaller one (in Section \ref{sec:majRT}) and can potentially have additional anomalies.

We start with the even $L$ lattice model. 
In Appendix \ref{app:evenL}, we review the order 2 anomaly in $T$ and $(-1)^F$, and we show that the symmetry generated by $\Pi$ and $(-1)^F$  is anomaly-free. This means that there must be a mixed anomaly between $T$ and $\Pi$, which we show to be of order 8 below.

This mixed anomaly can be diagnosed from the projective phases in the algebra of $T$ and $\Pi$. 
For even $L$ lattice, the algebra is in \eqref{eq:evenLTPi}: 
\ie 
\Pi \, T = (-1)^{\frac{N_f(N_f-1)}{2}}T^{-1} \,\Pi\,,~~~~\text{even}~L\,.
\fe 
This phase cannot be removed without altering the phase of $T^L=1$. 
When $N_f = 0 \bmod 4$, this phase becomes 1 and hence the mixed anomaly between $T$, $\Pi$ and $(-1)^F$ is at least of order 4. 
In fact, we argue below that this mixed anomaly is actually of order 8. 

Recall the algebra  \eqref{odd L algebra} for the odd $L$ (and even $N_f$):
\ie \label{PianomalyMaj}
\Pi^2=(-1)^{\frac{N_f(N_f-2)}{8}}, \quad \Pi(-1)^F=(-1)^{\frac{N_f}{2}}(-1)^F \Pi\,,\quad \text{odd}~L\,.
\fe
The two phases are 1 only when $N_f=0\bmod 8$. Therefore, the  anomaly between $(-1)^F$ and $\Pi$ is of order 8 when $L$ is odd. 

Now we offer another interpretation. 
One way to diagnose an anomaly is by first inserting a symmetry defect and then analyzing how the symmetry algebra is modified in its presence. 
The projective phases in this algebra signal the anomaly.  This fact is standard in continuum field theory, and has been used in lattice systems in various places including \cite{Yao:2020xcm,Cheng:2022sgb,Yao:2023bnj,Seifnashri:2023dpa}. 

However, this approach is more subtle for spacetime symmetries like lattice translations. In such cases, changing the system size from even to odd can sometimes be interpreted as a “lattice translation defect,” as discussed in \cite{Cheng:2015kce,Cho:2017fgz,Metlitski:2017fmd,ThorngrenPRX2018,Manjunath2020, SongPRR2021, Manjunath2021,Ye:2021nop,Cheng:2022sgb,Seiberg:2023cdc,Seifnashri:2023dpa,Seiberg:2024gek}. 
Indeed,   the odd $L$ Majorana chain can be viewed as the even $L$ one with a translation defect, whose continuum limit corresponds to the insertion of a chiral fermion parity $\sC$ defect \cite{Seiberg:2023cdc}.

With this interpretation in mind, we view  \eqref{PianomalyMaj} for odd $L$ as the projective symmetry algebra of $\Pi$ and $(-1)^F$ in the presence of a $T$-defect. These phases then imply a mixed mod 8 anomaly between   $T$, $\Pi$ and $(-1)^F$ for all $L$. 

\subsubsection {Matching the lattice anomalies with the continuum}

We now compare these lattice anomalies of the gapless Majorana chain \eqref{HNf} with their continuum counterparts. 
In the continuum field theory in \eqref{LNf}, the fermion parity $(-1)^F$ and the chiral fermion parity $\sC$, which emanates from lattice translation $T$, have an order 8 anomaly classified by $\text{Hom(}\, \Omega^\text{Spin}_3(B\mathbb{Z}_2),U(1))=\mathbb{Z}_8$\cite{Ryu:2012he,Gu:2012ib,Kapustin:2014dxa}.\footnote{See \cite{Gu:2013azn} for earlier work on 2+1d fermionic SPT phases with a unitary $\mathbb{Z}_2$ symmetry. See also \cite{2022CMaPh.395..405O,2023JMP....64i1901O}  for another approach to fermionic SPT phases using operator algebras on infinite lattices and the importance of the lattice CRT symmetry. }
On the lattice, however, we only find an order 2 anomaly between $T$ and $(-1)^F$. This difference arises because the symmetry algebras are not the same: the continuum $\mathbb{Z}_2$ symmetry is a quotient of the lattice translation group $\mathbb{Z}_L$ for even $L$. Consequently, the continuum $\bZ_2$ symmetry is an emanant symmetry and its anomaly is an emergent anomaly.  (See Section \ref{sec:emanaandem}.)  This leads to a natural question. Can we detect this mod 8 emergent anomaly on the lattice? In fact, it was argued in \cite{Jones:2019lwm} that this mod 8 anomaly cannot be realized exactly on the  Majorana chain by a unitary, locality-preserving, $\mathbb{Z}_2$ lattice operator. (See the same reference for a realization of such a lattice operator on a non-tensor product Hilbert space.)

Now, the  $\Theta=\sC\sR\sT$ symmetry serves as a useful crutch to probe this continuum anomaly. 
In Lorentzian continuum field theory with relativistic symmetry, there is a standard $\sC\sR\sT$ transformation, which becomes spacetime $\pi$-rotation in Euclidean signature. 
It follows that  $\Theta=\sC\sR\sT$  has no anomaly. 
However, as discussed in Section \ref{sec:introCPT}, the self-anomaly of the unitary $\mathbb{Z}_2$ symmetry $\sC$ is reflected in the projective algebra of $\Theta_\sC$ in the presence of a $\sC$ defect.\footnote{The $\sC$ symmetry plays the role of $\sU$ in the discussion around \eqref{ThetaUsq}.  Note, however, that here we study a fermionic theory, while the discussion around \eqref{ThetaUsq} is about a bosonic theory.}

On the lattice with odd $L$, the operator $\Pi$ flows to the continuum operator $\Theta_\sC$, and the projective algebra in \eqref{PianomalyMaj} is the fermionic counterpart of \eqref{ThetaUsq} on the lattice. 
Hence, the mod 8 anomaly we find between lattice translation $T$, $\Pi$, and $(-1)^F$ should be understood as a direct probe of the anomaly for $\sC$ and $(-1)^F$ in the continuum.

In Table \ref{tab:majshort}, we summarize the anomalies on the lattice and in the continuum that are discussed in this subsection. See more discussion about anomalies of other subgroups of the symmetry generated by $T$, $\T$, $\R$, and $(-1)^F$ in Appendix \ref{app:evenL}.

\subsection{Lattice realization of the Smith isomorphism}\label{SmithFermions}

There are two anomalies in continuum fermionic field theories that are related to   $\PP$. 
The first one is the anomaly of a \textit{unitary} $\mathbb{Z}_2$ global symmetry in 1+1d, which admits a $\mathbb{Z}_8$ classification and is related to SPTs in 2+1d. 
The second one is the anomaly of an \textit{anti-unitary}, time-reversal $\mathbb{Z}_2^\T$ symmetry in 0+1d quantum mechanics, which also admits a $\mathbb{Z}_8$ classification and is related to SPTs in 1+1d \cite{Fidkowski:2009dba}. 
Mathematically, the 1+1d and 0+1d anomalies are respectively classified by the cobordism group $\text{Hom}(\,\Omega^\text{Spin}_3 (B\mathbb{Z}_2),U(1))=\mathbb{Z}_8$
 and 
$\text{Hom}(\,\Omega^{\text{Pin}^-}_2 (pt),U(1))=\mathbb{Z}_8$, which are known to be isomorphic under the Smith homomorphism \cite{gilkey1989geometry}. (See \cite{Tachikawa:2018njr,Kapustin:2014dxa} for physicist-friendly expositions.)

Is it a coincidence that these two anomalies in different dimensions share the same classification? In \cite{Hason:2020yqf,Cordova:2019wpi}, a physical interpretation of this isomorphism was presented in terms of the $\sC\sR\sT$ symmetry of the continuum theory. (See also \cite{Wang:2019obe,Wan:2023nqe}.)
Here, we will present a closely related, but different setup in terms of interfaces in finite space, which is to be contrasted with the pictures using domain walls in \cite{Hason:2020yqf} and topological defects in \cite{Cordova:2019wpi}.

\begin{figure}
    \centering
    \input{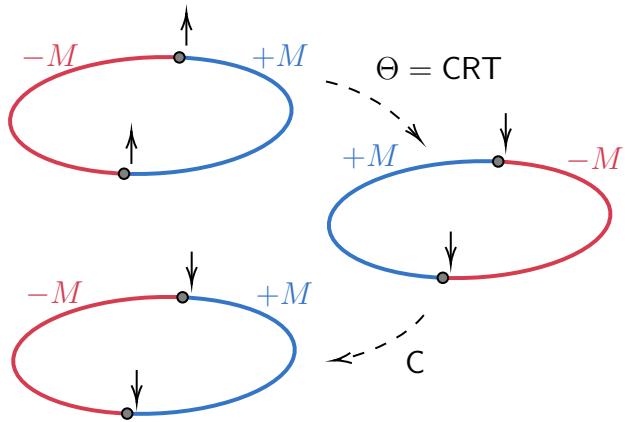}
    \caption{A physics realization of the Smith isomorphism.  In the continuum Majorana fermion field theory, we turn on opposite mass terms in two halves of space with localized zero modes at the interfaces.  This  setup relates  a unitary $\mathbb{Z}_2$ symmetry $\sC$ in 1+1d to an anti-unitary time-reversal symmetry $\sR\sT$ in quantum mechanics \cite{Hason:2020yqf,Cordova:2019wpi}.  The short arrows near the localized zero modes indicate how the time coordinate of the effective quantum mechanical model reflects under these operators. }
    \label{fig:smith}
\end{figure}

We start with the continuum field theory of $N_f$ massless Majorana fermions as in \eqref{LNf} on a spatial circle, with periodic boundary conditions for both the left- and right-movers. 
The chiral fermion parity $\sC$ is a unitary $\mathbb{Z}_2$ global symmetry with a modulo 8 anomaly. 
Next, we turn on a positive mass term $+iM \sum_A\chi^A_\text{L}\chi^A_\text{R}$ on half of the space, and a negative mass term $-iM\sum_A\chi^A_\text{L}\chi^A_\text{R}$ on the other half. 
This creates an interface at the locations where the mass changes sign. 
In the infinite mass limit, the two sides are described by two  invertible field theories that are exchanged by $\sC$.\footnote{For odd $N_f$, these are two distinct invertible phases whose difference is  the Arf invariant. These two invertible field theories are classified by Hom($\Omega_2^\text{Spin}(pt),U(1))=\mathbb{Z}_2$.} 
The low-energy limit is an effective 0+1d quantum mechanical system, with one  Majorana zero mode localized at each of the two interfaces.  
See Appendix \ref{app:RRDW} for the explicit solutions of these zero modes.  
While this mass configuration breaks $\sC$ and $\Theta=\sC\sR\sT$ separately, it preserves the following anti-unitary symmetry: 
\ie\label{smithcont}
\sR\sT = \sC  \Theta\,.
\fe
See Figure \ref{fig:smith}, for an illustration. 
This expression means that the $\Theta=\sC\sR\sT$ symmetry relates a unitary $\mathbb{Z}_2$ symmetry $\sC$ of a 1+1d system, to an anti-unitary time-reversal symmetry $\sR\sT$ of a 0+1d quantum mechanical system. 
The operator equation \eqref{smithcont} and the mass deformation therefore provide a physical realization of the Smith isomorphism between these two symmetries in different dimensions.

The situation is different if the left and right movers obey opposite boundary conditions. 
Now, there are only localized zero modes near $x=0$, but not near the opposite end of the spatial circle. See 
Appendix \ref{app:NSRDW}.

Next, we discuss the lattice counterpart of this continuum construction. 
We start with the gapless Hamiltonian \eqref{HNf} for the 1+1d Majorana chain, whose continuum limit is \eqref{LNf}, and deform it to the interface Hamiltonian \eqref{HDW}. 
In the continuum, this Hamiltonian describes a space-varying mass term.
It breaks $T$ and $\Pi$, but it is invariant under the product
\ie
\PP = T^{-1} \Pi\,.
\fe
This is the lattice counterpart of \eqref{smithcont}.

The odd and even $L$ systems correspond to the continuum theories with equal and opposite boundary conditions for the left and right-moving fermions, respectively. 
As discussed around \eqref{HDW} and Figure \ref{fig:hdw}, for odd $L$, there are $N_f$ zero modes at $\ell=0$, and for even $L$, there are additional $N_f$ zero modes at $\ell=\frac L2$.  
These match with the continuum discussion above and in Appendix \ref{MZM}.

For all $L$, these zero modes are robust and are protected by  $\PP$: they cannot be lifted by local, $\PP$-symmetric deformation of the Hamiltonian unless $N_f=0$ mod 8.\footnote{One might suspect that the protected Majorana zero modes arise solely because the two sides of the interfaces belong to different invertible phases of \cite{Kitaev:2000nmw}, and are thus unrelated to $\PP$. This, however, is not the case. Imposing $\PP$ leads to a stronger conclusion. First, the two sides correspond to distinct invertible phases only when $N_f$ is odd. Second, even in this case, the phase difference protects the number of zero modes only modulo 2, whereas $\PP$ protects this number modulo 8.} 
When $N_f=0$ mod 8, we can add an onsite,  $\PP$-symmetric quartic term $\sum_{A,B,C,D}\mu_{ABCD} \chi^A_\ell\chi^B_\ell\chi^C_\ell\chi^D_\ell$ at the interface to lift the zero modes.  
This is consistent with the discussion of the modulo 8 anomaly for $\PP$ in Section \ref{sec:majRT}. 

\section{The Heisenberg spin chain}\label{Spinchain}

In this section, we discuss an anomaly involving translation and a lattice CRT symmetry on a spin chain of qubits in 1+1d. 

\subsection{Lattice translation and CRT}
 
The space is a 1d periodic chain of $L$ sites with one qubit per site. 
We denote the  Pauli operators at site $j$ by $\vec S_j =(S_j^x,S_j^y , S_j^z) = \frac 12( X_j, Y_j,Z_j)$.  They are subject to periodic boundary conditions $\vec S_{j+L } =\vec S_j$. 

We define three  operators: lattice translation $T$, (site-centered) reflection $\R$, and time-reversal $\T$:
\ie\label{CRTLatticeS}
&T\, \vec S_j\, T^{-1} = \vec S_{j+1}\, ~~~~\R \,\vec S_j \,\R^{-1} = \vec S_{-j}\,~~~~
\T \, \vec S_j \,\T^{-1}  = - \vec S_j\,.
\fe
Here $T$ and $\R$ are unitary operators, while the time-reversal operator $\T$ is anti-unitary, i.e., $\T i \T^{-1}=-i$. 
Explicitly, $\T=  \left(\prod_{j=1}^L i Y_j \right)\mathcal{K}$, where $\mathcal{K}$ is the complex conjugation.\footnote{The anti-unitarity of $\T$ forces the minus sign in its action on $\vec S_j$, such that the commutation relations of $\vec S_j$ are preserved. Related to that, $\T$ acts on each qubit such that its square is $-1$, i.e., each qubit is a Kramers' doublet. \label{qubitKD}}
On the other hand, $T$ and $\R$ are products of the SWAP gates. 
These operators and their LSM constraints were discussed in \cite{Chen:2010zpc,2015PNAS..11214551W,Ogata:2020hry,Yao:2023bnj}.
Below, we will discuss a related, but different anomaly associated with lattice translation $T$ and   the following anti-unitary operator
\ie
\Pi =  T \R\T\,.
\fe
It acts as
\ie
\Pi \, \vec S_j \,  \Pi^{-1} = -\vec S_{-j+1}\,,\quad 
\Pi \, i \,  \Pi^{-1} = -i \,.
\fe
It reflects simultaneously the spatial coordinate around a link and the time coordinate. 
Later, we will argue that in the anti-ferromagnetic phase of the Heisenberg model, $\Pi$ flows to the standard CRT symmetry $\Theta$ of the continuum low-energy theory.

We will focus on the algebra generated by the lattice translation $T$ and the lattice CRT operator $\Pi$
\ie\label{TPi}
T^L=1 \,,~~~~\Pi^2=(-1)^{L}\,,~~~~T\Pi = \Pi T^{-1}\,.
\fe
The important $(-1)^L$ phase follows from the fact that each qubit is a Kramers' doublet (see footnote \ref{qubitKD}).
For completeness, we also record the larger algebra generated by $\R$, $\T$, and $T$:
\ie\label{RTalgebra}
&T^L=1\,,\quad &&\R^2=1\,, \quad&& \T^2=(-1)^L\,,\\
&T \R =\R T^{-1}\,, \quad && T\T = \T T\,,\quad &&\R \T = \T \R\,.
\fe

Finally, we define another anti-unitary operator 
\cite{Yao:2023bnj}:
\ie
\PP=  T^{-1 } \Pi = \R\T\,,
\fe
which acts on the spins as
\ie
\PP \, \vec S_j \, \PP^{-1}=  -\vec S_{-j}\,,\quad\PP \, i \, \PP^{-1} = -i\,.
\fe
This is an $\R\T$ symmetry that reflects simultaneously the spatial coordinate around a site and the time coordinate. 
It obeys 
\ie\label{Thetanomaly}
\PP^2= (-1)^L\,.
\fe
Note that for odd $L$, site-centered reflection and link-centered reflection are unitarily equivalent.  Therefore, the same is true for $\Pi$ and $\PP$.  More explicitly, $\PP=T^{-1}\Pi=T^{L-1\over 2}\Pi T^{-{L-1\over 2}}$.  This explains why they both satisfy $\Pi^2=\PP^2=-1$ and $T\PP=\PP T^{-1}$.

\subsection{Lattice Hamiltonians and the continuum limit}

\subsubsection{Anti-ferromagnetic phase}\label{antiferromagnetic}

Even though our discussion of the anomalies will be independent of the choice of the Hamiltonian, it is useful to have in mind some examples that commute with these symmetry operators. 
The anti-ferromagnetic $SO(3)$ Heisenberg spin chain Hamiltonian 
\ie\label{Heisenberg}
H = \sum_{j=1}^L \vec S_j\cdot \vec S_{j+1}\,,
\fe
is invariant under the lattice translation $T$ and the anti-unitary operator  $\Pi=T\R\T$. 
In fact, it is also invariant under $\R$, $\T$, and $SO(3)$. 
A class of  terms we can add to the  Hamiltonian that preserves $T,\Pi$ but breaks $\R,\T, SO(3)$ is
 \ie\label{Hch}
 H' =  \sum_{j=1}^L \sum_{\alpha,\beta,\gamma}\lambda_{[\alpha\beta ]\gamma}  
 S^\alpha_j S^\gamma_{j+1} S^\beta_{j+2}
 \,,
 \fe
where $\alpha,\beta,\gamma$ are summed over $x,y,z$, and  $\lambda_{[\alpha\beta ]\gamma}$ is a generic tensor that is antisymmetric in $\alpha$ and $\beta$.

It is well-known that the continuum limit of \eqref{Heisenberg} is the $su(2)_1$ Wess-Zumino-Witten (WZW) model, which is the $c=1$ CFT of a compact boson (also known as the Luttinger liquid) at the self-dual radius. 
The lattice translation  $T$ becomes an internal, unitary, emanant  $\mathbb{Z}_2$ global symmetry, which we denote by $\sC$, in the continuum \cite{Metlitski:2017fmd,Cheng:2022sgb}.

We would like to match the lattice symmetry operators $T$, $\R$, and $\T$ with the standardly-defined continuum operators $\sC$, $\sR$, and $\sT$.  To do that, we match with a continuum field theory that flows in the IR to the same CFT as the lattice model.  (For more details, see Appendix \ref{app:boson}.) 

One such theory is the $O(3)$ non-linear sigma model with a theta term.  Its Lagrangian is
\ie\label{O3Lagi}
&{\cal L} ={f\over 2}\partial_\mu \vec n\cdot \partial^\mu \vec n + \theta \Gamma\,,~~~~\vec n\cdot\vec n=1\\
&\Gamma ={\epsilon^{\mu\nu}\over 8\pi} \vec n\cdot (\partial_\mu \vec n\times \partial_\nu\vec n)\,.
\fe
Here, $f$ is a relevant coupling constant.
For $\theta \in \pi {\mathbb Z}$, the theory 
is invariant under  $\sC$, reflection $\sR $, and time-reversal $\sT$. They act as:
\ie\label{CRTonn}
&\sC \,\vec n(t,x)\, \sC^{-1}= -\vec n(t,x) \,,~~~~
\sR \,\vec n(t,x)\,\sR^{-1} = \vec n(t,-x)\,,~~~~
\sT  \,\vec n(t,x)\,\sT^{-1} = -\vec n(-t,x)\,.
\fe
The standard $\Theta=\sC\sR\sT$ symmetry, which becomes the spacetime $\pi$-rotation in Euclidean signature, acts on the fields as
\ie 
&\Theta \,\vec n(t,x)\, \Theta^{-1}= \vec n(-t,-x) \,.
\fe

In the quantum theory, the classical field $\vec n$ becomes an operator, which is not constrained by $\vec n^2=1$.  (In order not to clutter the equations, we will not distinguish between the classical field and its quantum counterpart.)
For $\theta\in \pi(1+2{\mathbb Z})$, this continuum theory flows to the $su(2)_1$ CFT, which is the same as the IR limit of the lattice model \eqref{Heisenberg}. 
The lattice operators $\vec S_j$ flow to the continuum operators \cite{Affleck:1988nt,Affleck:1988zj}
\ie\label{StonA}
&(-1)^j\vec S_j \to  \vec n\,.
\fe
This map is up to wave function renormalization.  The symmetry action of the lattice operators $T,\R,\T$ in \eqref{CRTLatticeS} is then matched with the symmetry action of the continuum operators $\sC,\sR,\sT$ in \eqref{CRTonn} with the simple identification in Table \ref{tab:spin}.

\begin{table}
    \centering
    \begin{tabular}{|c|c||c|c|}
    \hline
        \text{Lattice} &\text{Order of}  &\text{Continuum} & \text{Order of} \\
          \text{symmetries} &\text{anomalies}  &\text{symmetries} & \text{anomalies}\\
        \hline
       $T$  & 1 & $\sC$ & 2 \\
        $\R$ & 1  & $\sR$&1 \\
        $\T$ & 1  & $\sT$&1 \\
        $\Pi =T\R\T$ & 1  & $\Theta=\sC\sR\sT$&1 \\
        $T,\Pi$&2 & $\sC,\sC\sR\sT$ &2\\
        \hline
    \end{tabular}
    \caption{Dictionary  between the  lattice translation $T$, (site-centered) reflection $\R$, time-reversal $\T$ of the Heisenberg model \eqref{Heisenberg}, and the $\sC, \sR,\sT$ symmetries of the continuum $su(2)_1$ WZW model or the $O(3)$ non-linear sigma model at $\theta=\pi$. The second and fourth columns show the order of the lattice and continuum anomalies. (Order 1 means that there is no anomaly.) Here, we take $L$ to be even.}
    \label{tab:spin}
\end{table}

 \subsubsection{Ferromagnetic phase}\label{ferromagnetic}
 
Above, we studied the spin chain with the Hamiltonian \eqref{Heisenberg} and showed how the lattice symmetries $T$, $\R$, and $\T$ are mapped to the continuum symmetries $\sC$, $\sR$, and $\sT$.  And in particular, how the lattice symmetries $\Pi=T\R\T$ and $\PP=\R\T$ are mapped to the continuum symmetries $\Theta=\sC\sR\sT$ and $\sR\sT$ respectively.   

Here, we will see that by changing the Hamiltonian, without changing the lattice symmetries, they can act differently in the continuum limit. Specifically, consider the ferromagnetic phase of the same system.  It is obtained by changing the sign of \eqref{Heisenberg}.  The lattice symmetries are as above.  But the continuum theory is different.

The continuum limit again involves a field $\vec n$ with $\vec n^2=1$, but unlike \eqref{StonA}, now 
\ie
\vec S_j \to \vec n\,.
\fe
As is well-known, this change has profound consequences.  The continuum limits of the lattice symmetries
\ie
\R\to \sR\qquad ,\qquad \T\to \sT\,,
\fe
with the action in \eqref{CRTonn} 
\ie\label{CRTonnf}
\sR \,\vec n(t,x)\,\sR^{-1} = \vec n(t,-x)\,,~~~~
\sT  \,\vec n(t,x)\,\sT^{-1} = -\vec n(-t,x)
\fe
are unchanged.  However, the lattice translation symmetry $T$ does not lead to an emanant internal symmetry $\sC$.  Instead, it acts as a discrete translation symmetry of the continuum theory \cite{Seiberg:2024wgj,Seiberg:2024yig}.

The leading derivative terms in the continuum Lagrangian differ from those of the anti-ferromagnet \eqref{O3Lagi}.  In particular, the ferromagnetic Lagrangian includes a term, which is first-order in time-derivative, known as the Berry term or a Wess-Zumino term
\ie\label{Berryterm}
{\cal B}={k\over 2V}{n^1 \partial_t n^2-n^2\partial_t n^1\over 1+n^3}\,.
\fe
Let us make some comments about this term.
\begin{itemize}
    \item The expression for $\cal B$ has to be defined carefully, such that $\exp\left(i\int {\cal B}\right)$ is $SO(3)$ invariant and the singularity at $n^3=-1$ is harmless. This restricts the coefficient of $\cal B$.  $V$ is the volume of space, which in our one-dimensional problem is $V=\oint dx$, and $k$ should be an integer.  See \cite{Seiberg:2024wgj,Seiberg:2024yig} for a more detailed discussion and many earlier references.
    \item Starting from a one-dimensional lattice with spin $1\over 2$ in each site, the integer $k$ is the total number of sites $L$.
    \item $\exp\left(i\int {\cal B}\right)$ is invariant under $\sR$ and $\sT$.  The lattice symmetry $\PP$ still acts as $\sR\sT$ and is a standard symmetry.  In verifying the invariance under $\sT$, one needs to integrate by parts and use the proper definition of $\exp\left(i\int {\cal B}\right)$.
    \item Despite appearance, $\exp\left(i\int {\cal B}\right)$ is not invariant under continuous translations.  In our case of a one-dimensional lattice with spin $1\over 2$ in each site, it has only the underlying discrete $\bZ_L$ lattice translation symmetry \cite{Haldane:1986zz}.  See \cite{Seiberg:2024wgj,Seiberg:2024yig} for a more modern perspective.
    \item  $\cal B$ is odd under the internal transformation $\sC:\ \vec n\to -\vec n$. In the anti-ferromagnetic phase, this symmetry emanates from lattice translation.  Here, however, lattice translation remains a discrete symmetry.
     \item This continuum theory is not relativistic, and therefore it does not have a standard $\Theta$ symmetry. In more detail, $\cal B$ is a term with spin $1$ and hence $\Theta{\cal B} \Theta^{-1}=-{\cal B}$.  (See Appendix \ref{CPTreview}.) This is consistent with the fact that the lattice symmetry $\Pi$, which could lead to $\Theta$ in the continuum limit, involves lattice translation, which in this case, does not lead to an internal symmetry $\sC$.
\end{itemize}

Below, we will see how the fact that the continuum limit of the lattice symmetries is different in the anti-ferromagnetic phase and in the ferromagnetic phase affects the anomaly matching.

\subsection{A mod 2 anomaly involving CRT}\label{sec:spinanomaly}

As in Section \ref{sec:majCRT},
here, we discuss anomalies in the discrete group generated by $\Pi=T\R\T$ and $T$.  The Hamiltonian $H$ is important only insofar as it determines the actual symmetry of the model.  In particular, we will discuss different Hamiltonians with the same symmetry leading to different continuum limits.  They have the same lattice anomaly, but it is realized in the continuum differently.

\subsubsection{Anomaly for $\PP=\R\T$}\label{sec:spinRT}

We first discuss a mod 2 anomaly for $\PP$ itself when $L$ is odd. 
In this case, $\PP$ is unitarily related to $\Pi$. 
Hence, $\Pi$ and $\PP$  share the same anomaly.

From \eqref{Thetanomaly}, we have
\ie\label{Thetaanomaly2}
 \PP^2 = -1\,,\quad \text{odd}~L\,.
\fe
Therefore, all the states in the spectrum are in Kramers' doublets, i.e., they are in energy-degenerate pairs. This fact can be thought of as an anomaly in $\PP$.  Note that this anomaly is detected even without taking into account the locality constraints in 1+1d. Viewing the system as a 0+1d quantum system, $\PP$ acts as time-reversal with a mod 2 anomaly theory $\pi\int w_1^2$. 
This provides a lower bound of 2 on the order of the anomaly of $\PP$.

This reasoning does not apply to even $L$ because $\PP^2=1$. 
Hence, one might think that in that case, the system is anomaly-free.  However, following the discussion of a similar situation in Section \ref{sec:majRT}, we conclude that this anomaly is present also for even $L$.  Again, this depends on viewing the system as a 1+1d system rather than merely a collection of quantum mechanical spins without locality properties.

It is easy to see that the upper bound of the anomaly is also 2. 
In fact, all  lattice anomalies involving $T,\R,\T$ are at most of order 2, since if there are two qubits per site, there is a trivially gapped Hamiltonian respecting $T,\R,\T$:
 \ie
 \sum_{j=1}^L \vec S^{A=1}_j \cdot \vec S^{A=2}_j\,,
 \fe
where $A=1,2$  labels the two species of qubits.

We therefore conclude that the anti-unitary symmetries $\PP$  have a mod 2 anomaly for all $L$. 
This anomaly implies an LSM-type constraint: \textit{any 1+1d local Hamiltonian invariant under $\PP$ cannot be trivially gapped.}  
See \cite{Yao:2023bnj} for related discussions on the LSM constraint of $\PP$ from the perspective of 2+1d SPT phases. 
In particular, this anomaly predicts that the spin-$\frac12$ Heisenberg model \eqref{Heisenberg} deformed by \eqref{Hch},  $H+ H'$, cannot be trivially gapped for any $\lambda_{[\alpha\beta]\gamma}$.

We can generalize our local Hilbert space to  $2s+1$-dimensional so that it forms a spin-$s$ representation of $su(2)$. For integer $s$, we can take all $\vec S_j$ to be real and then $\T={\mathcal K}$.  For half-integer $s$ we can take $S^y_j$  imaginary with real $S^x_j$ and $S^z_j$, such that $\T = \left( \prod_{j=1}^L\exp(i\pi S^y_j)\right) \mathcal{K}$ acts as $\T \vec S_j \T^{-1} = -\vec S_j$. 
Then, for odd $L$, $\PP^2 = (-1)^{2s}$. This can be summarized as
\ie\label{ppsL}
\PP^2 = (-1)^{2sL}\,.
\fe
Hence, the anomaly in $\PP$ is present for half-integer $s$  and is absent for integer $s$.  

\subsubsection{Anomaly between translation and $\Pi$ (the lattice CRT)}

Now we discuss the anomaly of the lattice symmetry generated by translation $T$ and the anti-unitary operator $\Pi$, as well as its subgroups.

In Appendix \ref{app:spinanomaly}, we show that $T$ and $\Pi$ separately do not have any anomaly for even $L$.  On the other hand, in Section \ref{sec:spinRT} we discussed a mod 2 anomaly of their product $\PP=T^{-1}\Pi$.  This means that there is a mixed anomaly between $T$ and $\Pi$.

Let us discuss it in more detail.  First, for odd $L$, $\Pi$ is unitarily equivalent to $\PP$.  Therefore,  
\ie\label{Pianomaly}
\Pi^2=-1 \,,~~~~~\text{odd }L\,,
\fe
and we see that $\Pi$ is  anomalous for odd $L$.  If we view the odd $L$ system as an even $L$ system with a ``translation defect,'' this means that there is a mixed anomaly between $T$ and $\Pi$.

As above, this is trivially generalized to local spin $s$, showing that this anomaly is present if and only if $s$ is half-integer.

\subsubsection {Matching the lattice anomalies with the continuum}

We now relate this anomaly to the continuum in the context of the anti-ferromagnetic Heisenberg model \eqref{Heisenberg} and its continuum limit, the $su(2)_1$ WZW model.  (See Section \ref{antiferromagnetic}.)
Even though  $T$ has no intrinsic anomaly on the lattice, its continuum limit $\sC$  has an emergent, order 2 anomaly \cite{Gepner:1986wi}  classified by $H^3(\mathbb{Z}_2,U(1))=\mathbb{Z}_2$ \cite{Freed:1987qk}. 
(See \cite{Han:2017hdv,Metlitski:2017fmd,Bultinck:2017iff,Numasawa:2017crf,Chang:2018iay,Lin:2019kpn,Lin:2021udi,Cheng:2022sgb} for recent discussions of this anomaly.)

We claim that the anomalous algebra of $\Pi$ for odd $L$ \eqref{Pianomaly} is a lattice signal of this anomaly.  As discussed in Section \ref{sec:introCPT}, the self-anomaly of the unitary $\mathbb{Z}_2$ symmetry $\sC$ in the continuum can be probed by the projective algebra of $\Theta_\sC$ in the presence of a $\sC$ defect:
\ie\label{ThetaCsq}
\Theta_\sC^2=-1 \,.
\fe
(See \eqref{ThetaUsq}, where $\sC$ plays role of $\sU$ there.)
The lattice operator $\Pi$ flows to $\Theta$ in the continuum, and the projective algebra \eqref{Pianomaly} is the lattice counterpart of \eqref{ThetaCsq}.  
Therefore, we conclude that the mod 2 lattice anomaly between $T$ and $\Pi$ directly probes the pure $\sC$ anomaly in the continuum.

  The situation in the ferromagnetic phase is simpler.  As in Section \ref{ferromagnetic}, again, $\PP$ is mapped to the continuum symmetry $\sR\sT$.   On the lattice, $\PP^2=(-1)^L$.  In the continuum, because of the Berry term \eqref{Berryterm}, the ground state of the system has spin $k\over 2$.  For even $k$, this is a real $SO(3)$ representation.  And for odd $k$ it is a projective $SO(3)$ representation, which is pseudo-real.  Correspondingly, the anti-unitary transformation $\sR\sT$, which commutes with $SO(3)$ satisfies $(\sR\sT)^2=(-1)^k$.  The comparison with the lattice is $k=L$, and hence, the anomalies match.  More generally, if we have spin $s$ in every site of the lattice, $k=2Ls$, and the lattice anomaly \eqref{ppsL} matches with the continuum.  Finally, since the lattice translation $T$ is mapped to a discrete translation of the continuum theory, its lattice anomaly is essentially the same as its continuum version.

\subsection{Lattice realization of the Smith isomorphism}\label{SmithBosons}

In our discussion above, we encountered two anomalies in the continuum: (i) A mod 2 anomaly for a unitary $\mathbb{Z}_2$ symmetry in 1+1d. 
(ii) A mod 2 anomaly for an anti-unitary time-reversal $\mathbb{Z}_2^{\T}$ symmetry in 0+1d quantum mechanics, i.e., the Kramers' doublet. 
Mathematically, the two anomalies are respectively classified by  $\mathrm{Hom}(\Omega_3^\mathrm{SO}(B\mathbb{Z}_2),U(1)) \cong H^3(\mathbb{Z}_2,U(1)) \cong \mathbb{Z}_2$  
 and $\mathrm{Hom}(\Omega_2^\mathrm{O}(pt),U(1)) \cong \mathbb{Z}_2$ \cite{Kapustin:2014tfa}, which are known to be isomorphic.

The relation between these two anomalies was explained in \cite{Hason:2020yqf,Cordova:2019wpi} using the $\sC\sR\sT$ symmetry of the continuum theory. Here, we will discuss a closely related setup in terms of interfaces in a finite space. 
We start with the continuum $O(3)$ non-linear sigma model with a theta term $\theta=\pi$ \eqref{O3Lagi} on a spatial circle, where the unitary $\mathbb{Z}_2^\sC$ symmetry generated by charge conjugation is known to have an anomaly in 1+1d. 
Next, we deform the $\theta$-angle to be position-dependent: on the left half of the circle, we set $\theta = \pi-\delta$  ($0 < \delta <\pi$) and on the other half, we set $\theta= \pi +\delta$. This creates two interfaces where $\theta$ jumps by $2\delta$. The two regions are exchanged by $\sC$.\footnote{Similar interfaces where the value of $\theta$ changes from $\pi-\delta$ to $\pi+\delta$ often separate two different SPTs, leading to a non-trivial theory along the interface.  See e.g., \cite{Gaiotto:2014kfa,Gaiotto:2017tne,Hsin:2018vcg,Cordova:2019jnf,Cordova:2019uob}.}
(See Figure \ref{fig:smith} with $\pm M$  replaced by $\theta=\pi \pm \delta$.) 
Deep in each region, the physics is trivially gapped by the deformation away from $\theta=\pi$.\footnote{For $\theta=\pi$, the $O(3)$ non-linear sigma model flows to the $c=1$ compact boson CFT at the self-dual radius. Deforming away from $\theta=\pi$ corresponds to turning on a cosine potential $ \cos \tilde{\phi}$. See Appendix \ref{compactboson}.} The low-energy dynamics reduces to two decoupled effective 0+1d quantum mechanical systems, each is a Kramers' doublet localized at an interface. While the $\theta$ deformation breaks $\sC$ and $\Theta$ separately, it preserves the anti-unitary symmetry: 
\begin{equation}
\label{smithspin}
\sR\sT = \sC  \Theta\,.
\end{equation}
Thus, $\sC\sR\sT$ relates the unitary $\mathbb{Z}_2$ symmetry $\sC$ in 1+1d to the anti-unitary time-reversal symmetry $\sR\sT$ in 0+1d. Equation \eqref{smithspin} and the $\theta$ deformation  provide a physical realization of the Smith isomorphism between these symmetries across dimensions.

Next, we discuss the lattice counterpart of this interface construction. 
We start with the gapless Heisenberg model \eqref{Heisenberg}, which shares the same continuum limit as the $O(3)$ non-linear sigma model \eqref{O3Lagi} at $\theta=\pi$. 
Then, we deform the Hamiltonian to the following:
\ie\label{HDWs}
&H_\text{Interface}\\
&=\begin{cases}
    \sum_{m=1}^n \left(\vec{S}_{2m-1} \cdot\vec{S}_{2m} + \vec{S}_{-2m }\cdot\vec{S}_{-2m+1} \right) & L=4n+2\text{ or }4n+1\\
    \sum_{m=1}^{n-1} \left(\vec{S}_{2m-1}\cdot \vec{S}_{2m} + \vec{S}_{-2m }\cdot\vec{S}_{-2m+1} \right) + \vec{S}_{2n-1}\cdot\vec{S}_{-2n+1}& L=4n\text{ or }4n-1\\
\end{cases}
\fe
The local terms in the two halves of the circle are exchanged by translation $T$, the lattice counterpart of $\sC$.
Although  $H_\text{Interface}$ breaks $T$ and $\Pi$, it preserves the anti-unitary symmetry: 
\ie 
\PP = T^{-1} \Pi \,.
\fe 
This is the lattice counterpart of \eqref{smithspin}.

As discussed in Appendix \ref{app:spinanomaly}, this deformation gaps the bulk of the two halves of the circle. 
Indeed, under the identification $(-1)^j\vec S_j \cdot \vec S_ {j+1}\to \Gamma$ \cite{Affleck:1988nt,Affleck:1988zj} (see Appendix \ref{compactboson}), \eqref{HDWs} corresponds to changing the $\theta$-angle by the opposite amounts away from $\theta=\pi$ on the left and right halves of the circle.

For odd $L$, there is a single interface at $j=0$, where there is an unpaired qubit. 
The anti-unitary symmetry $\PP$, which obeys $\PP^2=-1$ as in \eqref{Thetanomaly}, acts projectively on this qubit, which forms a Kramers' doublet. 
For even  $L$, there are two interfaces at  $j=0$ and $j=L/2$.  
At low energies, the local Hilbert spaces at these points are decoupled and each of them hosts a Kramers' doublet. 
These Kramers' doublets are protected by $\PP$ --- there is no local $\PP$-symmetric term that can lift the degeneracy.\footnote{Note that the two Hamiltonians $\sum_m \vec S_{2m-1}\cdot \vec S_{2m}$ and $\sum_m \vec S_{2m} \cdot \vec S_{2m+1}$ are in two different SPT phases with respect to $SO(3)$ (or a $\mathbb{Z}_2\times \mathbb{Z}_2$ subgroup thereof). However, here we do not assume the $SO(3)$ global symmetry, or any of its subgroups, so this fact is not directly relevant to us. Relatedly, the degeneracies from the unpaired qubits on the interfaces are protected by $\PP$, not by $SO(3)$ or any other internal symmetries.}

Therefore, the deformation from the Heisenberg spin chain \eqref{Heisenberg} to \eqref{HDWs} continuously interpolates between a 1+1d gapless system to an effective 0+1d quantum mechanical model localized at the interfaces.
Correspondingly, the lattice CRT symmetry $\Pi$ relates the lattice translation $T$ in 1+1d to an anti-unitary time-reversal symmetry $\PP$ of the 0+1d quantum mechanical system. This provides the lattice realization of the Smith isomorphism.

\section{The Levin-Gu edge model}\label{Levin-Gu}

\subsection{The model and its symmetries}\label{Levin-Gusymme}

In this section, we study the Levin-Gu edge model \cite{Levin:2012yb}. 
As we will see, it exhibits similar phenomena to those discussed in the previous examples, but there are also some new ingredients.

Consider the Hamiltonian on a 1+1d spin chain with $L=0 \bmod 4$ sites:
\ie\label{LG}
H  =  \sum_{j=1}^L (X_j -Z_{j-1}X_j Z_{j+1} )\,.
\fe
This is the edge theory of a lattice model for the 2+1d $\mathbb{Z}_2$ SPT phase in  \cite{Levin:2012yb}.

We are interested in the following symmetries (see \cite{Cheng:2022sgb} for more details):
\begin{itemize}
\item A $\mathbb{Z}_2^\text{even}\times\mathbb{Z}^\text{odd}_2$ symmetry generated by
\ie
\X_\text{even} = \prod_{m=1}^{L/2} X_{2m} \,,\qquad
\X_\text{odd} = \prod_{m=1}^{L/2}X_{2m+1} \,. 
\fe
We also define the generator of the diagonal $\bZ_2^\X$ symmetry
\ie
\X= \X_\text{even}\X_\text{odd}=\prod_{j=1}^LX_j\,.
\fe
\item A $U(1)^\Q$ symmetry, whose charge is
\ie
\Q = \frac 14 \sum_{j=1}^L Z_j Z_{j+1}\,.
\fe
Note that while $\mathbb{Z}_2^\text{even}\times\mathbb{Z}^\text{odd}_2$ acts ``on-site,'' this is not the case for $U(1)^\Q$.  However, the model has a dual presentation in which $U(1)^\Q$ acts on-site, while $\mathbb{Z}_2^\text{even}\times\mathbb{Z}^\text{odd}_2$ does not \cite{Levin:2012yb}.

Of particular importance is $\mathbb{Z}_2^\Q\subset U(1)^\Q$, which is generated by
\ie
e^{i \pi \Q} = e^{ {i\pi \over4}\sum_{j=1}^L Z_jZ_{j+1}}=(-1)^{L\over4} \prod_{j=1}^LZ_j \,\mathrm{CZ}_{j,j+1}
 \,,
\fe
where $\mathrm{CZ}_{j,k} = {1+Z_j +Z_k -Z_jZ_k\over2}$.

\item Lattice translation by one site $T$.

\item Site-centered reflection $\R: \mathcal{O}_j \to \mathcal{O}_{-j}$.

\item Time-reversal\footnote{Note that $\T$ here is a different time-reversal operator compared to the one defined in \eqref{CRTLatticeS}. 
They differ by an internal transformation $X_j\to -X_j$, $Z_j\to -Z_j$.
} 
\ie \label{eq:LGtimereversal}
\T X_j \T^{-1} = X_j\,,
\quad
\T Y_j \T^{-1} = -Y_j\,,
\quad
\T Z_j \T^{-1} = Z_j\,.
\fe

\end{itemize}

The continuum limit is the $c=1$ compact boson CFT at radius $R=\sqrt{2}$.\footnote{Here, we use the convention, where the self-dual radius is $R=1$. 
This CFT is also the free Dirac fermion field theory with the gauged fermion parity, which is the reason this value of the radius is often referred to as the Dirac point.}
We can take the continuum limit of $\Q$ to be the momentum $U(1)$ charge $\mathsf{Q}_\text{m}$, and that of $\X$ to be the winding $\mathbb{Z}_2^\text{w}$ operator $e^{i\pi \mathsf{Q}_\text{w}}$.  The remaining $\mathbb{Z}_2^\text{even} $ lattice symmetry, which is generated by $\mathcal{X}_\text{even}$, flows in the continuum to the $\bZ_2$ symmetry that flips the signs of $\mathsf{Q}_\text{m}$ and $\mathsf{Q}_\text{w}$.

Next, we discuss some  anomalies of the lattice internal symmetries that we will diagnose in later subsections.
Each of the symmetry operators $\Q$, $\X_\text{even}$, and $\X_\text{odd}$ is separately anomaly-free.  However, there is a mixed anomaly between $\X$ and $e^{i\pi \Q}$,
which corresponds in the continuum to the mixed anomaly between $\exp(i \pi \mathsf{Q}_\text{m})$ and $\exp(i\pi \mathsf{Q}_\text{w})$.

We are particularly interested in a $\mathbb{Z}_2^\U$ subgroup of the internal symmetry, which is generated by\footnote{When $L=0$ mod 4, one can perform a unitary transformation $\prod_{j=1,2~\text{mod}~4}X_j$ to map $\U$ to $\prod_{j=1}^L \mathrm{CZ}_{j,j+1}\prod_{j=1}^L X_j$, which is sometimes known as the CZX symmetry \cite{Chen:2011bcp}.  } 
\ie
\U  = e^{ i \pi \Q} \X\,.
\fe

It is known that the lattice $\mathbb{Z}_2^\U$ symmetry has a self-anomaly \cite{Chen:2011bcp,Levin:2012yb,Else:2014vma,Wang:2017loc,KawagoeH32021}.  In the continuum, $\U$ flows to $\sU=\exp(i \pi \mathsf{Q}_\text{m}+i\pi \mathsf{Q}_\text{w})$, which generates a $\mathbb{Z}_2^\sU\subset U(1)^\text{L}$ symmetry. 
Here, $U(1)^\text{L}$ is a chiral symmetry generated by $\mathsf{Q}_\text{m}+ \mathsf{Q}_\text{w}$.
The lattice anomaly of $\mathbb{Z}_2^\U$ is the same as the continuum anomaly associated with the nontrivial element of $H^3(\mathbb{Z}_2,U(1))=\mathbb{Z}_2$.\footnote{More generally, the $\mathbb{Z}_2^3$ symmetry generated by $\mathcal{X}_\text{even}, \mathcal{X}_\text{odd}, e^{i\pi \Q}$ has a type III anomaly \cite{deWildPropitius:1995cf,Wang:2014tia} of the form $i \pi \int A_\text{even}\cup A_\text{odd} \cup A_Q$, where the $A$'s are their $\mathbb{Z}_2$ background gauge fields.  This anomaly manifests itself in the projective phases in the symmetry algebra in the presence of defects \cite{Seifnashri:2024dsd,Seiberg:2025bqy}.}

We now turn to the crystalline symmetries.
In the continuum limit, the lattice translation operator $T$ leads to an emanant $\mathbb{Z}_4$ symmetry generated by $e^{i\pi \mathsf{Q}_\text{w}/2}$. However, unlike other emanant symmetries, here, $T^2$ does not lead to a new symmetry. Instead, both $T^2$ and $\X$ flow to $e^{i\pi \mathsf{Q}_\mathrm{w}}$ in the continuum \cite{Cheng:2022sgb}.

\subsection{Identifying $\Pi$ that flows to $ \Theta$}

We claim that the lattice operator that flows to the canonical  CRT operator $\Theta$ in the continuum is\footnote{$\mathcal{X}_\text{even} $ and $\mathcal{X}_\text{odd} $ are conjugate to each other by $T$.  The symmetry between them was broken by the fact that we used $\R$, which is a site-centered reflection around an even site rather than around an odd site.  In other words, we can write $\Pi=\mathcal{X}_\text{odd}\R\T=T\mathcal{X}_\text{even}(T^{-1}\R T) \T T^{-1}$, where $(T^{-1}\R T)$ is reflection around an odd site.  The conjugation by $T$ simply changes the fixed point of the reflection.\label{evenvsodd}}
\ie
\Pi  = \mathcal{X}_\text{odd}\R\T\,.
\fe
It acts on the Pauli operators as
\ie
\Pi X_j \Pi^{-1} =  X_{-j}\,,\quad
\Pi Y_j \Pi^{-1} =  (-1)^{j+1} Y_{-j}\,,\quad 
\Pi Z_j \Pi^{-1} = (-1)^j Z_{-j}\,.
\fe
Note that, unlike  $\Pi$ in the preceding sections, here $\Pi$ involves a site-centered (rather than link-centered) spatial reflection.
As a consistency check, we note that $\Pi$ commutes with all the internal global symmetry operators:\footnote{The Hamiltonian \eqref{LG} also has an exactly conserved lattice winding charge which flows to $\mathsf{Q}_\text{w}$ \cite{Pace:2024oys}
\ie
\tilde \Q = \frac 12 \sum_{m=1}^{L/2}(Y_{2m}Z_{2m+1} -Z_{2m}Y_{2m+1})\,.
\fe
It generates another $U(1)$ lattice symmetry whose $\mathbb{Z}_2$ subgroup is generated by  $e^{i \pi \tilde \Q} =\X$. 
Under $\Pi$, it becomes
\ie
\Pi \tilde \Q\Pi^{-1} =  - T\tilde \Q T^{-1}\,.
\fe
This is consistent with the map from the lattice to the continuum $T\to e^{i\pi \mathsf{Q}_\text{w}/2},\Pi\to \Theta, \tilde \Q\to \mathsf{Q}_\text{w}$ and $\Theta \mathsf{Q}_\text{w} = -\mathsf{Q}_\text{w} \Theta$. Note that $\tilde \Q$ is broken by deformations such as $\sum_j Z_{j-1}Z_{j+1}$,  which preserve all the other symmetries discussed above, including those in \eqref{symmetryflowsum}, as well as $\R$ and $\T$.}
\ie
\Pi \mathcal{X}_\text{even} = \mathcal{X}_\text{even}\Pi\,,\quad
\Pi \mathcal{X}_\text{odd} = \mathcal{X}_\text{odd}\Pi\,,\quad
\Pi e^{i \alpha \Q}  = e^{i \alpha \Q}\Pi\,.
\fe
Importantly, $\Pi \Q = -\Q\Pi$. It also obeys
\ie
\Pi^2=1\,,\quad
T \Pi = \X \Pi T^{-1}\,.
\fe
We also define another anti-unitary operator $\PP$ as
\ie
\PP = \U \Pi\,.
\fe

As another consistency check of our identification of $\Pi$ as the lattice version of $\Theta$, note that the symmetry generated by $\Pi$ is anomaly-free and has no mixed anomaly with the other anomaly-free internal symmetries $ \Q, \mathcal{X}_\text{even}, \mathcal{X}_\text{odd}$.

For the symmetry generated by $\Pi$ and $\mathcal{X}_\text{even}$, the paramagnetic Hamiltonian 
\ie
\sum_{j=1}^L X_j
\fe
is symmetric and trivially gapped.  The same Hamiltonian can be used to argue that $\Pi$ and $\mathcal{X}_\text{odd}$ are anomaly-free. 

To show that the symmetry generated by $\Pi$ and $\Q$ is anomaly-free, consider the following Hamiltonian, which is symmetric and trivially gapped:\footnote{Following footnote \ref{evenvsodd}, one could ask how we identified $\Pi$ as the operator that flows to $\Theta$, rather than
$\Pi_\text{e}   = \mathcal{X}_\text{even} \R\T = X\Pi$, which flows to $\exp(i \pi \mathsf{Q}_\text{w})\Theta$. The key point is that the lattice anomaly between $\X$ and $\Q$ leads to a mixed anomaly between $\Pi_\text{e} $ and $\Q$. See the discussion in Subsection \ref{sec:XQanomaly}. On the other hand, there is no anomaly between $\Pi$ and $\Q$ as we show here.}
\ie
Z_0 +Z_{\frac L2}+\sum_{j=1}^{\frac L2-1} Z_j + \sum_{j=\frac L2+1}^{L-1} (-1)^jZ_j\,.
\fe
(Note that here it is important that $L=0$ mod $4$.)

Below we want to focus on a smaller set of operators and analyze their anomalies.  
These operators are the lattice translation $T$, the lattice CRT operator $\Pi$, and an internal $\mathbb{Z}_2\times \mathbb{Z}_2$ symmetry generated by $e^{i \pi \Q}$ and $\X$. They satisfy the algebra 
\ie 
& \Pi^2=\X^2=\left(e^{i \pi \Q}\right)^2=1\,,\quad T^L = e^{i\pi \Q}\,,\\
& T\X = \X T\,,\quad T e^{i\pi \Q} = e^{i\pi \Q}T\,, \quad \X  e^{i\pi \Q} =  e^{i\pi \Q} \X\,,\\
&  e^{i\pi \Q} \Pi = \Pi e^{i\pi \Q}\,,\quad \Pi \X = \X \Pi\,,\quad \Pi T = \X T^{-1}\Pi\,,
\fe 
with no projective phases. 

The dictionary between the lattice and continuum operators is summarized as:
\ie\label{symmetryflowsum}
&\X &&\rightarrow &&\exp( i \pi \mathsf{Q}_\text{w})  \\
 &\exp({i\pi \Q}) &&\rightarrow && \exp({i\pi \mathsf{Q}_\text{m}}) \\
 &\U = \exp({i \pi \Q}) \X&&\rightarrow &&\exp(i \pi \mathsf{Q}_\text{m} +i\pi \mathsf{Q}_\text{w})\\
 & T && \rightarrow && \exp\left({i \pi\over 2} \mathsf{Q}_\text{w}\right)\\
 &\Pi &&\rightarrow &&\Theta
\fe

In the following, we explore the anomalies of these symmetries. 

Before we dive into the details, we first note that with two copies of the model, there exists a trivially gapped Hamiltonian preserving all these symmetries,
\ie \label{HpLG}
H' = -\sum_{j} Z_j^{A=1}Z_j^{A=2} - \sum_j \left(X_j^{A=1}X_j^{A=2} + \prod_{A=1,2} Z_{j-1}^{A} X_j^{A} Z_{j+1}^{A}\right).
\fe 
The ground state is the product of local Bell states with $Z_j^{A=1}Z_j^{A=2} = X_j^{A=1}X_j^{A=2} = 1$. Therefore, the order of any anomaly involving these symmetries is at most 2.

\subsection{Self-anomaly of $\U$}\label{sec:selfU}

Next, we would like to detect the anomaly of the lattice symmetry $\U$. The standard way to do it uses the algorithm in \cite{Else:2014vma,KawagoeH32021,Seifnashri:2023dpa} in terms of the F-symbols.  Instead, here, we will take another approach.

In the continuum limit, $\U$ becomes the anomalous $\mathbb{Z}_2^\sU$ symmetry generated by $\sU=e^{i\pi \mathsf{Q}_\text{m}+i\pi \mathsf{Q}_\text{w}}$. 
As discussed in Section \ref{sec:introCPT}, the self-anomaly of $\sU$ can be probed by the projective algebra of $\Theta_\sU$ in the presence of a $\sU$ defect as in \eqref{ThetaUsq}:
\ie\label{ThetaUsqLG}
\Theta_\sU^2=-1\,.
\fe
Since the lattice $\U$ and $\Pi$ flow to the continuum $\sU$ and $\Theta$, we can identify the same algebra on the lattice.  In turn, this will provide an alternative anomaly indicator on the lattice.

To this end, we insert a $\U$ defect around site 0. This is described by the Hamiltonian \cite{Cheng:2022sgb}
\ie 
H_\U = &\sum_{j\neq -1,0,1}
(X_j-Z_{j-1}X_{j}Z_{j+1})\\
&+(X_{-1}-Z_{-2}X_{-1}Z_{0})
+(Z_{-1}Y_{0}+Y_{0}Z_{1})+(X_1+Z_{0}X_1Z_{2})\,.
\fe 

In the presence of the $\U$ defect,  $\Pi$ is modified to 
\ie\label{PiUla}
\Pi_\U = Y_0 \Pi =  Y_0 \mathcal{X}_\text{odd}\R\T\,.
\fe
It satisfies
\ie\label{PiUsq}
(\Pi_\U)^2 = Y_0 \mathcal{X}_\text{odd}\R\T Y_0 \mathcal{X}_\text{odd}\R\T 
=  -1\,.
\fe
We conclude that $\Pi$ is anomalous in the presence of a $\U$ defect, implying a mixed anomaly between them. 
In the continuum, $\Pi$ becomes $\Theta$, and \eqref{PiUsq} is the lattice counterpart of \eqref{ThetaUsqLG}. 
Hence, this lattice anomaly becomes the mod 2 self-anomaly of $\sU=e^{i\pi \mathsf{Q}_\text{m}+i\pi \mathsf{Q}_\text{w}}$, classified by $H^3(\bZ_2,U(1))\cong\bZ_2$.

In the presence of the $\U$ defect, the lattice symmetry generated by $\U$ is modified to 
\ie\label{UUdef} 
\U_\U = Z_0Z_1e^{i \pi \Q}\X\,.
\fe 
Then, using \eqref{PiUla}, 
\ie\label{UUrelat}
\U_\U^2=1\qquad,\qquad \Pi_\U\U_\U=\U_\U\Pi_\U\,.
\fe 
Similarly,  $\PP$ is modified to 
\ie 
\PP_\U = \U_\U \Pi_\U=Z_{-1}Z_0Y_0\PP\,,
\fe 
which is unitarily equivalent to $\Pi_\U$ using conjugation by $T_\U^{\frac{L}{2}}$, where $T_\U^L=\U_\U$. Therefore, it obeys the same algebra as $\Pi_\U$, 
\ie 
\PP_\U^2=-1\,,
\fe 
which also probes the self-anomaly of $\U$. This is similar to the previous fermionic and bosonic cases in Subsections \ref{sec:majRT} and \ref{sec:spinRT}, where in the presence of ``lattice translation defect," the operator $\tilde{\Pi}$ becomes anomalous. See more comparisons in Section \ref{sec:conclusions}.

Finally, to match with the continuum, it is natural to take $\U_\U\to i\sU_\sU$ such that the continuum version of the two relations in \eqref{UUrelat} pick a minus sign, as in \eqref{Uanomal}.

\subsection{Anomaly for $\PP$} \label{PPLevinGu}
Given that there is a mixed anomaly between $\Pi$ and $\U$, it is natural to ask if $\PP$ itself is anomalous. 
We will follow the discussion in Section \ref{sec:spinRT} to identify a mod 2 anomaly for $\PP$. 
We will restrict ourselves to the even $L$ case, and separate the discussion into two cases: $L=0,2$ mod 4.\footnote{In the discussion in Section \ref{Levin-Gusymme}, we limited ourselves to $L=0\bmod 4$. This was necessary for the properties of $\Q$.  However, here we do not discuss $\Q$, so we can study the model for any even $L$.} It will become clear that these two cases are analogous to the even and odd $L$ cases in Section \ref{sec:spinRT}.

First, as commented earlier, the upper bound on any anomaly involving $\Pi$ and $\U$, and therefore $\PP$, is 2 because of the Hamiltonian \eqref{HpLG}.

To find a lower bound for the anomaly, we look for a new basis where $\PP$ acts in a simple way. 
We conjugate the operators by the following unitary transformation
\ie 
\prod_{j} \mathrm{CZ}_{2j,2j+1}\,.
\fe 
The Pauli operators in the new basis are
\ie 
(\text {even } j) & \quad X'_j= X_jZ_{j+1}\,, \quad Y'_j= Y_jZ_{j+1}\,, \quad Z'_j=Z_j\,, \\
(\text {odd } j) & \quad X'_j=Z_{j-1} X_j\,, \quad Y'_j=Z_{j-1} Y_j\,, \quad Z'_j=Z_j \,.
\fe 
$\PP$ acts on these operators as
\ie 
\PP \, X'_j\, \PP^{-1} = (-1)^{j+1} X'_{-j}\,, \quad \PP\, Y'_j \,\PP^{-1}= -Y'_{-j}\,, \quad \PP\, Z'_j \,\PP^{-1}= (-1)^{j+1} Z'_{-j}\,.
\fe 
When $L$ is even, there are two fixed points localized at $j=0$ and $j=L/2$. 
The qubits away from these two fixed points can be paired by the action of $\PP$, such that $\PP$ acts non-projectively on them. 
We can then view $\PP$ as a time-reversal operator acting on the two 0+1d quantum mechanical systems at $j=0$ and $j=L/2$.

When $L = 2 \bmod 4$, 
\ie 
& \PP\vec{S}'_0 \PP^{-1} = - \vec{S}'_0\,, \quad \PP \vec{S}'_{L/2} \PP^{-1} = \T\vec{S}'_{L/2}\T^{-1}\,,
\fe
where we used the fact that $\T$ acts as complex conjugation in this basis, in the same way as in \eqref{eq:LGtimereversal}. 
Furthermore, 
\ie\label{Lis2mod4}
\PP^2=-1\,,~~~~L=2~\text{mod}~4\,,
\fe
which implies a time-reversal anomaly for this quantum mechanical system. 
This establishes a lower bound of 2 on the order of the anomaly when $L=2$ mod 4.

When $L=0 \bmod 4$, $\PP$ acts   at on these two qubits as
\ie\label{Kramers0L2} 
\PP \vec{S}'_0 \PP^{-1} = - \vec{S}'_0\,,\quad \PP \vec{S}'_{L/2} \PP^{-1} = - \vec{S}'_{L/2}\,.
\fe
Furthermore, $\PP^2=+1$, and there is no time-reversal anomaly when viewed as a 0+1d quantum mechanical system. 
Nonetheless, as discussed in Sections \ref{sec:majRT} and \ref{sec:spinRT}, using the locality properties of the systems, these two far-separated qubits  at $j=0$ and at $j={L\over 2}$ cannot be coupled in a local way, and \eqref{Kramers0L2}
signal the existence of a 1+1d anomaly also for $L=0\bmod 4$.

We conclude that $\PP$ has a mod 2 anomaly. This implies an LSM-type constraint:  every local Hamiltonian that is invariant under $\PP$ cannot be trivially gapped.

\subsection{More anomalies}\label{sec:moreanomalies}

In Section \ref{sec:selfU}, we used the lattice CRT operator $\Pi$ to detect the self-anomaly of $\U$. We now extend this approach to diagnose additional anomalies.

In preparation for this analysis, we discuss some defects.

The Hamiltonian with an $e^{i \pi \Q}$ defect is \cite{Cheng:2022sgb}
\ie\label{HQdefect}
H_{\Q} = \sum_{j\neq 0} (X_j -Z_{j-1}X_jZ_{j+1})
+(Z_{-1}Y_0 -Y_0 Z_1)\,.
\fe
The modified $\X$, $T$ and $\Pi$ are
\ie
\X_\Q = \X\,,\quad T_\Q =e^{i\frac{\pi}{4}Z_0Z_1}T \,,\quad \Pi_\Q =  Z_0\Pi\,.
\fe
They obey the algebra
\ie \label{eq:algebraQdefct}
& \Pi_\Q^2=\X_\Q^2=\left(e^{i \pi \Q}\right)^2=1\,,\quad T_\Q^L = e^{i\pi \Q}\,,\\
& T_\Q\X_\Q = \X_\Q T_\Q\,,\quad T_\Q e^{i\pi \Q} = e^{i\pi \Q}T_\Q\,, \quad \X  e^{i\pi \Q} =  e^{i\pi \Q} \X\,,\\
&  e^{i\pi \Q} \Pi_\Q = \Pi_\Q e^{i\pi \Q}\,,\quad \Pi_\Q \X_\Q = - \X_\Q \Pi_\Q\,,\quad \Pi_\Q T_\Q = -i \X_\Q T_\Q^{-1}\Pi_\Q\,.  
\fe

We also consider the Hamiltonian with an $\X$ defect:
\ie
H_\X =
&\sum_{j\neq 0,1}(X_{j} - Z_{j-1} X_{j}Z_{j+1})
+(X_0 +Z_{-1} X_0Z_1)+(X_1 +Z_0 X_1 Z_2)
\,.
\fe 
The modified $\X$, $T$ and $\Pi$ are
\ie
e^{i \pi \Q_\X} = Z_0Z_1e^{i \pi \Q}\,,\quad T_\X =X_1T \,,\quad \Pi_\X =  X_0\Pi\,.
\fe
They obey the algebra
\ie \label{eq:algebraXdefct}
& \Pi_\X^2=\X^2=(e^{i\pi\Q_\X})^2=1\,,\quad T_\X^L = \X\,,\\
& T_\X \X = \X T_\X\,, \quad T_\X e^{i\pi \Q_\X}=e^{i\pi \Q_\X}T_\X\,,\quad \X e^{i\pi \Q} = e^{i\pi \Q}\X \\
& e^{i\pi \Q_\X}\Pi_\X = -\Pi_\X e^{i\pi \Q_\X}\,,\quad  \Pi_\X \X = \X \Pi_\X\,,\quad \Pi_\X T_\X = \X T_\X^{-1}\Pi_\X
\fe 

Below, we will interpret the projective phases in \eqref{eq:algebraQdefct} and \eqref{eq:algebraXdefct} as various lattice anomalies involving $\Pi$ and compare them to the continuum.

\subsubsection{The lattice anomaly between $\X$ and $e^{i\pi \Q}$}\label{sec:XQanomaly}

Following the discussion in Section \ref{sec:selfU}, here, we use $\Pi$ to probe the mixed anomaly between the lattice operators $\X$ and $e^{i\pi \Q}$.  See \cite{ZhangLevinDQCP,Cheng:2022sgb} for related discussions of this anomaly. 

In the presence of an $\X$ defect \eqref{eq:algebraXdefct},
\ie 
\left(e^{i\pi \Q_\X}\Pi_\X\right)^2=-1\,.
\fe 
This means that $e^{i\pi \Q}\Pi$ is anomalous in the presence of an $\X$-defect. Alternatively, with a $e^{i\pi \Q}$ defect \eqref{eq:algebraQdefct}, 
\ie 
\left(\X_\Q \Pi_\Q\right)^2 = -1\,.
\fe 
This means that $\X\Pi$ is anomalous in the presence of an $e^{i\pi \Q}$-defect. Therefore, $\Pi$ probes the mod 2 mixed anomaly between $\X$ and $e^{i\pi \Q}$. In the continuum, this becomes the mod 2 mixed anomaly between $e^{i \pi \sQ_\mathrm{w}}$ and $e^{i \pi \sQ_\mathrm{m}}$. 

\subsubsection{The emergent anomaly between $e^{i \frac{\pi}{2}\sQ_\mathrm{w}}$ and $e^{i \pi \sQ_\mathrm{m}}$} \label{sec:anomalyTQ}

Here, we use $\Pi$ to probe the emergent anomaly between $e^{i \frac{\pi}{2}\sQ_\mathrm{w}}$ and $e^{i \pi \sQ_\mathrm{m}}$. (See Section \ref{sec:emanaandem}.)

On the lattice, there is no mixed anomaly between the translation symmetry generated by $T$ and the internal symmetry generated by $e^{i\pi \Q}$. In fact, the paramagnetic Hamiltonian
\ie 
\sum_{j=1}^L Z_j
\fe 
is symmetric under $T$ and $e^{i\pi \Q}$, and is trivially gapped. 

However, in the continuum, $T\to e^{i \frac{\pi}{2}\sQ_\mathrm{w}}$ and $e^{i\pi \Q} \to e^{i \pi \sQ_\mathrm{m}}$, and these two continuum symmetries have a mod 2 emergent anomaly. To probe this emergent anomaly on the lattice, we can use the operator $\Pi$.  (This is similar to the examples analyzed in Sections \ref{sec:majCRT} and \ref{sec:spinanomaly}.)

In the presence of an $e^{i\pi \Q}$ defect \eqref{eq:algebraQdefct}, we find
\ie 
\left(T_\Q^{-1}\Pi_\Q\right)^4 = -1\,.
\fe 
This means that $T^{-1}\Pi$ is anomalous in the presence of a $e^{i\pi \Q}$-defect.  Therefore, $\Pi$ probes the emergent, mod 2 mixed anomaly between $e^{i \frac{\pi}{2}\sQ_\mathrm{w}}$ and $e^{i \pi \sQ_\mathrm{m}}$.

\section{Non-invertible RT}\label{sec:noninvertible}

In Section \ref{sec:maj}, we discussed the $\PP$ operator on a Majorana chain. 
Here, we discuss its counterpart on the spin chain under bosonization. 
We will see that $\PP$ becomes a non-invertible operator that reflects both space and time. Other examples of non-invertible time-reversal and spatial-reflection symmetries have been discussed in \cite{Kaidi:2021xfk,Choi:2022rfe,Moy:2025esw} and \cite{Seiberg:2024gek,Pace:2024tgk,Gorantla:2024ocs,Graf:2024itd,Pace:2025hpb} respectively. 
See \cite{Schafer-Nameki:2023jdn,Shao:2023gho} for more general reviews on non-invertible symmetries.

Consider a periodic spin chain of $N$ qubits. 
Even though our discussion below will be independent of the choice of the Hamiltonian, it is useful to have the critical transverse-field Ising model in mind:
\ie\label{Ising}
H_\text{Ising} = -\sum_{j=1}^N X_j - \sum_{j=1}^N Z_jZ_{j+1}\,.
\fe
This Hamiltonian has the following symmetries
\begin{itemize}
\item Lattice translation $T_\text{Ising}$, which acts as $T_\text{Ising}O_jT^{-1}_\text{Ising} = O_{j+1}$.
\item An invertible, unitary $\mathbb{Z}_2$ symmetry generated by $\eta= \prod_{j=1}^N X_j$, which acts on the Pauli operators as
\ie
\eta X_j \eta^{-1} = X_j\,,\quad \eta Z_j \eta^{-1} = -Z_j\,.
\fe

\item A non-invertible operator $\D$ which implements the Kramers-Wannier transformation 
\ie
\D X_j  =Z_{j}Z_{j+1} \D\,,~~~~\D Z_{j}Z_{j+1}  = X_{j+1} \D\,.
\fe
The operators $T_\text{Ising},\eta,\D$ obey the non-invertible algebra \cite{Seiberg:2023cdc,Seiberg:2024gek}:
\ie\label{noinD}
 \D^2=  (1+\eta)T_\text{Ising}\,,\quad
\D \eta = \eta\D =\D \,,\quad \eta^2=1 \,,\quad T_\text{Ising}^N=1\,.
\fe
Crucially, unlike its continuum limit counterpart, it involves the lattice translation operator $T_\text{Ising}$, and therefore, the number of elements in the algebra grows with the system size $N.$

\item Site-centered reflection $\R_\text{Ising}$ , which acts on the local operators as
\ie\label{Rsite}
\R_\text{Ising} O_j \R_\text{Ising}^{-1} = O_{-j}\,.
\fe

\item An anti-unitary, time-reversal operator $\T_\text{Ising}$, which in our basis simply acts by complex conjugation: 
\ie\label{TIsing}
\T_\text{Ising} X_j \T_\text{Ising}^{-1} =X_j\,,~~~\T_\text{Ising} Y_j \T_\text{Ising}^{-1} = -Y_j \,,~~~\T_\text{Ising} Z_j \T_\text{Ising}^{-1} = Z_j\,,~~~\T_\text{Ising} i \T_\text{Ising}^{-1} = -i\,.
\fe

\end{itemize}
(The operators $T_\text{Ising}$ and $\R_\text{Ising}$ are the same as the translation and the site-centered reflection in the Heisenberg model, defined in \eqref{CRTLatticeS} and in the Levin-Gu edge model in Section \ref{Levin-Gu}. 
The time-reversal operator $\T_\text{Ising}$, on the other hand, is the same as the time-reversal operator  in the Levin-Gu edge model in  \eqref{eq:LGtimereversal}. 
We added the subscript ``Ising" to distinguish these operators from $T$, $\R$, and $\T$ of the Majorana chain.  
This will become important below when we discuss the Jordan-Wigner transformation and in  Table \ref{table:Ising}.)

We define an anti-linear operator
\ie
\PP_\text{Ising} =  \D \R_\text{Ising}\T_\text{Ising}\,,
\fe
which acts on the local operators as
\ie
\PP_\text{Ising} X_j   = Z_{-j }Z_{-j+1} \PP_\text{Ising}\,,~~~~
\PP_\text{Ising} Z_jZ_{j+1} = X_{-j} \PP_\text{Ising}\,.
\fe
Since $\PP_\text{Ising} \prod_{j=1}^N X_j  = \prod_{j=1}^N (Z_{-j}Z_{-j+1})\PP_\text{Ising} =\PP_\text{Ising}$, this operator is non-invertible.\footnote{In particular,  $\PP_\text{Ising}$ is anti-linear, but not anti-unitary.}
Using $\T_\text{Ising} \D =\D\T_\text{Ising}$ and $\R_\text{Ising} \D = \D^\dagger \R_\text{Ising}$ \cite{Seiberg:2024gek}, we find the following non-invertible algebra:
\ie
&(\PP_\text{Ising})^2= 1+ \eta\,,~~~~\PP_\text{Ising} \eta = \eta\PP_\text{Ising} = \PP_\text{Ising}\,,\quad \eta^2=1\,.
\fe
Note that, unlike the non-invertible algebra of $\D$ \eqref{noinD}, this algebra of does not involve lattice translation.  

Now, we relate these operators on the spin chain to those on the Majorana chain. 
The spin chain of $N$ sites is related to the Majorana chain of $L=2N$ sites (and $N_f=1$) by the Jordan-Wigner transformation. In particular, translation by two Majorana sites in the Majorana chain becomes translation by one site in the spin chain:
\ie
T^2  \to T_\text{Ising}\,. 
\fe

Locally, the $(-1)^F$-even operators of the Majorana chain are mapped to the Pauli operators  as 
\ie
i \chi_{2j-1}\chi_{2j} = X_j \,,~~~~
i\chi_{2j}\chi_{2j+1} = Z_j Z_{j+1}\,.
\fe
The gapless Majorana Hamiltonian \eqref{HNf} is locally equivalent to the critical Ising Hamiltonian \eqref{Ising} under this map. 
However, globally, the Jordan-Wigner transformation is not a local unitary transformation on a closed chain. 
The gapless Majorana Hamiltonian is not equivalent to the Ising Hamiltonian as they have different spectra. 
Rather, they are related by gauging $(-1)^F$.  
In particular, the nature of a symmetry operator changes under gauging.

Consider for instance the lattice translation $T$ on the Majorana chain. 
Since $T$ has an anomaly with $(-1)^F$, i.e., $T(-1)^F= - (-1)^FT$, after gauging $(-1)^F$, it becomes the non-invertible operator $\D$ \cite{Seiberg:2023cdc}
\ie
T \to \D\,.
\fe
(See \cite{Thorngren:2018bhj,Ji:2019ugf} for the corresponding discussion in the continuum.)

The site-centered reflection in the Majorana chain is also modified.  The Jordan-Wigner transformation maps two Majorana fermions to one spin. Therefore, a site-center reflection in the spin chain $\R_\text{Ising}$ originates from a link-centered reflection in the Majorana chain
\ie
T^{-1}\R  \to \R_\text{Ising}\,.
\fe

Finally, the time-reversal $\T$ defined in \eqref{eq:Taction}, which acts as $\chi_\ell \to (-1)^\ell \chi_\ell$ on the Majorana chain, becomes $\T_\text{Ising}$ on the spin chain
\ie
\T \to \T_\text{Ising}\,.
\fe

Next, consider the anti-unitary operator $\PP=\R\T$  in \eqref{Thetadef} in the Majorana chain, which acts as $\chi_\ell \to \chi_{-\ell}$. 
It has an anomaly with $(-1)^F$ as in \eqref{even L algebraT}:
\ie
\PP (-1)^F = -(-1)^F \PP\qquad (\text{even $L$ and $N_f=1$)}\,.
\fe
Even though we will not do it explicitly, following the steps in \cite{Seiberg:2023cdc,Seiberg:2024gek,Pace:2024oys},  $\PP$ becomes a non-invertible, anti-linear operator  $\PP_\text{Ising}$
\ie
\PP\to \PP_\text{Ising}\,.
\fe
Indeed, it is straightforward to check that the action of   $\PP_\text{Ising}$ on the spin chain and the action of $\PP$  on the Majorana chain are compatible with the Jordan-Wigner map. 

On the other hand, the anti-unitary operator $\Pi$, which acts as $\chi_\ell \to \chi_{-\ell+1}$  on the Majorana chain, has no  anomaly with $(-1)^F$. 
It becomes an (invertible) anti-unitary operator on the spin chain $\Pi_\text{Ising}= T_\text{Ising}\R_\text{Ising}\T_\text{Ising}$ which acts on the Pauli operators as  
\ie
\Pi_\text{Ising} X_j \Pi_\text{Ising}^{-1} = X_{-j+1}  \,, \quad
\Pi_\text{Ising} Y_j \Pi_\text{Ising}^{-1} = -Y_{-j+1}\,,\quad
\Pi_\text{Ising} Z_j \Pi_\text{Ising}^{-1} = Z_{-j+1}\,.
\fe

In Table \ref{table:Ising}, we summarize how various transformations on the Majorana chain are mapped to transformations on the spin chain.
Note that the operators $T^2$, $T^{-1}\R$,  $\T$, and $\Pi=T\R\T$, their mappings, and their algebra are completely standard and follow from the Jordan-Wigner bosonization transformation.  The non-invertible operator $\D$ and its algebra also follow from bosonization along the lines of \cite{Seiberg:2023cdc,Seiberg:2024gek,Pace:2024oys}.  The only novelty is in the composite operator $\PP\to \PP_\text{Ising}$, which is non-invertible, because it involves $\D$.

\begin{table}
    \centering
    \begin{tabular}{|c||c|c|c|}
    \hline
        Majorana chain & Spin chain  & Anti-linear?  & Invertible? \\
        \hline
        $T$ & $\D$ & $\times$ & $\times$ \\
        $T^2$ & $T_\text{Ising}$  & $\times$  & $\checkmark$\\
        $T^{-1}\R$ &$\R_\text{Ising}$  &$\times$  &$\checkmark$ \\
         $\T$ & $\T_\text{Ising}$  & $\checkmark$ &$\checkmark$ \\ $\Pi=T\R\T=T^2(T^{-1}\R)\T$& ~~$\Pi_\text{Ising} = T_\text{Ising}\R_\text{Ising}\T_\text{Ising}$ ~~ & $\checkmark$ &$\checkmark$ \\
         $\PP = \R\T=T(T^{-1}\R)\T$& $\PP_\text{Ising} = \D\R_\text{Ising}\T_\text{Ising}$  &$\checkmark$  & $\times$ \\
         \hline
    \end{tabular}
    \caption{Operators on the Majorana chain and their counterparts on the spin chain under bosonization. All the operators on the Majorana chain in this table are invertible, while some of their counterparts are non-invertible. }
    \label{table:Ising}
\end{table}

A deformation  of the Ising Hamiltonian \eqref{Ising} that breaks $\T_\text{Ising}$ and $\R_\text{Ising}$, but preserves $\PP_\text{Ising}$ (as well as  $T_\text{Ising},\eta,\D$)  is the Dzyaloshinskii–Moriya interaction \cite{RevModPhys.25.166, PhysRev.120.91,DZYALOSHINSKY1958241}
\ie
H_\text{DM}=  -\sum_{j=1}^N X_j - \sum_{j=1}^N Z_jZ_{j+1} 
 + \lambda \sum_{j=1}^N (Z_{j}Y_{j+1}-  Y_jZ_{j+1}) \,.
\fe
See \cite{SISKENS1975259,SISKENS1975296,PhysRevB.73.214407,Zou:2019dnc,PhysRevB.99.205159,Chatterjee:2024ych} for discussions of this model. 
In fact, under the Jordan-Wigner transformation, it is locally mapped to the $N_f=1$ version of \eqref{eq:HNNN},  $H_\text{NNN}= i\sum_\ell \chi_{\ell}\chi_{\ell+2}$. 
A more nontrivial deformation is locally the bosonized version of \eqref{eq:majint}:
\ie
\sum_{j=1}^N \left[
\alpha_\text{e} (Z_jY_{j+2} +Z_j Y_{j+1}X_{j+2})
+\alpha_\text{o} (Y_j Z_{j+2}+ X_jY_{j+1}Z_{j+2} )\right]\,,
\fe
which for generic $\alpha_\text{e},\alpha_\text{o}$ preserves only $\PP_\text{Ising}$, $\eta$, $T_\text{Ising}$, and breaks $\R_\text{Ising},\T_\text{Ising}, \D$.\footnote{Here, the real parameters $\alpha_\text{e,o}$ are related to $\alpha_\pm$ in \eqref{eq:majint} linearly.}

\section{Conclusions}\label{sec:conclusions}

In this note, we analyzed 1+1d lattice models with anti-unitary symmetries that act as a combination of space-reflection and time-reversal. The anomalies in these symmetries lead to LSM-type constraints.

We saw several distinct situations (see Table \ref{tab:summary}):
\begin{itemize}
   \item The model has an anomaly-free order-2 anti-unitary symmetry operator $\Pi$, and it flows to the continuum $\Theta$.  In that case, we saw several phenomena:
   \begin{itemize}
   \item $\Pi$ has a mixed anomaly with a lattice internal symmetry generated by $\U$. (This mixed anomaly can be identified as an anomaly in $\PP=\U\Pi$.) This mixed anomaly signals a self-anomaly of $\U$.  The latter exists both on the lattice and in the continuum limit.  See examples in Sections \ref{sec:selfU} and \ref{PPLevinGu}. 
   \item $\Pi$ has a mixed anomaly with lattice translation $T$. (This mixed anomaly can be identified as an anomaly in $\PP=T^{-1}\Pi$.) In the continuum, $T$ leads to an emanant symmetry generated by $\sC$ and the lattice anomaly becomes an emergent self-anomaly of that symmetry.  See examples in Sections \ref{sec:majCRT} and  \ref{sec:spinanomaly} for fermionic and bosonic systems, respectively. 
   \item $\Pi$ has a mixed anomaly with two lattice internal symmetries generated by $\V_1$ and $\V_2$. (This mixed anomaly can be identified as an anomaly in $\PP=\V_2\Pi$ in the presence of a $\V_1$ defect.) This probes a mixed anomaly between $\V_1$ and $\V_2$, which exists both on the lattice and in the continuum. See an example in Section \ref{sec:XQanomaly}.
   \item $\Pi$ has a mixed anomaly with a lattice internal symmetry generated by $\V_1$ and lattice translation $T$. (This mixed anomaly can be identified as an anomaly in $\PP=T^{-1}\Pi$ in the presence of a $\V_1$ defect.)  
   In the continuum limit, $T$ leads to an emanant internal symmetry, and the lattice anomaly becomes an emergent mixed anomaly between this emanant symmetry and an exact symmetry. See an example in Section \ref{sec:anomalyTQ}. 
   \end{itemize}
   \item While a $\Pi$-invariant lattice model can have a phase described by a relativistic continuum theory with $\Pi$ flowing to $\Theta$, it can also have another phase (with a different $\Pi$-invariant Hamiltonian), where the low-energy theory is non-relativistic and does not have $\Theta$.
   In that case, the action of $\Pi$ in the low-energy theory and its various anomalies are more subtle.  See Section \ref{ferromagnetic}.
   \item The lattice model has no such $\Pi$ symmetry, but it does have an anomalous $\PP$ symmetry.  Here, the anomaly of $\PP$ leads to an LSM-type constraint. In this case, if the continuum theory is relativistic, its $\Theta$ symmetry is an emergent symmetry.  An example in a fermionic theory is given in Section
   \ref{sec:majCRT}.  It is straightforward to find bosonic examples, e.g., using appropriate modifications of the Hamiltonian in Section \ref{Spinchain}. 
\end{itemize}

\begin{table}[t]
\centering
{\renewcommand{\arraystretch}{1.3}
\begin{tabular}{|c||c|c|c|c|}
\hline
 \makecell{Lattice \\Model} & \makecell{Majorana\\Chain} & \makecell{Heisenberg \\ Chain (AFM)} & \makecell{Heisenberg \\ Chain (FM)} & \makecell{Levin-Gu\\Edge Model} \\
\hline\hline
\makecell{Bosonic \\ or Fermionic} 
  & F 
  & \multicolumn{2}{c|}{B} 
  & B \\
\hline
$\Pi$ 
  & $\chi_\ell\to \chi_{-\ell+1}$ 
  & \multicolumn{2}{c|}{$\vec{S}_j \to - \vec{S}_{-j+1}$}
  & \makecell{$\begin{aligned}
      X_j & \to X_{-j}\\
      Z_j &\to (-1)^j Z_{-j}
  \end{aligned}$} \\
\hline
$\PP$ 
  & $\chi_\ell\to \chi_{-\ell}$ 
  & \multicolumn{2}{c|}{$\vec{S}_j \to - \vec{S}_{-j}$}
  & \makecell{$\begin{aligned}
      X_j & \to -Z_{-j-1}X_{-j}Z_{-j+1}\\
      Z_j &\to (-1)^{j+1} Z_{-j}
  \end{aligned}$} \\
\hline
\makecell{Order of the \\ $\PP$ Anomaly} 
  & mod 8 
  & \multicolumn{2}{c|}{mod 2} 
  & mod 2 \\
\hline
$\Pi  \PP^{-1}$
  & $T: \chi_\ell \to \chi_{\ell+1}$
  & \multicolumn{2}{c|}{$T: \vec S_j \to \vec S_{j+1}$}
  &$\U:$ \makecell{$\begin{aligned}
      X_j & \to -Z_{j-1}X_{j}Z_{j+1}\\
      Z_j &\to  -Z_{j}
  \end{aligned}$} \\
\hline
\makecell{Order of the \\ $\Pi\PP^{-1}$\\ Anomaly} & mod 2 & \multicolumn{2}{c|}{mod 1 (anomaly-free)} & mod 2\\ 
\hline
\hline
Continuum $\Theta$ 
& $\Pi \to \Theta$ 
& $\Pi \to \Theta$  
& does not exist 
& $\Pi \to \Theta$ \\
\hline 
\makecell{Anomalous \\ Continuum $\bZ_2$} 
  & \makecell{emanant \\   $T\to \sC$}
  & \makecell{emanant \\   $T\to \sC$}
  & does not exist
  & \makecell{exact \\ $\U\to \sU$ }\\
\hline
\end{tabular}}
 \caption{
Summary of various 1+1d lattice models studied in Sections \ref{sec:maj}, \ref{Spinchain}, and \ref{Levin-Gu}.  The second to seventh rows (separated by double lines) are properties of the lattice model.  (In particular, they are the same in the two columns representing the two different limits of the Heisenberg chain.) The anti-unitary operator $\Pi$, which is anomaly-free, flows to the continuum CRT operator $\Theta$. In contrast, the anti-unitary operator $\PP$ is anomalous and implies LSM-type constraints.  The order of its anomaly is shown in the fifth row.  The product $\Pi \PP^{-1}$ is a unitary operator, which has a mixed anomaly with $\Pi$.  In the continuum limit, it flows to an internal $\mathbb{Z}_2$ symmetry whose anomaly can be probed using $\Theta$.  The ferromagnetic Heisenberg chain is an exception, whose continuum limit does not have a $\Theta$ symmetry or an anomalous internal $\mathbb{Z}_2$ symmetry. As we discussed in Section \ref{sec:moreanomalies}, in the case of the Levin-Gu edge model, there are also other anomalies involving $\Pi$.}
 \label{tab:summary}
\end{table}

Even though we have found several different phenomena associated with such space-reflecting and time-reversing symmetries, we think it would be interesting to look for additional ones and to explore them further.  Here are some possible extensions of our study:
\begin{itemize}
    \item Given an anti-unitary, order-2 operator $\Pi$ on the lattice that reflects the spatial coordinate, what conditions must be satisfied for there to exist a $\Pi$-symmetric Hamiltonian whose continuum limit realizes $\Pi$ as the  CRT operator $\Theta$ of a relativistic field theory?
    \item Generalizing the structure seen in Section \ref{Levin-Gu}, consider a lattice model with a symmetry $\Pi$ that flows to $\Theta$, and a lattice translation symmetry generated by $T$, which leads to an emanant internal $\bZ_n$ symmetry generated by $\mathsf{W}$. Assume the lattice relation $\Pi T = \V T^{-1}\Pi$ with an internal $\mathbb{Z}_m$ symmetry generated by $\V$. In the continuum limit, $\Pi\to \Theta$ and $\V\to \sV$.  Then,  the relation $\Theta\mathsf{W}= \mathsf{W} \Theta$ implies that in the continuum theory, $\mathsf{W}^2=\sV$. This suggests the following questions: 
    \begin{itemize}
        \item What is the most general structure of emanant symmetries? 
    \item Does this kind of reasoning provide constraints on the existence of a lattice symmetry $\Pi$ that flows to $\Theta$ and on the nature of the emanant symmetry? 
    \end{itemize}
    \item The authors of \cite{Kobayashi:2024bts} studied 2+1d fermionic SPT phases protected by RT symmetries and found a mod 8  classification. It would be interesting to explore the connection to the mod 8 anomaly of $\PP$ in the 1+1d Majorana chain discussed in Section \ref{sec:majRT}.
    \item More generally, can we phrase our results using an invertible continuum field theory? Related to that, how do our results fit into the conjectured equivalence principle of \cite{ThorngrenPRX2018}?  
    \item What is the higher-dimensional generalization of our discussion?  This is interesting because in higher dimensions, a more intricate interplay between crystalline symmetries and internal symmetries is possible.  In a forthcoming work \cite{staggeredFermion2p1d}, we will analyze 2+1d staggered fermion on various manifolds and will use similar techniques to analyze the anomalies and their consequences. In this context, see also \cite{Barkeshli:2025cjs,Pace:2025rfu,Gioia:2025bhl} for recent discussions of anomalies involving spacetime reflections for 2+1d lattice fermions.
\end{itemize}

\section*{Acknowledgments}

We are grateful to Maissam Barkeshli, Boris Bulatovic, Meng Cheng, Lukasz Fidkowski, Tarun Grover,  Ryohei Kobayashi, Zohar Komargodski, Ho Tat Lam, Max Metlitski, Greg Moore, Salvatore Pace, Abhinav Prem, Sahand Seifnashri, Nikita Sopenko, Xincheng Zhang, and Yifan Zhang for interesting discussions.
The work of NS was supported in part by DOE grant DE-SC0009988. SHS was supported in part by NSF grant PHY-2449936. WZ was supported in part by the Princeton University Department of Physics. This work was also supported by the Simons Collaboration on Ultra-Quantum Matter, which is a grant from the Simons Foundation (651444, NS, SHS).

\appendix

\section{Review of CRT}\label{CPTreview}

A fundamental fact about Lorentz-invariant quantum field theory is the CPT theorem.  This deep result has a long history, going back to the 1950s \cite{Luders:1954zz,Pauli1988,Bell:1955djs,jost1957bemerkung,Haag:book,streater2000pct}.
For recent discussions from a physics perspective and a mathematics perspective, see e.g., \cite{Witten:2015aba,Hason:2020yqf,Harlow:2023hjb,Witten:2025ayw} and \cite{Lawson:1998yr,bar2011spin,Freed:2016rqq} respectively.  These papers also provide many relevant references.

In this appendix, we will review some well-known aspects of the CPT theorem. We will denote the spatial coordinates by $x^i$ with $i=1,\cdots, d$ and Euclidean time by $x^{d+1}$.

We will start with non-spin theories and will extend to spin theories below.

\subsection{Non-spin theories}

First, we need some group theory facts, which are independent of the application to quantum field theory.  The Lorentz algebra is $so(1,d)$.  Exponentiating it leads to the connected group
\ie
SO^+(1,d) \subset SO(1,d)\,.
\fe
Note that $SO(1,d)$ is not connected.  It is an extension of $SO^+(1,d)$ by an order 2 element $\varTheta$.\footnote{Although this does not affect the discussion here, it will be important below that we take $\varTheta$  to be anti-unitary.}

A simple example to keep in mind, which is relevant to this paper, is the case of $d=1$.  Here, $SO^+(1,1)$ is represented by the matrices $B(\theta)$ and the element that extends it to $SO(1,1)$ is the matrix $\varTheta$
\ie\label{SO11}
&B(\theta)=\begin{pmatrix}\cosh\theta&\sinh \theta\\ \sinh \theta &\cosh\theta\end{pmatrix}\in SO^+(1,1)\\
&\varTheta=\begin{pmatrix}-1&0\\ 0 &-1\end{pmatrix}\in SO(1,1)\,.
\fe
(Here and below, we suppress a factor of the complex conjugation operator that makes $\varTheta$ anti-unitary.)
Viewing it as Lorentz transformations in 1+1 dimensions, $B(\theta)$ is a boost and $\varTheta$ is a combined reflection of space and time.

As a group, $SO(1,1)\cong SO^+(1,1)\times \bZ_2^\varTheta $ with $SO^+(1,1)\cong \bR$.  Its representations are one-dimensional and real.  They are labeled by $s\in \bR$ and $\varTheta=\pm1$.  (We use $\varTheta$ both for the group element and its eigenvalue in a representation.)  The matrices in \eqref{SO11} correspond to the reducible representation $(s=1, \varTheta=-1)\oplus (s=-1,\varTheta=-1)$.

An important subgroup of $SO^+(1,d)$ is the group of spatial rotations
\ie
SO(d)\subset SO^+(1,d)\,.
\fe
It is generated by
\ie
R_{ij}(\alpha)\in SO(d) \qquad, \qquad i\ne j\qquad ,\qquad i,j=1,\cdots, d\,,
\fe
corresponding to rotation by $\alpha$ in the $(i,j)$ plane.  In the example of $d=1$ \eqref{SO11}, $SO(1)$ includes only the identity.

We will discuss an anti-unitary time-reversal transformation $\scT$ and linear transformations $\scR^i$ that reverse the sign of $x^i$. These transformations are not in $SO(1,d)$.  However, the orientation-preserving bilinears $\scR^i\scR^j$ are rotation elements
\ie
\scR^i\scR^j=R_{ij}(\pi) \in SO(d)\subset SO^+(1,d)\qquad, \qquad i\ne j\qquad ,\qquad i,j=1,\cdots, d\,.
\fe
And while $\scR^i\scT \not\in SO^+(1,d)$,
\ie
\scR^i\scT\in SO(1,d)\,.
\fe
Picking one of the spatial reflection elements, say
\ie
\scR^1=\scR\,,
\fe
we can take the order-two group element
\ie
\varTheta=\scR\scT
\fe
to be the one that extends $SO^+(1,d)$ to $SO(1,d)$.\footnote{We denote the geometric transformations on the coordinates by $\scR$,  $\scT$, and $\varTheta=\scR\scT$.  When these geometric transformations act on fields in a field theory, they can be accompanied by additional action. We will denote the corresponding operators as $\sR$, $\sT$, and $\Theta=\sC\sR\sT$ with an appropriate $\sC$ to be discussed below.\label{CRTnotation}}

Trivially, every Lorentz-invariant theory is $SO^+(1,d)$ invariant.   The CPT theorem states that a Lorentz invariant field theory is in fact invariant under the larger group $SO(1,d)$, including the invariance under the anti-unitary transformation $\varTheta=\scR\scT$.  This transformation is often referred to as $\sC\sR\sT$.  (See the discussion below about $\sC$ and $\sC\sP\sT$.)

A simple way to think about it is to consider the Euclidean version of the theory.  (This is non-trivial, as one needs to show that there exists a consistent analytic continuation to a Euclidean signature, which is known as Wick rotation.)  Here, the Lorentzian time direction becomes $x^{d+1}$, and $SO(1,d)$ becomes $SO(d+1)$.  Unlike $SO(1,d)$, the group $SO(d+1)$ is connected. Therefore, a Lorentz invariant theory in Euclidean signature is $SO(d+1)$ invariant.  

In more detail, the anti-unitary transformation $\scT$ becomes the linear reflection transformation $\scR^{d+1}$.  As in Lorentzian signature, the reflections $\scR^i$ with $i=1,\cdots, d+1$ are not in $SO(d+1)$. But unlike the Lorentzian situation, here, all the bilinears $\scR^i\scR^j$, including the Euclidean version of $\scR^i\scT$, are in $SO(d+1)$.  In particular
\ie
\scR^i\scR^{d+1}=R_{i,d+1}(\pi)\in SO(d+1)\,.
\fe

Returning to Lorentzian signature, this means that the theory must be invariant under $\varTheta=\scR\scT\in SO(1,d)$.  Equivalently, $SO^+(1,d)$ must be extended to $SO(1,d)$.  The Euclidean picture identifies the extending element $\varTheta=\scR\scT$ as the Lorentzian version of $\scR\scR^{d+1}=R_{1,d+1}(\pi)$.
The latter is a special case of $R_{1,d+1}(\alpha)$, which is continuously connected to the identity.

Another important lesson from the Euclidean picture is a restriction on the $SO(1,d)$ representations of local operators.  We will demonstrate it below.

In the simple case of $d=1$ \eqref{SO11}, the Euclidean group is $SO(2)$, and it is realized as
\ie\label{SO2}
&R_{12}(\alpha)=\begin{pmatrix}\cos\alpha&\sin \alpha\\ -\sin \alpha &\cos\alpha\end{pmatrix}\in SO(2)\,.
\fe
This group is manifestly connected.  Its representations are labeled by $s\in \bZ$.  And the matrices in \eqref{SO2} are in a direct sum of two complex one-dimensional representations $(s=1)\oplus (s=-1)$.

The analytic continuation to Lorentzian signature  \eqref{SO11} maps the spin $s$ representation of $SO(2)$ to the spin $s$ representation of $SO(1,1)$.  This is achieved by conjugating $R_{12}$ and analytically continuing $\alpha=i\theta $
\ie\label{SO2toSO11}
&B(\theta)=UR_{12}(i\theta)U^{-1}\\
&\varTheta=UR_{12}(\pi)U^{-1}\\
&U=\begin{pmatrix}1&0\\0&i\end{pmatrix}\,.
\fe

As we mentioned after \eqref{SO11}, the representations of $SO(1,1)$ are labeled by $s\in \bR$ and $\varTheta=\pm 1$. The analytic continuation from $SO(2)$ leads only to $s\in \bZ$ and $\varTheta=(-1)^s$.  Therefore, the operators in (bosonic) quantum field theory are labeled by the integer $s$.

Let us apply this reasoning to a field theory. As mentioned in footnote \ref{CRTnotation}, now we should use $\scR\to \sR$, $\scT\to \sT$, and $\varTheta \to \Theta$. In addition to the invariance under $\Theta$, the theory might also have additional symmetries.  These can be internal symmetries, which do not act on the spacetime coordinate, as well as spacetime symmetries like space-reflection and time-reversal.  We would like to make a few comments about the consequences of that. 
\begin{itemize}

\item If there is an internal symmetry with symmetry operator $\sU$, then $\sU$ commutes with the Euclidean $R_{1,d+1}(\pi)$ and with its Lorentzian image $\Theta$.  If $\sU$ is associated with a continuous symmetry, it is constructed out of a charge $\sQ$ as ${\sU}=e^{i\alpha \sQ}$.  Then, the anti-unitary $\Theta$ anti-commutes with $\sQ$.  
That is,
\ie
&\Theta {\sU}  = {\sU} \Theta\,,\\
&\Theta \sQ  = - \sQ \Theta\,.
\fe
This is the reason that $\Theta$ is usually referred to as $\sC\sR\sT$.  See more about that below.
\item It is sometimes stated that $\Theta$ can be redefined by combining it with an arbitrary $\bZ_2$ internal symmetry operator.
However, we emphasize that the $\Theta$ transformation, which is always a symmetry, has a canonical definition as the Lorentzian image of $R_{1,d+1}(\pi)$.
\item In addition to internal symmetries, the theory might also have spatial-reflection or time-reversal symmetries.  In fact, in a relativistic system, if we have one of them, say spatial-reflection $\sR$, we also have a time-reversal symmetry $\Theta \sR$.  
\item In this case with spatial and time-reversal symmetries, it is customary to write $\Theta=\sC\sR\sT$, where $\sC$ is a transformation that reverses some of the internal global symmetry charges $\sQ$ in ${\sU}=e^{i\alpha \sQ}$, such that $\sR$ and $\sT$ commute with $\sQ$.  We note, however, that such $\sC$ might not exist as a global symmetry, and furthermore, if there is such an internal $\bZ_2^\sC$ symmetry, it might not reverse all the global symmetry charges $\sQ$. 
\item Above, we discussed $\scR$ and $\scT$ that are characterized by the way they act on the spacetime coordinates $x^1$ and $t$. Therefore, if there are $\sR$ and $\sT$ symmetries, their symmetry operators should include factors of $\scR$ and $\scT$ respectively.  These factors should be multiplied by additional operators that determine how the fields transform. 
\item In this case, with $\sR$ and $\sT$, these symmetry operators are ambiguous.  Even if we write $\Theta=\sC\sR\sT$ with a particular $\sC$, we can still redefine $\sR$ and $\sT$ by another internal symmetry transformation, such that $\Theta$ is not affected.
\end{itemize}

For odd spatial dimensions, e.g., $d=3$, it is also common to define parity
\ie\label{CPTdef}
\scP=\scR^1\scR^2\cdots\scR^d\,.
\fe
Unlike the reflection $\scR$, it commutes with the spatial rotation group $SO(d)$ and for this reason, it is sometimes more convenient to use $\scP$ rather than $\scR$.

Just like $\scR^i$, the parity transformation $\scP$ is also not included in $SO(d)$.  But since $\scR^i\scR^j\in SO(d)$, the anti-unitary transformation $\scP\scT$ is included in $ SO(1,d)$ and must be a symmetry of the physical system.  And again, even though it has a canonical definition as a Euclidean rotation, it is often phrased as $\sC\sP\sT$.

\subsection{Spin theories}

Next, we lift this discussion to spin theories. The standard way to analyze this problem uses Clifford algebras.  Here, we will analytically continue from the known results in Euclidean signature.

The lift uses the operator $(-1)^F$, which lifts the spatial rotation group $SO(d)$ to $Spin(d)$ such that
\ie\label{liftspace}
R_{ij}(2\pi)=(-1)^F\qquad, \qquad i\ne j\qquad ,\qquad i,j=1,\cdots d\,.
\fe
(Hoping not to create confusion, we denote $R_{ij}(\alpha)\in SO(d)$ the same as its lift $R_{ij}(\alpha)\in Spin(d)$.) Similarly, it lifts $SO(1,d)$ to $Spin(1,d)$.

In Euclidean signature, we lift $SO(d+1)$ to $Spin(d+1)$ and again $R_{ij}(2\pi)=(-1)^F$ with $i,j=1,\cdots,d+1$ and $i\ne j$.\footnote{Even though we do not focus on the individual reflection elements $\scR^i$, but only on their bilinears $\scR^i\scR^j=R_{ij}(\pi)$, we note that $R_{ij}(\pi)=(-1)^FR_{ji}(\pi)$.  And therefore, for $i\ne j$, $\scR^i\scR^j=(-1)^F \scR^j\scR^i$.  (This is regardless of whether $(\scR^i)^2$ is $1$ or $(-1)^F$, i.e., regardless of whether they extend $Spin(d+1)$ to $Pin^+(d+1)$ or to $Pin^-(d+1)$.)\label{RiRj}}   This means that after the lift $R_{ij}(\pi)^2=(-1)^F$, including
\ie
(\scR\scR^{d+1})^2=(-1)^F\,.
\fe

Rotating this back to Lorentzian signature, $\scR^{d+1}$ becomes the anti-unitary transformation $\scT$.  Therefore, $\scR\scR^{d+1}$ becomes the anti-unitary transformation $\varTheta=\scR\scT$ and its square has another factor of $(-1)^F$ to give\footnote{As in footnote \ref{RiRj}, $\scR\scT=(-1)^F\scT\scR$, and using \eqref{RTsquare}, $\scR^2=(-1)^F\scT^2$.  Also, $\Theta\scT=(-1)^F\scT\Theta$, $\Theta\scR=(-1)^F\scR\Theta$, and $\Theta\scR^i=\scR^i\Theta$ for $i\ne 1$.   (Again, this is regardless of whether $\scR^2$ is $1$ or $(-1)^F$.) \label{RsTs}}
\ie\label{RTsquare}
(\scR\scT)^2=\varTheta^2=1\,.
\fe
This crucial relation will be discussed in detail below for the special case of $d=1$.  The extension to $d>1$ is straightforward.

A peculiar fact about $d=1$, which is not present for higher $d$, is that the spatial rotation group is trivial, and we cannot use \eqref{liftspace} to find $Spin(1,1)$.  Instead, we will start with the Euclidean problem and analytically continue it back to Lorentzian signature. 

In Euclidean signature, the situation is similar to higher $d$.  $SO(2)$ is lifted to $Spin(2)$, which is connected and is realized as
\ie\label{Spin2}
R_{12}(\alpha)=\begin{pmatrix}\cos{\alpha\over 2}&\sin {\alpha\over 2}\\ -\sin{\alpha\over 2} &\cos{\alpha\over 2}\end{pmatrix}\in Spin(2)\,,
\fe
with the special elements
\ie\label{spin2s}
&R_{12}(\pi)=\begin{pmatrix}0&1\\-1 &0\end{pmatrix}\\
&R_{12}(2\pi)=R_{12}(\pi)^2=(-1)^F=\begin{pmatrix}-1&0\\ 0 &-1\end{pmatrix}\in Spin(2)\,.
\fe
Note that $R_{12}(\pi)$ is an order-four element.  It generates $\bZ_4^{R_{12}(\pi)}\subset Spin(2)$.

The representations of $Spin(2)$ are labeled by $s\in {\bZ\over 2}$ with $R_{12}(\pi)$ eigenvalues $e^{i\pi s}$ and $(-1)^F=R_{12}(\pi)^2$ eigenvalues $(-1)^{2s}$.  
And the representation in \eqref{Spin2} is a direct sum of two complex one-dimensional representations $(s={1\over 2})\oplus(s=-{1\over 2})$.

As groups $SO(2)\cong Spin(2)$. 
However, when viewed as Lorentz groups in Euclidean signature, they act differently on the spacetime coordinates. This makes the analytic continuations to Lorentzian signature different in these two cases.

The analytic continuation from the Euclidean group $Spin(2)$ \eqref{Spin2} to Lorentzian signature proceeds as in \eqref{SO2toSO11}:
\ie\label{Spin11p}
&B(\theta)=UR_{12}(i\theta)U^{-1} =\begin{pmatrix}\cosh{\theta \over 2} &\sinh {\theta \over 2}\\ \sinh {\theta \over 2} &\cosh{\theta \over 2}\end{pmatrix} \,,\\
&(-1)^F=UR_{12}(2\pi)U^{-1}= \begin{pmatrix}-1 &0\\ 0&-1\end{pmatrix}\,,\\
&U=\begin{pmatrix}1&0\\0&i\end{pmatrix}\,.
\fe

We should also map the Euclidean symmetry $R_{12}(\pi)$ to the Lorentzian symmetry $\varTheta$.  As a first attempt, we could consider   $UR_{12}(\pi)U^{-1}=\begin{pmatrix}0&-i\\-i&0\end{pmatrix}$.  However, this is not the right answer.  In Euclidean signature, the spin $s$ representation is a one-dimensional complex representation.  And in Lorentzian signature, it is a one-dimensional real representation.  Therefore, the eigenvalues of  $\varTheta$  should be real.  
This motivates us to take
\ie\label{cRspin11}
& \varTheta =iUR_{12}(\pi)U^{-1} = \begin{pmatrix}0&1\\1&0\end{pmatrix}\,.
\fe
In addition, $\varTheta$ is anti-unitary.  (We will soon explain this crucial factor of $i$ from another perspective.)  The expression \eqref{cRspin11} for $\varTheta$ means that it generates 
$\bZ_2^\varTheta$, i.e., $\varTheta^2=1$. 

To summarize, the analytic continuation of $Spin(2)$ to Lorentzian signature is
\ie
Spin(1,1)\cong\bR\times\bZ_2^F\times\bZ_2^\varTheta\,.
\fe
In particular, $\bZ_4^{R_{12}(\pi)}\subset Spin(2)$ is rotated to $\bZ_2^F\times \bZ_2^\varTheta \subset Spin(1,1)$.

The representations of $Spin(1,1)$  are labeled in an obvious way by $s\in \bR$, $(-1)^F=\pm 1$, $\varTheta=\pm 1$.  In particular,  the representation \eqref{Spin11p}, \eqref{cRspin11} is reducible and is the direct sum of two real irreducible representations $(s=+{1\over 2}, (-1)^F=-1,\varTheta=+ 1)$ and $(s=-{1\over 2}, (-1)^F=-1,\varTheta=- 1)$.\footnote{These two representations correspond to the right- and left-moving Majorana-Weyl fermions $\chi_\text{R}$ and $\chi_\text{L}$ in Section \ref{sec:majcont}. Indeed, $\varTheta$ acts as $\pm1$ on these fields as in \eqref{Thetachi}.} 

As in the non-spin case, the analytic continuation from Euclidean signature shows that only some of the $Spin(1,1)$ representations are present, and they are labeled by $s\in {\bZ\over 2}$.  

We claim that a Hermitian operator $\cal O$ with spin $s$ has\footnote{The anti-unitary operator $\varTheta$ acts on a local operator $\mathcal{O}(t,x)$ as $\varTheta \mathcal{O}(t,x ) \varTheta^{-1} = \xi \mathcal{O}(-t,-x)^\dagger$ with a phase $\xi$. (We have suppressed possible Lorentz indices of the local operator.) Note that the phase $\xi$ is invariant under operator redefinitions because of the adjoint operation on the right-hand side. To define the eigenvalues of $\varTheta$, we need to further assume that the operator $\mathcal{O}$ is Hermitian. This fact is at the root of the $\varTheta$ assignment for spinors in \eqref{Rof O}. In this case we denote the eigenvalue as $\xi=\varTheta(\mathcal{O})$. 
\label{actonH}}
\ie\label{Rof O}
&(-1)^F({\cal O})=(-1)^{2s}\,,\\
&\varTheta({\cal O})=(-1)^{s(2s-1)}=\begin{cases}(-1)^s& s\in \bZ\\ (-1)^{s-{1\over 2}} & s\in \bZ+{1\over 2} \end{cases}\,.
\fe
For integer $s$, this is as in the non-spin problem.  However, for half-integer $s$, the situation is more subtle because of the fermionic nature of the operators and the anti-unitarity of $\varTheta$. In particular, the factor of $(-1)^{-{1\over 2}}$ makes the eigenvalue real and is correlated with the factor of $i$ in \eqref{cRspin11}. (The asymmetry between $s$ and $-s$ for half-integer $s$ can be absorbed in redefining $\varTheta\to (-1)^F\varTheta$.)

In order to understand the significance of the assignment of $\varTheta({\cal O})$, let us discuss the multiplication of these representations.  Consider two Hermitian operators, ${\cal O}_i $ with spins $s_i$ and $\varTheta$ and $(-1)^F$ as in \eqref{Rof O}.  If $s_1$ is an integer, ${\cal O}_1{\cal O}_2$ is Hermitian.  Its total spin is $s_1+s_2$ and $\varTheta({\cal O}_1{\cal O}_2)=(-1)^{s_1}(-1)^{s_2(2s_2-1)}=(-1)^{(s_1+s_2)(2s_1+2s_2-1)}$.  The same conclusion applies if $s_2$ is an integer.  However, if both $s_1$ and $s_2$ are half-integers, these two operators are fermions.  Then, to determine the $\varTheta$-eigenvalue of their product, we should consider the Hermitian operator $i{\cal O}_1{\cal O}_2$. (See footnote \ref{actonH}.) Its spin is $s_1+s_2$ and $\varTheta(i{\cal O}_1{\cal O}_2)=-(-1)^{s_1(2s_1-1)}(-1)^{s_2(2s_2-1)}=(-1)^{s_1+s_2}$.  (This reasoning provides a derivation of the $\varTheta$-eigenvalue assignments in \eqref{Rof O} by multiplying representations with $s=\pm {1\over 2}$.)

We see that the product of two real representations labeled by $s_1$ and $s_2$ is the real representation with $s_1+s_2$.  This is exactly as in the Euclidean problem.

Let us summarize.  We started with a non-spin theory with the connected part of the Lorentz group $SO^+(1,1)$.  The $\sC\sR\sT$ theorem told us that we have a larger symmetry, $SO(1,1)=SO^+(1,1)\times \bZ_2^\varTheta$.  We lifted it using $(-1)^F$ to a spin theory. The discussion through the Euclidean group $Spin(2)$ led us to $Spin(1,1)$.  $SO^+(1,1)$ is lifted to  $Spin^+(1,1)\cong \bR\times \bZ_2^F$ and $SO(1,1)$ is lifted to $Spin(1,1)\cong\bR\times\bZ_2^F\times\bZ_2^\varTheta$.  Clearly, the latter has four connected components.

As above, in field theory, the canonical operator associated with  $\varTheta$ is denoted as $\Theta=\sC\sR\sT$.  It  satisfies
\ie
\Theta^2=(\sC\sR\sT)^2=1\,.
\fe

The individual transformations $\sC$, $\sR$, and $\sT$ might not be symmetries, and even if they are, they are subject to redefinitions using internal symmetries.  However, regardless of such redefinitions, when they exist,
\ie\label{TThetaThetaT}
\sT\Theta=(-1)^F\Theta\sT\quad, \quad \sR\Theta=(-1)^F\Theta\sR\quad, \quad \sR^i\Theta=\Theta\sR^i\qquad {\rm for}\ i\ne 1\,.
\fe
where $\sR^i$ acts as $\sR$ except that it reflects $x^i$. 
Geometrically, this follows from the expressions in footnote \ref{RsTs}. (Intuitively, for $i\ne 1$, the symmetry $\sR^i$ acts as an internal symmetry as far as $x^1$ and $t$ are concerned, and therefore, as in \eqref{ThetaUc}, it commutes with $\Theta$.)  For example, these relations can be verified explicitly in the example of the 1+1d Majorana fermion by using \eqref{eq:contCPT}.  Note that the geometric relation in footnote \ref{RsTs} $\scR^2=(-1)^F\scT^2$, might not lead to a similar relation for $\sR$ and $\sT$ because their redefinitions using internal transformations might change it.  Indeed, in \eqref{eq:contCPT}, we have $\sR^2=\sT^2=1$.

Finally, repeating the discussion about odd spatial dimensions in \eqref{CPTdef}, we have
\ie
(\sC\sP\sT)^2=\left((-1)^F\right)^{d-1\over 2}\,.
\fe
And if we have $\sR$ and $\sT$ symmetries, \eqref{TThetaThetaT} is replaced with
\ie
\sT(\sC\sP\sT)=(-1)^F(\sC\sP\sT)\sT\quad, \quad \sR^i(\sC\sP\sT)=(-1)^F(\sC\sP\sT)\sR^i\quad, \quad \forall i\,.
\fe

\section{Review of time-reversal anomalies in quantum mechanics of fermions}\label{QMA}

In this Appendix, we review a modulo 8 time-reversal anomaly in quantum mechanics, which is related to SPT phases in 1+1d  \cite{Fidkowski:2009dba}. 
See also \cite{Elitzur:1985xj,Witten:1985mj} for related discussions on anomalies in quantum mechanics and \cite{Witten:2015aba,Stanford:2019vob,Kaidi:2019tyf,Delmastro:2021xox,Witten:2023snr} for reviews of this anomaly.

Consider the quantum mechanics of $\nu$ Majorana fermions $\psi^a\,(a=1,\dots,\nu)$. They are hermitian operators satisfying the Clifford algebra
\ie
\{\psi^a,\psi^b\}=2\delta^{ab}\,.
\fe

We are going to look for two anti-unitary (time-reversal) $\bZ_2$ operators $\T$ and $\T'$ and a unitary (fermion parity) $\bZ_2$ symmetry operator $(-1)^F$, related via $\T'=\T(-1)^F$.  They act on the fermions as
\ie\label{QMdefinitions}
(-1)^F\psi^a&=-\psi^a(-1)^F,\\
\T\psi^a&=+\psi^a\T,\\
\T'\psi^a&=-\psi^a\T'\,.
\fe
These operators are analogous to time-reversal $\T$, particle-hole symmetry $\T'$, and chiral symmetry $(-1)^F$ of the famous tenfold classification of topological insulators and superconductors \cite{PhysRevB.55.1142, PhysRevB.78.195125, 10.1063/1.3149495}.  
Note that both $\T^2$ and $(\T')^2$ act on the operators trivially
\ie
\T^2 {\cal O} \T^{-2} ={\cal O}\,,\quad
(\T')^2 {\cal O} (\T')^{-2} ={\cal O}
\fe
Therefore, both $\T^2$ and $(\T')^2$ are c-numbers.

As we will review, depending on $\nu$, not all of these operators exist, and when they exist, their algebra depends on $\nu\bmod 8$.

Rather than proceeding abstractly, we can adopt a concrete irreducible representation of the Clifford algebra.

For even $\nu=2m$, they are represented as
\ie\label{fermionseven}
&\psi^{2 j-1}=\left(\sigma^3\right)^{\otimes(j-1)} \otimes \sigma^1 \otimes \mathbf{1}^{\otimes(m-j)}\quad, \qquad 1 \leq j \leq m\\
&\psi^{2 j}=\left(\sigma^3\right)^{\otimes(j-1)} \otimes \sigma^2 \otimes \mathbf{1}^{\otimes(m-j)}\qquad, \qquad 1 \leq j \leq m\,.
\fe
And for odd $\nu=2m+1$, the last fermion is
\ie\label{fermionsodd}
\psi^{2m+1}=(\sigma^3)^{\otimes m}\,.
\fe

For even $\nu=2m$, we have the fermion parity operator
\ie\label{moneFp}
(-1)^F=\left(-\sigma^3\right)^{\otimes m}=i^{m}\psi^1\psi^2\cdots \psi^\nu \qquad , \qquad \nu=2m\,,
\fe
where the phase is set such that $\left((-1)^F\right)^2=1$.

For odd $\nu$, the operator $\psi^1\psi^2\cdots \psi^\nu$ is central and there is no fermion parity operator $(-1)^F$.  Consequently, the Hilbert space is not graded.  However, as in \cite{Seiberg:2023cdc}, we can still study it using the fermions $\psi^a$ in \eqref{fermionseven} and \eqref{fermionsodd}.  Importantly, the path integral representation of this system is problematic, and one might want to ignore these odd $\nu$ systems as anomalous (see e.g., \cite{Stanford:2019vob,Delmastro:2021xox,Witten:2023snr,Freed:2024apc}).  We will not ignore them here, and instead, we will discuss only the Hilbert space of these systems and ignore the path integral description.

Next, we look for the time-reversal operators $\T$ and $\T'$.  Since they are anti-unitary, their form depends on the representation.  We are going to use the representation \eqref{fermionseven}, \eqref{fermionsodd} and then, the complex conjugate operator $\cal K$ acts as
\ie
{\cal K}\psi^a =(-1)^{a+1}\psi^a{\cal K}\,.
\fe
Demanding that $\T$ and $\T'$ act as in \eqref{QMdefinitions}, we find
\ie
&\T={\cal K}(\sigma^1\otimes\sigma^2)^{\otimes n} &&\qquad, \qquad &&\nu=4n, 4n+1\\
&\T'={\cal K}(\sigma^2\otimes\sigma^1)^{\otimes n} &&\qquad, \qquad &&\nu=4n\\
&\T={\cal K}(\sigma^1\otimes\sigma^2)^{\otimes n}\otimes\sigma^1 &&\qquad, \qquad &&\nu= 4n+2\\
&\T'=-i{\cal K}(\sigma^2\otimes\sigma^1)^{\otimes n}\otimes\sigma^2 && \qquad, \qquad && \nu= 4n+2, 4n+3\,.
\fe

We see that for even $\nu$, we have both $\T$ and $\T'=\T(-1)^F$.  (The phase was set such that this relation is satisfied.)  And for odd $\nu$, we have either $\T$ or $\T'$, but not both.\footnote{Most authors have limited themselves to even $\nu$.  Then, it is enough to consider $\T$ and $(-1)^F$, because $\T'$ is derived from them.  Here, we also discuss odd $\nu$ and therefore both $\T$ and $\T'$ should be considered.\label{TTp}}

 Let us examine their algebra.  
For odd $\nu$, we have only one operator satisfying
\ie
\T^2 & = (-1)^{\frac{\nu-1}{4}} \qquad, \qquad \nu=1 \bmod 4\,,\\
\T'^2 & = (-1)^{\frac{\nu+1}{4}}\qquad, \qquad \nu=3\bmod 4\,.\\
\fe 
For even $\nu$, we have  
\ie
& [(-1)^F]^2=1\,,~~~
\T^2 = (-1)^{\frac{\nu(\nu-2)}{8}}\,,~~~
\T (-1)^F = (-1)^{\frac{\nu}{2}} (-1)^F\T\,,\quad \nu=0\bmod 2 \,.
\fe
Using $\T'=\T(-1)^F$, it follows that:
\ie
&\T'^2 = (-1)^{\frac{\nu(\nu+2)}{8}}\,,~~~\T' (-1)^F = (-1)^{\frac{\nu}{2}} (-1)^F\T' 
\,,\quad \nu=0\bmod 2\,.
\fe
 For $\nu \neq 0$ mod 8, these projective signs cannot be removed by redefining the operators.\footnote{We can redefine $(-1)^F$ by $i^{\nu\over 2}$ to remove the phases in its commutation relations with $\T$ and $\T'$, but then, its square would be $(-1)^\nu$. (See \eqref{moneFp} and the following comment.)}
 In this sense, the symmetry has a modulo 8 anomaly.

Our Hilbert space is in a spinor representation of $Spin(\nu)$.  These are known to have a modulo 8 periodicity.  The reality properties of the representations are compatible with the square of the anti-unitary operators.  They are real for $\nu=0,\pm 1\bmod 8$ ($\T^2$ and/or $(\T')^2$ are $+1$), pseudo-real for $\nu=\pm 3,4\bmod 8$ ($\T^2$ and/or $(\T')^2$ are $-1$) and complex conjugate representations for $\nu=\pm 2\bmod 8$ ($\T^2=-(\T')^2=\pm 1$).

We have just seen that the system has at least a modulo 8 anomaly.  Can there be a more subtle anomaly making this integer larger?  

Let us start with a vanishing Hamiltonian.  Then, in addition to the three symmetries $\T$, $\T'$, and $(-1)^F$, there is also a global $O(\nu)$ symmetry.  (More precisely, for even $\nu$, the operator $(-1)^F$ is included in $SO(\nu)$, and for odd $\nu$, this operator is absent and the symmetry is only $SO(\nu)$.) Since the Hilbert space is in a spinor representation of this symmetry, it realizes it projectively and therefore, for every $\nu$, there is a nontrivial anomaly.  

But what if we explicitly break this symmetry and preserve only $\T$, $\T'$, and $(-1)^F$? 

To preserve $(-1)^F$, the deformation of the Hamiltonian should be a linear combination of terms, each with an even number of fermions.  Every quadratic term violates $\T$ (and therefore also $\T'$).  For $\nu=4$, we can consider the invariant term $\psi^1\psi^2\psi^3\psi^4$, but it is clear that this deformation leaves a two-fold degenerate ground state. This is consistent with the fact that the $\nu=4$ system still has an anomaly.

Next, for $\nu=8$, every generic quartic coupling lifts the degeneracy, leaving the system with a single invariant ground state.\footnote{Fidkowski and Kitaev exhibited a special term that preserves $Spin(7)\subset O(8)$ \cite{Fidkowski:2009dba}.\label{FKterm}} This demonstrates that the anomaly in $\T$, $\T'$, and $(-1)^F$ is indeed modulo $8$, rather than modulo a larger integer.

Finally, we would like to make one more comment, which was used in the body of the paper.  Consider a system with two kinds of fermions, $\nu_+$ fermions $\psi_+^a$ transforming as above, and $ \nu_-$ fermions $ \psi_-^{\tilde a}$ transforming under $\T$ and $\T'$ with the opposite signs.  In that case, we can turn on quadratic ``mass terms'' $im_{a\tilde a}\psi_+^a \psi_-^{\tilde a}$, which preserve all our symmetries.  This decouples some of the fermions and leaves us with $|\nu_+ - \nu_-|$ fermions, all with the same action of $\T$ and $\T'$.  Therefore, the anomaly is labeled by $(\nu_+-\nu_-)\bmod 8$.

\section{1+1d Majorana fermion CFT}\label{continuumMajorana}

We consider $N_f$ non-chiral, massless, free Majorana fermion fields in the continuum \eqref{LNf}, with their left- and right-moving components denoted by $\chi_\text{L,R}^A$ and $A=1,\dots,N_f$. The global symmetries include chiral fermion parity $\sC$ (sometimes denoted as $(-1)^{F_\text{L}}$), reflection $\sR$, time-reversal $\sT$, and  fermion parity $(-1)^F$. 
See \eqref{eq:contCPT}. 
They form the group $D_4 \times \mathbb{Z}_2$.  Here, $D_4$ is the dihedral generated by $a=\sC\sR$ and $b=\sR$ with $a^4=1$, $b^2=1$, and $b ab^{-1}= a^{-1}$.  The fermion parity operator is $(-1)^F=a^2$. And the $\bZ_2$ factor is generated by the anti-unitary transformation $c=\sR\sT$, which commutes with $a$ and $b$ and $c^2=1$.\footnote{In Appendix \ref{CPTreview}, we discussed the geometric transformations of space-reflection $\scR$ and time-reversal $\scT$, and in footnote \ref{RsTs}, we noted that $\scR\scT=(-1)^F\scT\scR$ and $\scR^2=(-1)^F\scT^2$. As we explained in Appendix \ref{CPTreview}, the field theory transformations $\sR$ and $\sT$ are combinations of the geometric actions $\scR$ and $\scT$ with some action on the fields.  Therefore, their relations can differ from the geometric relations.}   As we will see, this group can be realized projectively.

\subsection{Parity-preserving boundary condition}\label{app:RR}

With periodic boundary conditions for both the left- and the right-movers (Ramond-Ramond boundary conditions), there are $2N_f$ Majorana zero-modes, denoted as $\tilde{\chi}_\text{L,R}^A$.

When we discuss fermions, we should view the system with anti-periodic boundary conditions for both left- and right-movers (Neveu-Schwarz-Neveu-Schwarz boundary conditions) as the untwisted theory, and the system with periodic boundary conditions (Ramond-Ramond) as twisted by $(-1)^F$.\footnote{This is particularly clear in the context of CFTs.  Here, the operator-state correspondence maps the identity operator to the ground state of the theory with Neveu-Schwarz-Neveu-Schwarz boundary conditions. And the states with Ramond-Ramond boundary conditions correspond to spin fields.}
For this reason, when we discuss the system with periodic boundary conditions, the algebra is not the same as the untwisted algebra mentioned above.  To follow our notation, we should have written the symmetry elements as $\sR_{(-1)^F}$, $\sT_{(-1)^F}$, etc.  But in order not to clutter the equations, we will suppress this subscript. 

The effective theory of the zero modes is the same as the quantum mechanics in Appendix \ref{QMA} with $\nu= 2N_f$.  Following that appendix, we represented the fermions as
\ie\label{fermionRR}
&\tilde{\chi}_\text{L}^A=\left(\sigma^3\right)^{\otimes(j-1)} \otimes \sigma^1 \otimes \mathbf{1}^{\otimes(N_f-j)}\,, \qquad 1 \leq j \leq N_f\\
&\tilde{\chi}_\text{R}^A=\left(\sigma^3\right)^{\otimes(j-1)} \otimes \sigma^2 \otimes \mathbf{1}^{\otimes(N_f-j)}\,, \qquad 1 \leq j \leq N_f\,.
\fe

The action of the symmetry operators in the space of the zero modes is as follows. The chiral fermion parity $\sC$ acts as 
\ie 
\sC & = (\sigma^2 \otimes \sigma^1)^{\otimes k}, \quad \quad && N_f =2k,\\
\sC & = (\sigma^2 \otimes \sigma^1)^{\otimes k}\otimes \sigma^2, \quad \quad && N_f =2k+1.\\
\fe 
Reflection is 
\ie  
\sR & = \frac{1}{2^{\frac{N_f}{2}}}\left[(\sigma^1+\sigma^2) \otimes (\sigma^1-\sigma^2)\right]^{\otimes k} , \quad \quad && N_f =2k,\\
\sR & = \frac{1}{2^{\frac{N_f}{2}}}\left[(\sigma^1+\sigma^2) \otimes (\sigma^1-\sigma^2)\right]^{\otimes k}\otimes (\sigma^1+\sigma^2) , \quad \quad && N_f =2k+1.\\
\fe
Time-reversal is 
\ie  
\sT & = \frac{1}{2^{\frac{N_f}{2}}}\left[(1-i\sigma^3)\right]^{\otimes N_f} \mathcal{K}
\fe
And finally fermion parity is $(-1)^F = (-\sigma^3)^{\otimes N_f}$. 

They obey the algebra\footnote{This is the projective representation of the group $D_4\times \bZ_2$ and there are only 6  independent relations of the generator $a=\sC\sR$, $b=\sR$, and $c=\sR\sT$ (where $c$ is anti-unitary):
\ie 
& a^4 = (-1)^{N_f}\,,\quad b^2=1\,, \quad bab^{-1} = a^{-1}\,,\\
& c^2=(-1)^{\frac{N_f(N_f-1)}{2}}\,, \quad ca = ac\,, \quad cb =(-1)^{\frac{N_f(N_f-1)}{2}} bc \,.
\fe
And we also have $(-1)^F=i^{-N_f} a^2$. We see that the projective representation of $D_4$ uses a $\bZ_2$ central extension. Adding the anti-unitary $\bZ_2$ generated by $c$, the projective presentation becomes mod 4 in $N_f$.}
\ie  \label{RRallalg}
& [(-1)^F]^2=1,\quad \sC^2=1, \quad \sR^2=1,\quad (-1)^F\sC = (-1)^{N_f}\sC(-1)^F,  \\
& \sR(-1)^F = (-1)^{N_f}(-1)^F\sR,\quad \sR \sC = i^{N_f}\sC (-1)^F\sR,\\
& \sT^2=1, \quad \sT(-1)^F = (-1)^F\sT, \quad (\sT \sR)^2=(-1)^{\frac{N_f(N_f-1)}{2}}, \quad \sT \sC = i^{N_f^2}\sC (-1)^F \sT.    \fe 
This generalizes the continuum algebra in \cite{Seiberg:2023cdc} for $N_f=1$. 
(We can redefine fermion parity by a phase $i^{N_f}$, then all the projective phases   involving $(-1)^F,\sC,\sR$ are  signs.)

The algebra involving  $\Theta=\sC\sR\sT$ is\footnote{As we said above, we should view this system as twisted, and these $\Theta$ and $\sC$ are not the ones in the untwisted theory.  Therefore, the phase in the third equation does not contradict \eqref{ThetaUc} in the untwisted theory. Also, we can redefine $\sC \to i^{\frac{N_f(N_f-1)}{2}}\sC$ so that $\Theta \sC = \sC \Theta$ coincides with \eqref{ThetaUc}, but then $\sC^2=(-1)^{\frac{N_f(N_f-1)}{2}}$.}
\ie \label{app:CPTalg}
\Theta^2=1, \quad (-1)^F \Theta = \Theta(-1)^F, \quad \Theta \sC = (-1)^{\frac{N_f(N_f-1)}{2}}  \sC \Theta\,. 
\fe 
And the algebra involving $\sR\sT$ is 
\ie \label{app:PTalg}
(\sR\sT)^2=(-1)^{\frac{N_f(N_f-1)}{2}}, \quad (-1)^F (\sR\sT) = (-1)^{N_f}(\sR\sT)(-1)^F, \quad (\sR\sT) \sC = (-1)^{\frac{N_f(N_f-1)}{2}}  \sC (\sR\sT). 
\fe 

\subsection{Parity-violating boundary condition}\label{app:NSR}

With antiperiodic boundary conditions for the left-movers and the periodic boundary conditions for the right-movers (Neveu-Schwarz-Ramond boundary conditions), there are only $N_f$ zero modes from the right-movers.  In this case, reflection $\sR$ and time-reversal $\sT$ are broken. 
Chiral fermion parity $\sC$ acts trivially on the zero modes, and commutes with every other symmetry operator.

The representation of $(-1)^F$, $\sR\sT$ and $(-1)^F\sR\sT$ can be reduced to the quantum mechanics in Appendix \ref{QMA} with $\nu= N_f$.

When $N_f$ is even
\ie \label{contevenNfRT}
[(-1)^F]^2=1\,,\quad (\sR\sT)^2 = (-1)^{\frac{N_f(N_f-2)}{8}}\,,\quad (\sR\sT) (-1)^F = (-1)^{\frac{N_f}{2}} (-1)^F(\sR\sT)\,.
\fe 

When $N_f$ is odd, there is no $(-1)^F$, and we  have either $\sR\sT$ or $\sR\sT(-1)^F$, but not both
\ie \label{contoddNfRT}
&(\sR\sT)^2  = (-1)^{\frac{N_f-1}{4}} , \quad && N_f=4k+1,\\
&(\sR\sT(-1)^F)^2  = (-1)^{\frac{N_f+1}{4}} , \quad && N_f=4k+3.\\
\fe   

We present the algebra involving $\Theta=\sC\sR\sT$ as well. When $N_f$ is even,
\ie \label{contevenNfTheta}
[(-1)^F]^2=1\,,\quad \Theta^2 = (-1)^{\frac{N_f(N_f-2)}{8}}\,,\quad 
\Theta (-1)^F = (-1)^{\frac{N_f}{2}} (-1)^F\Theta\,. 
\fe 
When $N_f$ is odd, we have either $\sC\sR\sT$ or $\sC\sR\sT(-1)^F$, but not both
\ie \label{contoddNfTheta}
&\Theta^2  = (-1)^{\frac{N_f-1}{4}} \,, \quad &&N_f=4k+1\,,\\
&(\Theta(-1)^F)^2  = (-1)^{\frac{N_f+1}{4}} \,, \quad &&N_f=4k+3\,.\\
\fe

\section{Symmetries and anomalies of the Majorana chain}\label{appsyman}

We consider the Majorana chain of $L$ sites with $N_f$ Majorana fermions $\chi_\ell^A$ on each site, obeying the Clifford algebra $\left\{\chi_{\ell}^A, \chi_{\ell^{\prime}}^B\right\}=2 \delta_{\ell, \ell^{\prime}}\delta_{AB}$ and periodic boundary condition $\chi_{\ell+L}^A=\chi_\ell^A$. 
We focus on the gapless Hamiltonian in \eqref{HNf}. 
The global symmetries are generated by translation $T$, fermion parity $(-1)^F$ and spatial reflection $\R $, which are unitary, as well as time-reversal $\T$, which is anti-unitary. 
They can be expressed in terms of the local fermion operators and are realized projectively on the Hilbert space of dimension $2^{\lfloor \frac{L}{2}\rfloor}$.
Below, we discuss these projective phases and the corresponding anomalies.

\subsection{Even $L$}\label{app:evenL}
Below, we present the explicit expressions for the global symmetry operators for the Hamiltonian \eqref{HNf}, with even $L=2N$.

\begin{itemize}
\item Fermion parity 
\ie
(-1)^F = i^{NN_f} \prod_{A=1}^{N_f} \chi_1^A\chi_2^A\cdots \chi_\text{L}^A\,,
\fe
which acts on the fermions as
$(-1)^F \chi_\ell^A (-1)^F = -\chi_\ell^A$. 
\item Majorana translation $T = i^{\frac{N_f(N_f-1)}{2}} \left((-1)^F\right)^{N_f-1}
(\TRR_1 \TRR_2\cdots \TRR_{N_f})$, with
\ie
\TRR_A = \frac{e^{\frac{2 \pi i(N-1)}{8}}}{2^{\frac{2 N-1}{2}}}
\chi_0^A (1+\chi_0^A\chi_1^A)(1+\chi_1^A\chi_2^A)\cdots (1+\chi_{L-2}^A\chi_{L-1}^A)\,.
\fe
Since $T_A$ is a fermionic operator (i.e., $T_A(-1)^F= -(-1)^FT_A$), the factor $((-1)^F)^{N_f-1}$ is needed to ensure the action  $T\chi_\ell^A T^{-1} = \chi_{\ell+1}^A$. 
The phase $i^{N_f(N_f-1)\over2}$ is included so that $T^L=1$.\footnote{Note that $T$ is not simply a product of the translation operators of the individual Majorana chains. 
This is because the theory with periodic boundary conditions should really be viewed as the $(-1)^F$-twisted theory of the one with anti-periodic boundary conditions. 
In the continuum limit, this is related to the fact that the Ramond-Ramond theory should be viewed as a twisted theory as discussed in Appendix \ref{app:RR}. 
This is to be contrasted with the discussion in Section \ref{sec:introanomaly} for an internal global symmetry without any twist/defect. Similar comments apply to the reflection operator $\R$ below.}

\item Parity/reflection $\R  = i^{\frac{N_f(N_f-1)}{2}} \left((-1)^F\right)^{N_f-1} \prod_{A=1}^{N_f} \R_A$, with
\ie 
\R_A=&\frac{e^{\frac{2 \pi i N(N-1)}{8}}}{2^{\frac{N-1}{2}}}  \\ 
&\times \left\{\begin{array}{l}
\chi_0^A\left(\chi_1^A-\chi_{-1}^A\right)\left(\chi_2^A+\chi_{-2}^A\right) \cdots\left(\chi_{N-1}^A-\chi_{-N+1}^A\right) \chi_N^A\, , \, \text { even } N\\
\chi_0^A\left(\chi_1^A-\chi_{-1}^A\right)\left(\chi_2^A+\chi_{-2}^A\right) \cdots\left(\chi_{N-1}^A+\chi_{-N+1}^A\right)\,,\ \quad \,\text { odd } N\,,
\end{array}\right.
\fe
which acts as $\R\chi_\ell^A \R^{-1} = (-1)^\ell\chi_{-\ell}^A$.  Similar to the expression for $T$, here $((-1)^F)^{N_f-1}$ is included to ensure the action $\R\chi_\ell^A \R^{-1} = (-1)^\ell\chi_{-\ell}^A$ and the phase $i^{\frac{N_f(N_f-1)}{2}}$ is included so that $\R^2=1$.

\item Anti-unitary time-reversal $\T$ acts as $\T\chi_\ell^A \T^{-1} = (-1)^\ell\chi_{\ell}^A$. The explicit expression for $\T$ depends on the choice of the basis; there is a basis such that $\T$ is just the complex conjugate $\mathcal{K}$  \cite{Seiberg:2023cdc}.\footnote{ For $N_f=1$, the time-reversal operator $\T$ here differs from that in \cite{Seiberg:2023cdc} by $(-1)^F$.}
\end{itemize}

With the explicit expression of these operators, we derive the algebra between them:
\ie \label{evenLallalg}
& \left((-1)^F\right)^2=1, \quad \TRR^L=1, \quad \R^2=1, \quad (-1)^F \TRR=(-1)^{N_f}\TRR(-1)^F, \\
& (-1)^F \R=(-1)^{N_f}\R(-1)^F, \quad \R \TRR=i^{N_f} \TRR^{-1} (-1)^F\R \\
& \T^2=1, \quad (-1)^F \T = \T(-1)^F, \quad \T \R=(-1)^{\frac{N_f(N_f-1)}{2}}\, \R\T, \quad \T \TRR=i^{N_f^2} \TRR(-1)^F \T .
\fe
For $N_f=1$, the projective sign between $(-1)^F$ and $T$  was discussed in \cite{2015PhRvB..92w5123R,Hsieh:2016emq}, and this algebra  generalizes the result in \cite{Seiberg:2023cdc}.\footnote{See a related discussion in \cite{Li:2024dpq}.} 
(We can redefine the fermion parity by a phase $i^{N_f}(-1)^F$, then all the projective phases not involving time-reversal $\T$ are $(-1)^{N_f}$ and hence are modulo 2.) 

The algebra involving $\Pi=T\R\T$ (which is the lattice counterpart of $\Theta=\sC\sR\sT$) is
\ie \label{evenLPialg}
\Pi^2=1\,,~~~~(-1)^F \Pi  = \Pi (-1)^F\,,~~~~ 
\Pi \, T = (-1)^{\frac{N_f(N_f-1)}{2}}T^{-1} \,\Pi
\fe

The algebra involving the anomalous anti-unitary symmetry $\PP=\R\T$ is 
\ie\label{app:hTalgebra}
& \PP^2=(-1)^{\frac{N_f(N_f-1)}{2}}, ~~~~
(-1)^F \PP=(-1)^{N_f}\PP(-1)^F\,,~~~~
 \PP T = (-1)^{\frac{N_f(N_f-1)}{2}}T^{-1}\PP \,.
\fe 

The continuum limit of the theory based on the lattice Hamiltonian $H$ in \eqref{HNf} with even $L=2N$ is $N_f$ massless (non-chiral) Majorana fermions with the Lagrangian \eqref{LNf} and periodic boundary conditions for both the left- and the right-movers. 
Using the map of the symmetry operators from the lattice to the continuum
\ie
T \to e^{i\frac{2\pi}{L}P}\sC\,, \quad \R \to \sR\,, \quad \T \to \sT\, ,\label{eq:latcont}
\fe
the projective phases on the lattice \eqref{evenLallalg}, \eqref{evenLPialg}, and \eqref{app:hTalgebra} are matched with those in the continuum \eqref{RRallalg}, \eqref{app:CPTalg}, and \eqref{app:PTalg}, respectively. 
 
In Section \ref{sec:majRT}, we discussed the anomaly of $\PP=\R\T$, which can be seen from \eqref{app:hTalgebra}. Below, we will discuss the anomalies of various subgroups generated by $T, \R,\T$ (which become $\sC,\sR,\sT$ in the continuum) and the fermion parity $(-1)^F$. 
\begin{itemize}
\item Lattice translation $T$ and $(-1)^F$ have an order 2 anomaly \cite{Seiberg:2023cdc}. This is suggested from the projective sign in  $(-1)^FT  = (-1)^{N_f}T(-1)^F$. To demonstrate that the anomaly is indeed of order 2, we construct a local $T$-symmetric Hamiltonian for $N_f=2$ in a trivially gapped phase. This is given by
\ie\label{H1}
i \sum_{\ell=0}^{L-1} \chi^1_\ell \chi^2_\ell\,,
\fe
where we pair up the Majorana fermions between the two chains. The continuum limit of $T$ is a unitary $\mathbb{Z}_2$ symmetry generated by the chiral fermion parity $\sC$, which together with $(-1)^F$ has an order 8 anomaly classified by \cite{Kapustin:2014dxa}:
\ie
\text{Hom(}\, \Omega^\text{Spin}_3(B\mathbb{Z}_2),U(1))=\mathbb{Z}_8\,.
\fe
This order-8 anomaly is an example of an emergent anomaly.  It does not follow from the anomaly between $T$ and $(-1)^F$, which is of order 2.
\item The site-centered parity/reflection $\R$ and $(-1)^F$ have an order 2 anomaly, which is seen from the projective sign in $(-1)^F \R = (-1)^{N_f} \R(-1)^F$. For $N_f=2$, the same trivially gapped Hamiltonian \eqref{H1} is also $\R$-symmetric.  
The continuum counterparts of these symmetries, $\sR$ and $(-1)^F$, also have an order-2 anomaly classified by \cite{Kapustin:2014dxa}
\ie\label{pin+}
\text{Hom(}\, \Omega^{\text{Pin}^+}_3(pt),U(1))=\mathbb{Z}_2\,.
\fe
There is also a link-centered parity/reflection $\R' = T\R$, which acts as $\R' \chi_\ell \R^{'-1}  = (-1)^\ell \chi_{-\ell+1}$ and obeys $(\R')^2=i^{N_f}(-1)^F$.  This phase can be redefined away, and therefore, it does not signal an anomaly.  Indeed, the following $\R'$-symmetric Hamiltonian is trivially gapped for $N_f=1$:
\ie\label{app:latticem}
i\sum_{m=1}^{\frac L2} \chi_{2m-1} \chi_{2m}\,.
\fe
The continuum counterpart of $\R'$ is $\sC\sR$.  It is anomaly-free since the anomaly classification is trivial \cite{Kapustin:2014dxa}:
\ie\label{pin-}
\text{Hom(}\, \Omega^{\text{Pin}^-}_3(pt),U(1))=0\,.
\fe
\item Time-reversal $\T$ and $(-1)^F$ have no anomaly. 
To show this, we note that the trivially gapped Hamiltonian \eqref{app:latticem} is also $\T$-symmetric. As another check, there is no phase in $(-1)^F \T = \T(-1)^F$. The continuum counterpart of $\T$ is $\sT$, which is also anomaly-free because of \eqref{pin-}. In fact, since the continuum theory is relativistic, the lack of anomaly in $\sC\sR $ and the lack of anomaly in $\sT$ are related through the anomaly-free $\sC\sR\sT$.

We can also consider the symmetry generated by $T\T$ and $(-1)^F$. The relation $(T\T)(-1)^F = (-1)^{N_f}(-1)^F (T\T)$ points to a lower bound of 2 on the order of the anomaly.\footnote{The relation $(T\T)^L = (i^{N_f^2}(-1)^F)^{\frac{L}{2}}$ appears to suggest that the anomaly is at least of order 4. However, we can redefine the fermion parity operator as $i^{N_f^2}(-1)^F$, to turn this relation into $(T\T)^L = ((-1)^F)^{\frac{L}{2}}$ and $((-1)^F)^2=(-1)^{N_f}$, and the all the phases are signs. Hence, the anomaly is at least of order 2.} The upper bound is also 2, as can be seen by noting that the symmetric  Hamiltonian for $N_f=2$:
\ie\label{H2}
i\sum_{\ell=0}^{L-1} \chi_\ell^1 \chi_{\ell+1}^2\,
\fe
has a trivially gapped spectrum. 
The continuum counterpart is $\sC\sT$, which has a mod 2 anomaly classified by \eqref{pin+}.

\item $\Pi= T\R\T$ and $(-1)^F$ have no anomaly. 
Indeed, there is no anomalous phase in $\Pi^2=1$ and $\Pi(-1)^F = (-1)^F\Pi$.
Even for $N_f=1$, the same trivially gapped Hamiltonian \eqref{app:latticem} also preserves $\Pi$, which proves the absence of an anomaly.  
The continuum counterpart of $\Pi$ is $\sC\sR\sT$, which is anomaly-free.

\item The symmetry generated by $T$, $\R$, and $(-1)^F$ contains some anomalous subgroups discussed above. The order of the anomaly is 2 because the Hamiltonian \eqref{H1} for $N_f=2$ is both $T$ and $\R$ symmetric. In the continuum, the corresponding symmetry generated by $\sC,\sR,(-1)^F$ has a mod 8 anomaly classified by \cite{Kaidi:2019pzj,Kaidi:2019tyf}:
\ie\label{dpin}
\text{Hom(}\, \Omega^\text{DPin}_3(pt),U(1))=\mathbb{Z}_8\,.
\fe
This anomaly is an emergent anomaly.  Similarly, the symmetry generated by $T,\T$ and $(-1)^F$ has a mod 2 anomaly. Indeed, the trivially gapped Hamiltonian \eqref{H2} for $N_f=2$ is symmetric. 
The continuum counterpart of this symmetry is generated by $\sC$, $\sT$, and $(-1)^F$, with a mod 8 emergent anomaly classified again by \eqref{dpin}.

\end{itemize}

\subsection{Odd $L$}\label{app:oddL}

With odd $L=2N+1$, reflection and time-reversal, acting as \eqref{eq:Paction} and \eqref{eq:Taction}, are not well-defined. Therefore, we analyze the remaining symmetry generated by $\PP=\R\T$, $(-1)^F$, and $T$. 

Translation is represented as  
\ie  
\Todd = e^{i \pi(x N+y)} \prod_{A=1}^{N_f}\frac{e^{\frac{-2 \pi i(2 N+1)}{16}}}{2^N}\left(1-\chi_0^A \chi_1^A\right)\left(1-\chi_1^A \chi_2^A\right) \cdots\left(1-\chi_{2 N-1}^A \chi_{2 N}^A\right)
\fe 
where $x=-\frac{n_1}{2}, y=\frac{3 n_1}{4}+n_2$ and $n_{1,2}$ are integers. Hence, we find the algebra of translation 
\ie 
\Todd^L = \exp \left(\frac{2 \pi i n}{16}\right), \quad  n= 6 n_1+8 n_2-N_f \label{eq:Toddphase}
\fe 
independent of the system size $L$. 
Using local phase redefinition via different $n_{1,2}$, we can change the integer $n$ by 2, but we always have $n \equiv N_f \bmod 2$. 
When $N_f$ is even we can remove this phase by choosing $(n_1,n_2)= (-N_f/2,N_f/2)$, and we have $T^L=1$. 
When $N_f$ is odd, this phase can not be removed in a local way. Therefore, \eqref{eq:Toddphase} probes a modulo 2 anomaly in 1+1d \cite{Seiberg:2023cdc}.

Since $\PP$ leaves $\chi_0^A$ invariant and acts on $\frac{1}{\sqrt{2}}(\chi_\ell^A\pm \chi_{-\ell}^A ),\,\ell=1,\dots,N$ with positive and negative signs, the representation of $\PP$ and $(-1)^F$ can be reduced to the quantum mechanic analysis in Appendix \ref{QMA} with $\nu_+-\nu_-=N_f$. When $N_f$ is even,
\ie \label{latevenNfPP}
[(-1)^F]^2=1,\quad \PP^2 = (-1)^{\frac{N_f(N_f-2)}{8}},\quad \PP (-1)^F = (-1)^{\frac{N_f}{2}} (-1)^F\PP.
\fe 
When $N_f$ is odd, one can only have either $\PP$ or $\PP' = \PP (-1)^F$ and cannot have $(-1)^F$, 
\ie 
\PP^2 & = (-1)^{\frac{N_f-1}{4}} , \quad N_f=4k+1,\\
\PP'^2 & = (-1)^{\frac{N_f+1}{4}} , \quad N_f=4k+3.\\
\fe 
Finally we evaluate how $(-1)^F$, $\PP$ or $\PP'$ act on $T$ if they exist. When $N_f$ is even, 
\ie 
(-1)^F T = T (-1)^F,\quad \PP T =  T^{-1} \PP
\fe 
When $N_f$ is odd,  
\ie \label{latoddNfPPT}
\PP T &= T^{-1} \PP, \quad N_f=4k+1,\\
\PP' T &=  T^{-1} \PP', \quad N_f=4k+3.
\fe 
There are no phases between $T$ and $(-1)^F$, $\PP$ or $\PP'$.

Since $\PP$ and $\Pi$ are related by conjugation by a unitary operator, $T^{\frac{L-1}{2}}$, which commutes with $(-1)^F$, $\Pi$ obeys the same algebra as $\PP$.

In the continuum limit of $H$ with odd $L=2N+1$, the left and right movers obey anti-periodic and periodic boundary conditions, respectively. These projective phases on the lattice \eqref{latevenNfPP}--\eqref{latoddNfPPT} are matched with those in the continuum in \eqref{contevenNfRT}--\eqref{contoddNfTheta} through $\PP \to \sR\sT$ and $T\rightarrow \sC  $.

\subsection{A redundant deformation}\label{app:deformation}

Here we consider another interesting deformation of the gapless Majorana Hamiltonian $H$ \eqref{HNf} for even $L$.\footnote{For odd $L=2N+1$, we can consider $H_\text{NNN}'=i\sum_{A=1}^{N_f}\sum_{\ell=-N}^{N-2}(-1)^{\ell}\chi_\ell^A\chi_{\ell+2}^A$,  It breaks $T^r$ for any $r \neq 0 \bmod L$. But it is still odd under $\R$ and $\T$ while preserving $\PP$ and $(-1)^F$.\label{oddLprime}}  
It is the staggered version of next-to-nearest-neighbor deformation $H_\text{NNN}$ in \eqref{eq:HNNN}:
\ie \label{eq:HNNNm}
H_\text{NNN}'=i\sum_{A=1}^{N_f}\sum_{\ell=0}^{L-1}(-1)^{\ell}\chi_\ell^A\chi_{\ell+2}^A\,,
\fe 
which is odd under $T$, $\T$ and $\R$, while preserving only $\PP$ and $(-1)^F$. 
One might think that it flows in the continuum limit to the operator $i\sum_{A=1}^{N_f}\chi^A_{\text L}\partial_x\chi^A_{\text R}$, which is odd under $\sC$, $\sT$, and  $\sR$.  However, this operator is redundant in the sense that it vanishes on shell, i.e., by using the equations of motion. As a second attempt, one might try to argue that $H_\text{NNN}'$ flows to $i\sum_{A=1}^{N_f}\chi^A_{\text L}\partial_x^n\chi^A_{\text R}$ for some odd $n$.  But this operator is also redundant.  Below, we will see the lattice counterpart of this continuum redundancy.

To see the effect of the deformations by $H_\text{NNN}$ and $H'_\text{NNN}$, let us focus on $N_f=1$ and even $L=2N$, and consider the Hamiltonian
\ie\label{HNNNs} 
H+\lambda  H_\text{NNN} + \lambda'H_\text{NNN}'\,.
\fe 
For generic $\lambda,\lambda'$, it preserves the symmetry generated by $(-1)^F$, $T^2$, and $\PP$.\footnote{Since the model is quadratic in the fermions, it has infinitely many other symmetries that are not strictly locality preserving. We will discuss one such example below.} 

It is straightforward to diagonalize the Hamiltonian \eqref{HNNNs}.  Even though $T$ is not a symmetry of the Hamiltonian, we go to momentum space and write 
\ie 
&\chi_{\ell}=\frac{1}{\sqrt{N}} \sum_k \exp \left(\pi i \frac{\ell k}{N}\right) d_k\,, \qquad d_{k+2 N}=d_k\\
&d_k=d_{-k}^{\dagger}\,, \\
&\left\{d_k, d_{k^{\prime}}\right\}=\delta_{k,-k^{\prime}}\,.
\fe
In particular, $d_0$ and $ d_N$ are Hermitian. The Hamiltonian \eqref{HNNNs} is diagonalized as 
\ie \label{eq:HNNNsE}
\sum_{k=1}^{\lfloor \frac{N}{2}\rfloor}\left( E_+(k)\psi_{k,+}^\dagger\psi_{k,+} + E_-(k)\psi_{-k,-}^\dagger\psi_{-k,-}\right)+ \text{constant,}
\fe 
with the eigenmodes
\ie\label{psipmd} 
&\psi_{k,+}= \cos\frac{\theta_k}{2} d_{-k} + \sin \frac{\theta_k}{2} d_{-k+N}\,,\quad &&\psi_{-k,-}= -\sin\frac{\theta_k}{2} d_{k} + \cos \frac{\theta_k}{2} d_{k+N}\,, \\
&\psi_{k,+}^\dagger=\psi_{-k,+}\,,\quad &&\psi_{-k,-}^\dagger=\psi_{k,-}\,,\\
&\tan\theta_k = \lambda' \cos(\pi k/N)\,.
\fe
and the energy spectrum
\ie \label{eq:spectrum}
E_\pm (k) = 2\sin\left(\frac{\pi k}{N}\right)\left[\sqrt{1+\lambda'^2\cos^2\left(\frac{\pi k}{N}\right)} \pm \lambda \cos\left(\frac{\pi k}{N}\right)\right]\,. 
\fe
For small  $\lambda,\lambda'$, there are two Majorana zero modes, which generate a two-dimensional space of ground states. For larger $\lambda, \lambda'$, there are phase transitions to phases with several species of massless fermions.

As a check, the Hamiltonian \eqref{HNNNs} is not invariant under $T$, but it is invariant under $T^2$.  Therefore, the mixing in \eqref{psipmd} is only between momentum $k$ and momentum $k+N\bmod 2N$.

In the continuum limit of large $L=2N$, we focus on the low-lying modes
\ie \label{eq:NNNcont} 
H\approx \frac{2\pi}{N}\sum_{k=1}^{N_0} \left(\sqrt{1+\lambda'^2}+\lambda\right)k\psi_{k,+}^\dagger\psi_{k,+} + \frac{2\pi}{N}\sum_{k=1}^{N_0} \left(\sqrt{1+\lambda'^2}-\lambda \right)k\psi_{-k,-}^\dagger\psi_{-k,-}\,,
\fe 
where we used a UV cutoff $N_0$ with $1\ll N_0\ll N$. 

The continuum limit of the theory \eqref{eq:NNNcont} is gapless with left- and right-moving fermions. However, their velocities are different, and therefore it is not a Lorentzian invariant field theory. 
The lattice $\PP=\R\T$ becomes the continuum $\sR\sT$.   In addition, the low-energy theory is invariant under a chiral fermion parity $\sC$, which acts as $\pm 1$ on $\psi_{k,\pm}$ and consequently, under $\Theta=\sC\sR\sT$. Below, we will discuss the action of these continuum symmetries in more detail and will relate them to exact lattice symmetries.

The low-energy Hamiltonian \eqref{eq:NNNcont} shows that this model is gapless.  So far, we discussed the even $L$ version of the model.  In footnote \ref{oddLprime} we mentioned an odd $L$ version of it. Since it differs from the even-$L$ one only by local terms, it remains gapless as well. This is consistent with the constraint from the anomaly of $\PP$ and $(-1)^F$ in Section \ref{sec:majRT}.

The deformation by $H_\text{NNN}$, which flows to $T_{++}-T_{--}$, changes the spectrum at leading order in $\lambda $.  This is to be contrasted with the deformation by $H_\text{NNN}'$, which affects the spectrum only at second order in $\lambda'$.  This signals the fact that $H_\text{NNN}'$ leads to a redundant operator, and its leading order effect can be absorbed in the redefinition \eqref{psipmd}. Also, since the spectrum is quadratic in $\lambda'$, even though $H_\text{NNN}'$ is odd under $\Pi$  and $T$, it does not  lead to breaking of these symmetries.  This is related to the redundancy of the continuum limit of $H_\text{NNN}'$.

Let us discuss it in more detail. We will show that although \eqref{HNNNs} is not $T$-invariant, it is still symmetric under another deformed nonlocal lattice translation operator $T(\lambda')$. We construct this operator through the momentum space operators \eqref{eq:HNNNsE}. We define
\ie \label{eq:Tlambdadef}
&T(\lambda') \psi_{k,+} T(\lambda')^{-1}
= e^{-{i\pi k \over N} }\psi_{k,+}\,,\\
&T(\lambda') \psi_{k,-} T(\lambda')^{-1}
=- e^{-{i\pi k \over N} }\psi_{k,-}\,.
\fe
This operator leaves the Hamiltonian invariant for any $\lambda'$ and is of order $L=2N$. Using \eqref{eq:Tlambdadef}, we find the action of $T(\lambda')$  on $d_k$ 
\ie \label{eq:Tlambda}
T(\lambda' ) d_k T(\lambda' )^{-1}= e^{\frac{i\pi k}{N}}\left(\cos\theta_k d_k+ \sin\theta_k d_{k+N}\right)\,.
\fe
Through Fourier transformation, it acts on $\chi_\ell$ as 
\ie 
T(\lambda') \chi_{\ell} T(\lambda')^{-1}& = \frac{1}{2N}\sum_{\ell'}\left(\sum_k e^{i\frac{\pi k(\ell-\ell'+1)}{N}}\cos\theta_k+(-1)^{\ell'}\sum_k e^{i\frac{\pi k(\ell-\ell'+1)}{N}}\sin\theta_k\right)\chi_{\ell'}\,.
\fe 
We see that $T(\lambda')$ does not implement a locality-preserving transformation, i.e., it does not map a local operator with finite support to another local operator. Specifically, for small $\lambda'$,
\ie 
T(\lambda') \chi_{\ell} T(\lambda')^{-1} = & \chi_{\ell+1} + \frac{\lambda'}{2} (-1)^{\ell}(\chi_{\ell}+\chi_{\ell+2}) - \frac{\lambda'^2}{8} \left(2\chi_{\ell+1}+\chi_{\ell+3}+\chi_{\ell-2}\right)+O(\lambda'^3), 
\fe 
and therefore, at higher orders in $\lambda'$, $\chi_\ell$ is mapped to a linear combination of $\chi_{\ell'}$ with progressively wider range of $\ell'$. 

As a check, $T(0)=T$ is the original translation operator before the deformation. And as another check, $T(\lambda')^2=T^2$ is independent of $\lambda'$, which is consistent with that the deformation is invariant under the original $T^2$.

Finally, let us comment on the continuum limit \eqref{eq:NNNcont} of small $k$.  In this limit, $T(\lambda')$ acts as $+1$ and $-1$ on $\psi_{k,+}$ and $\psi_{k,-}$, and becomes the deformed chiral fermion parity $\sC(\lambda')$ in the continuum. Also, we can deform $\Pi(\lambda')=T(\lambda')\R\T$, which becomes $\sC(\lambda')\sR\sT$ in the continuum.

\section{Continuum Majorana zero modes on interfaces}\label{MZM}

Here, we consider a free  Majorana fermion in the continuum with the space-varying mass deformation discussed in Section \ref{SmithFermions}.  The resulting interfaces are associated with localized fermion zero modes.

The Lagrangian  is
\ie
{\cal L} = {i\over 2}\chi_\text{L} (\partial_t -\partial _x )\chi_\text{L}
+ {i\over 2}\chi_\text{R} (\partial_t +\partial _x )\chi_\text{R}
+ im(x) \chi_\text{L}\chi_\text{R}
\,.
\fe
We assume the space is a circle with $x$ identified with  $ x+2\pi$. 
The anti-unitary symmetry $\PP=\R\T$ acts on the fermions as
\ie
\PP \chi_\text{L} (t,x) \PP^{-1} =   \chi_\text{L} (-t,-x)\,,~~~~
\PP \chi_\text{R} (t,x) \PP^{-1} =   \chi_\text{R} (-t,-x)\,.
\fe
Therefore, the mass term is $\PP$-symmetric if $m(x) = m(-x)$. 

\subsection{Parity-preserving boundary condition}\label{app:RRDW}

We start by imposing periodic boundary conditions for both the left- and the right-movers. If we lift the spatial circle $S^1$ to the covering space $\mathbb{R}$, this means
\ie
\chi_\text{L} (t,x+2\pi )=\chi_\text{L} (t,x)\,,~~~
\chi_\text{R} (t,x+2\pi )=\chi_\text{R} (t,x)\,.
\fe

For this boundary condition, we choose our mass to be 
\ie\label{app:mass}
m(x) =\begin{cases}
-M \,,~~~~\text{for}~~~ -\pi <x < 0\,,\\
+M \,,~~~~\text{for}~~~ 0 <x <\pi\,,
\end{cases}
\fe
with positive $M$. 
In the covering space, we extend $m(x)$ periodically, i.e., $m(x+2\pi)=m(x)$. 
This is the continuum limit of the lattice Hamiltonian in \eqref{HDW}.

Denote the zero modes for $\chi_\text{L}(t,x),\chi_\text{R}(t,x)$ by $L(x),R(x)$, respectively. 
The zero modes obey the following equations:
\ie
&L'(x) =-M  R(x) \,,~~~R'(x) = -ML(x) \,,~~~~ && -\pi <x< 0\,,\\
&L'(x) =+M R(x) \,,~~~R'(x) = +M L(x) \,,~~~~ &&0 <x\le \pi \,.
\fe
There are two linearly independent solutions. For $|x|<\pi$, they are
\ie\label{LRMd}
&L_0 ( x ) = e^{-M  |x|}\,,~~~
&&R_0 ( x ) = - e^{- M  |x|}\,,\\
&L_\pi ( x ) = e^{-M (\pi - |x|)}\,,~~~
&&R_\pi ( x ) = e^{-M (\pi -|x|)}\,.
\fe
And they are extended periodically in $x$ beyond that range.   The first solution $(L_0(x),R_0(x))$ is localized at $x=0$ and is exponentially small in $M$ at $x=\pi$. This corresponds to a localized Majorana zero mode at $x=0$. Similarly, the second solution $(L_\pi(x),R_\pi(x))$ corresponds to a localized Majorana zero mode at $x=\pi$. 

We can also choose a smooth function, such as $m(x) = M\sin(x)$ with positive $M$. Then the two solutions are
\ie\label{LRMc}
&L_0 ( x ) = e^{M (-1+ \cos x)}\,,~~~
R_0 ( x ) = - e^{ M (-1+ \cos x)}\,,\\
&L_\pi ( x ) = e^{-M (1+\cos x)}\,,~~~
R_\pi ( x ) = e^{-M (1+\cos x)}\,.
\fe

To conclude, with parity-preserving boundary conditions, we find two Majorana zero-modes, localized at $x=0$ and at $x=\pi$. This is the continuum version of the even $L$ case in Section \ref{SmithFermions}.

\subsection{Parity-violating boundary condition}\label{app:NSRDW}

Next, we impose anti-periodic and periodic boundary conditions for the left- and the right-movers, respectively. If we lift the spatial circle $S^1$ to the covering space $\mathbb{R}$, this means\footnote{One way to think about it is to work on the two-fold cover $-2\pi<x\le 2\pi$ with  $\chi_\text{L}(x+4\pi)=\chi_\text{L}(x)$, $\chi_\text{R}(x+4\pi)=\chi_\text{R}(x)$, and $m(x+2\pi)=-m(x)$. Then, we can restrict to configurations satisfying \eqref{Tboundarc}.} 
\ie\label{Tboundarc}
\chi_\text{L} (t,x+2\pi )=-\chi_\text{L} (t,x)\,,~~~
\chi_\text{R} (t,x+2\pi )=\chi_\text{R} (t,x)\,.
\fe

In the fundamental domain $-\pi<x\le \pi$, we again choose the same mass function as in \eqref{app:mass}, but
 we extend this function to the covering space by $m(x+2\pi ) = - m(x)$, so that the mass term $im(x)\chi_\text{L}\chi_\text{R}$ is compatible with the boundary condition. (This $m(x)$ is still discontinuous at $x=0$, but it is continuous at $x=\pi$.) 
This is the continuum limit of the lattice Hamiltonian in \eqref{HDW}. 

Now we solve for the zero modes. 
Before imposing the boundary conditions, the most general solution is a linear combination of $(L_0(x), R_0(x))$ and $(L_\pi(x), R_\pi(x))$, of \eqref{LRMd}, which are both even functions of $x$. 
The boundary condition $L(-\pi ) = - L(\pi)$ eliminates one solution and we find for $|x|<\pi$
\ie
L(x) = \sinh(M(\pi -|x|)) \,,~~~~R(x) = -\cosh(M(\pi - |x|))\,,
\fe
which is localized at $x=0$. 

We can also choose a smooth function, such as $m(x) = M\sin x$, for $-\pi <x \le \pi$. (The extension to $\bR$ is not smooth at $x=\pi$.) Then \eqref{LRMc} leads to a single solution 
\ie
&L ( x ) = \sinh(M (1+  \cos x))\,,~~~
R ( x ) = -\cosh(M (1+  \cos x))\,,~~~-\pi  < x\le \pi\,.
\fe

To conclude, with parity-violating boundary conditions, we find one Majorana zero-mode, localized near $x=0$. This is the continuum version of the odd $L$ case in Section \ref{SmithFermions}.

\section{More about CRT in the Heisenberg model and its continuum limit}\label{app:boson}

In this Appendix, we elaborate on the discussion in Section \ref{Spinchain} about the symmetry action in the anti-ferromagnetic Heisenberg model and its continuum limit.

\subsection{$O(3)$ sigma model}

Our continuum model is based on the Lagrangian \eqref{O3Lagi}
\ie\label{O3Lagia}
&{\cal L} ={f\over 2}\partial_\mu \vec n\cdot \partial^\mu \vec n + \theta \Gamma\,,~~~\vec n\cdot\vec n=1\\
&\Gamma ={\epsilon^{\mu\nu}\over 8\pi} \vec n\cdot (\partial_\mu \vec n\times \partial_\nu\vec n)\,,
\fe
with $f$ a relevant coupling constant.

For $\theta$ an integer multiple of $\pi$, the theory is invariant under charge-conjugation $\sC$, reflection $\sR $, and time-reversal $\sT$. They act as in \eqref{CRTonn}
\ie\label{CRTonna}
&\sC \,\vec n(t,x)\, \sC^{-1}= -\vec n(t,x) \,,~~~~
\sR \,\vec n(t,x)\,\sR^{-1} = \vec n(t,-x)\,,~~~~
\sT  \,\vec n(t,x)\,\sT^{-1} = -\vec n(-t,x)\,.
\fe
Using these, we also have 
\ie\label{CRTonGammaa}
&\sC \,\Gamma (t,x)\, \sC^{-1}= -\Gamma(t,x) \,,~~~~
\sR \,\Gamma(t,x)\,\sR^{-1} = -\Gamma(t,-x)\,,~~~~
\sT  \,\Gamma(t,x)\,\sT^{-1} = \Gamma(-t,x)\,.
\fe
These actions commute with the global $SO(3)$ symmetry, whose currents are
\ie
\vec J_\mu = f \vec n\times\partial_\mu \vec n\,.
\fe
They transform under $\sC$,  $\sR $, and $\sT$ as
\ie\label{CRTonJa} 
\sC \,\vec J_\mu (t,x)\, \sC^{-1}=J_\mu(t,x) \,,~~~~
&\sR \,\vec J_t(t,x)\,\sR^{-1} = \vec J_t(t,-x)\,,~~&&
\sT  \,\vec J_t(t,x)\,\sT^{-1} =- \vec J_t({-}t,x)\\
&\sR \,\vec J_x(t,x)\,\sR^{-1} = -\vec J_x(t,{-}x)\,,~~&&
\sT  \,\vec J_x(t,x)\,\sT^{-1} = \vec J_x(-t,x)\,.
\fe
As a check, comparing \eqref{CRTonna}, \eqref{CRTonGammaa}, and \eqref{CRTonJa}, we see that in addition to reversing the signs of $x$ and $t$, the operator $\Theta=\sC\sR\sT$ leads to a factor of $(-1)^s$ with $s$ the spin of the operator. (See Appendix \ref{CPTreview}.)

\subsection{$c=1$ compact boson}\label{compactboson}

For $\theta\in \pi(1+2{\mathbb Z})$, the model flows to the $su(2)_1$ WZW model, which is a special case of the $c=1$ compact boson CFT.  This family of models is described by a free compact boson $\phi \sim \phi+2\pi$.  It is common to use also the dual boson $\tilde \phi$ such that $\partial_t\phi=\partial_x \tilde \phi$ and $\partial_t\tilde\phi=\partial_x \phi$.\footnote{In the Condensed Matter literature, this model is referred to as the Luttinger liquid, and the common notation is that our $\phi$ is $\theta$ and our $\tilde \phi$ is $\varphi$. \label{Luttinger}}  Since these two bosons are not relatively local, expressions that involve both of them should be treated with care.

To find the relation between the operators constructed out of $\vec n$ and the operators constructed out of $\phi$ and $\tilde \phi$, we compare some operators in the low-energy spectrum.

The lowest dimension operator is the identity.  Next, we have four degenerate operators, all with dimensions $({1\over 4},{1\over 4})$.  They have the quantum numbers of $\vec n$  and $\Gamma$.  In terms of $\phi$ and $\tilde \phi$:
\ie\label{nphiphitd}
&n^1\sim\cos \phi\,,\quad n^2\sim\sin \phi \,,\quad n^3\sim \sin \tilde \phi\,,\quad \Gamma \sim \cos \tilde \phi\,.
\fe
The symbol $\sim$ reminds us that the classical fields $\vec n$ and $\Gamma$ are not the same as the quantum operators with the same quantum numbers.\footnote{In particular, even though in the nonlinear sigma model \eqref{O3Lagia}, $\vec n^2=1$, this is not the case here.  Also, we cannot use trigonometric identities to simplify $(n^1)^2+(n^2)^2$ or $(n^3)^2+\Gamma^2$.}

At the next level, we find the currents, whose dimensions are $(1,0)$ and $(0,1)$.  We can study the following linear combinations of them 
\ie\label{currents continuum}
&J^1_t=-{2\over 2\pi }\sin\phi\sin \tilde \phi\,,\quad 
&&J^2_t={2\over 2\pi }\cos\phi\sin \tilde \phi\,,\quad
&&J^3_t={1\over2 \pi }\partial_t\phi={1\over2 \pi }\partial_x\tilde \phi\\
&J^1_x={2\over 2\pi }\cos\phi\cos \tilde \phi\,,\quad
&&J^2_x={2\over 2\pi }\sin\phi\cos \tilde \phi\,,\quad
&&J^3_x={1\over 2\pi }\partial_x\phi={1\over2 \pi }\partial_t\tilde \phi\,.
\fe

A simple consistency check of the expressions \eqref{currents continuum} is the following. In the conformal field theory, the currents $\vec J_\mu$ can be found in the operator product expansion of the operators $\vec n$ and $\Gamma$ of \eqref{nphiphitd}.  Ignoring the needed regularization, which will be important below, we can write 
\ie\label{nnto J}
&J^1_t\sim n^2 n^3 \,,\quad &&J^2_t\sim -n^1n^3\,,\\
&J^1_x\sim -n^1\Gamma\,,\quad
&&J^2_x\sim-n^2\Gamma \,.
\fe
Expressing the right-hand-side using $\phi$ and $\tilde \phi$ as in \eqref{nphiphitd}, reproduces the $\phi$ and $\tilde \phi$ dependence in \eqref{currents continuum}.

Next, we write the action of the continuum discrete symmetries on these free boson fields:
\ie\label{phiphittf}
&\sC \phi(x,t) \sC^{-1} =\phi(x,t)+\pi\,,\quad 
&&\sC \tilde \phi(x,t) \sC^{-1}= \tilde \phi(x,t)+\pi\\
&\sR\phi(x,t) \sR^{-1}= \phi(-x,t) \,,\quad 
&&\sR \tilde \phi(x,t)\sR^{-1}= -\tilde \phi(-x,t)+\pi\\
&\sT\phi(x,t) \sT^{-1} =\phi(x,-t) +\pi \,,\quad
&&\sT\tilde \phi(x,t)\sT^{-1}= -\tilde \phi(x,-t)\,.
\fe
Similar expressions appeared in various places, including \cite{Fuji:2014ila,Alavirad:2019iea,Yao:2023bnj}.  Note that $\Theta=\sC\sR\sT$ reflects $x$ and $t$, but other than that, it does not affect the fields. 

These actions on $\vec n$ and $\Gamma$ in terms of $\phi$ and $\tilde \phi$ \eqref{nphiphitd} reproduce \eqref{CRTonna} and \eqref{CRTonGammaa}.  Similarly, these actions on $J_\mu^3$ in terms of $\phi$ and $\tilde \phi$ \eqref{currents continuum} reproduce \eqref{CRTonJa}.  

The actions on $J_\mu^{1,2}$ in \eqref{currents continuum} are more subtle because they depend both on $\phi$ and $\tilde \phi$, which are not relatively local.  Therefore, these operators have to be defined carefully. As in \eqref{nnto J}, we can view these operators as a result of fusing two operators with dimensions $({1\over 4},{1\over 4})$ into operators with dimension $(1,0)$ and $(0,1)$.
Such a product has to be regularized.  If we use point-splitting in time, i.e., one operator is at $t$ and the other at time $t+\epsilon$, the product of the operators includes a factor of ${\epsilon\over |\epsilon|}$.  Therefore, the action of $\sT$ includes an additional minus sign beyond the transformation of the fields in \eqref{phiphittf}.\footnote{Another way to understand this minus sign is the following.  The conformal field theory has a global $SO(4)$ symmetry, under which $(\vec n, \Gamma)$ are in the vector representation.  The six currents are in an antisymmetric product of these vectors \eqref{nnto J}.  Therefore, the action of time-reversal $\sT$, which changes the order of the operators, leads to an additional minus sign.}  With this additional sign, the actions in \eqref{phiphittf} on $J_\mu^{1,2}$ in \eqref{currents continuum} reproduce \eqref{CRTonJa}.\footnote{We thank Max Metlitski for a useful discussion about this point.}

\subsection{Comparing with the lattice}

Here we compare the lattice Heisenberg model with the continuum theory discussed above. The lattice model is based on a one-dimensional periodic lattice with sites labeled by $j \sim j+L$ and the 
Hamiltonian \eqref{Heisenberg}
\ie\label{HeisenbergA}
H = \sum_{j=1}^L \vec S_j\cdot \vec S_{j+1}\,,
\fe 
where $\vec S_j$ are given by the Pauli matrices at the site $j$, $\vec S_j = \frac 12( X_j, Y_j,Z_j)$. 

The Hamiltonian \eqref{HeisenbergA} is in an anti-ferromagnetic phase.  In that phase, the lattice operators $\vec S_j$ do not have a smooth continuum limit.  Instead, $(-1)^j\vec S_j$ is smooth.  It is natural to identify the continuum limit as \cite{Affleck:1988nt,Affleck:1988zj}
\ie\label{sigmaSl}
(-1)^j\vec S_j \to \vec n\,.
\fe
As we said above, the operator on the right-hand side is not the same as the classical field $\vec n$.  It is a quantum operator with the same quantum numbers.  Also, this relation is up to an infinite renormalization factor.
We also have \cite{Affleck:1988nt,Affleck:1988zj}
\ie\label{sigmaGammal}
(-1)^j\vec S_j \cdot \vec S_ {j+1}\to \Gamma\,.
\fe
Again, the limit is up to an infinite renormalization factor.  These continuum operators $\vec n$ and $\Gamma$ can be expressed in terms of the free fields $\phi$ and $\tilde \phi$ as in \eqref{nphiphitd}. 

Importantly, the action of the lattice symmetry operators:
\ie\label{latticesymea}
&T \vec S_j T^{-1}=\vec S_{j+1}\,,\quad \R \vec S\R^{-1} =\vec S_{-j}\,,\quad
\T \vec S_j\T^{-1} =-\vec S_{j}
\fe
is compatible with the continuum limit \eqref{sigmaSl} and \eqref{sigmaGammal} and the action of the continuum operators \eqref{CRTonna} and \eqref{CRTonGammaa} provided
\ie\label{latticeconto}
&T\to \sC\,,\quad \R \to \sR\,\quad \T \to \sT\,.
\fe

Even though the $SO(3)$ global symmetry of these models is not essential to our main point, we would like to examine it.

It is straightforward to find
\ie\label{eq:latcontinuity}
{d\over dt}\vec S_j=i\left[H,\vec S_j\right]= \vec S_{j-1}\times \vec S_j-\vec S_j\times \vec S_{j+1}\,.
\fe
This equation can be interpreted as a lattice conservation equation for the $SO(3)$ current, whose time component is $\vec S_j$ and its spatial component is $-\vec S_{j-1}\times \vec S_j$. 
For our anti-ferromagnetic system, the operators that have a good continuum limit are linear combinations of these 
\ie\label{Latticecurrentsc}
&\vec S_j+\vec S_{j+1} \to \vec J_{t}\\
&-\vec S_{j-1}\times \vec S_{j}-\vec S_{j}\times \vec S_{j+1} \to \vec J_x\,.
\fe
Again, these limits need an infinite renormalization constant.

Given these lattice expressions for the currents, we can use \eqref{latticesymea} to find how they transform under our discrete symmetries.  As a check, this symmetry action is consistent with the continuum action \eqref{CRTonJa} with the identification between the lattice and the continuum symmetry operators \eqref{latticeconto}.

\subsection{Anomalies involving reflection and time-reversal on the spin chain}\label{app:spinanomaly}

In Section \ref{sec:spinanomaly}, we discussed an anomaly of $\PP = \R\T$, and an anomaly between lattice translation $T$ and $\Pi=T\R\T$. 
Here we discuss other anomalies involving lattice translation $T$, reflection $\R$, and time-reversal $\T$.

Lattice translation $T$ by itself is anomaly-free. To see that, note that the paramagnetic Hamiltonian 
\ie\label{para}
H_\text{paramagnetic} = \sum_j X_j\,.
\fe
is trivially gapped and $T$-invariant.

The lattice CRT operator $\Pi$ is also anomaly-free. 
To see this, consider the  Hamiltonian 
\ie\label{dim}
H_\text{dimerized} = \sum_{m=0}^{\frac L2-1} \vec S_{2m} \cdot \vec S_{2m+1} \,.
\fe
It is invariant under $T^2$, but not under $T$. 
Each local term is related to the total spin   $S_\text{tot}$ of the pair $(2m,2m+1)$ as $\vec S_{2m} \cdot \vec S_{2m+1} = \frac 12 S_\text{tot}(S_\text{tot}+1) -\frac 34$. Hence, $\vec S_{2m} \cdot \vec S_{2m+1}$ takes values in $-\frac34$ and $\frac 14$ for the singlet and triplet, respectively. 
The ground state of this Hamiltonian is the tensor product of the singlets from each pair $(2m,2m+1)$. 
Therefore, the ground state is non-degenerate with an order-one gap.  This tells us that the symmetry generated by $T^2$ and $\Pi$ is anomaly-free.

Next, we discuss some other anomalies involving $T$, $\R$, and $\T$. 
We first note that the lattice reflection $\R$ and time-reversal $\T$ are separately free of anomalies. 
Indeed, the trivially gapped Hamiltonians \eqref{para} and \eqref{dim} are respectively $\R$- and $\T$-symmetric. 
In  1+1d bosonic continuum field theory,  the anomalies of reflection and time-reversal  are both classified by the cobordism group \cite{Kapustin:2014tfa} 
\ie
\text{Hom}(\Omega_3^\text{O}(pt),U(1))=0\,,
\fe
which is trivial. This follows from the fact that on all closed 3-manifolds, $w_3=w_1w_2=w_1^3=0$, where $w_i$ is the $i$-th Stiefel-Whitney class of the tangent bundle of spacetime. Hence, there is no nontrivial 2+1d  bosonic SPT for time-reversal symmetries. (See also \cite{Chen:2011pg} for the classification in terms of the group cohomology.) Therefore, the corresponding continuum symmetries $\sR$ and $\sT$ are also separately anomaly-free.  

The symmetry generated by translation $T$ and time-reversal $\T$ has a mod 2 anomaly \cite{Chen:2010zpc,2015PNAS..11214551W,Ogata:2020hry,Yao:2023bnj} as can be seen from the operator algebra \eqref{RTalgebra}. The corresponding anomaly in the continuum between $\sC$ and $\sT$ was discussed in \cite{Sulejmanpasic:2018upi}. It is captured by the 2+1d SPT action $\pi\int w_1 ^2A$, where $A$ is the background $\mathbb{Z}_2$ gauge field for $\sC$. 
There is a more subtle mod 2 anomaly between reflection $\R$ and time-reversal $\T$  \cite{Ogata:2020hry}. It is not captured by any projective phases in the algebra. 
On the other hand, the symmetry generated by translation $T$ and reflection $\R$ is anomaly-free, since the paramagnetic Hamiltonian \eqref{para} preserves these symmetries. 
The lattice symmetry generated by $T$, $\R$, and $\T$ has a mod 2 anomaly as it contains $T$, and $\Pi$, or more simply, because it contains $\PP$. This lattice anomaly matches with the anomaly of $\sC$, $\sR$, and $\sT$ in the continuum discussed in \cite{Sulejmanpasic:2018upi}.

We summarize some of these anomalies on the lattice and in the continuum in Table \ref{tab:spin}.

\bibliographystyle{JHEP}
\bibliography{That}
\end{document}